\documentclass[11pt,english]{article}
\usepackage{ae,aecompl}
\usepackage[T1]{fontenc}
\usepackage[latin9]{inputenc}
\usepackage{color}
\usepackage{babel}
\usepackage{array}
\usepackage{verbatim}
\usepackage{float}
\usepackage{bm}
\usepackage{amsmath}
\usepackage{amssymb}
\usepackage{graphicx}
\usepackage{breakurl}
\usepackage{hyperref}

\makeatletter

\providecommand{\tabularnewline}{\\}

\@ifundefined{definecolor}{\@ifundefined{definecolor}
 {\usepackage{color}}{}
}{}
\@ifundefined{definecolor}{\@ifundefined{definecolor}
 {\usepackage{color}}{}
}{}
\@ifundefined{definecolor}{\@ifundefined{definecolor}
 {\@ifundefined{definecolor}
 {\@ifundefined{definecolor}
 {\usepackage{color}}{}
}{}
}{}
}{}
\@ifundefined{definecolor}{\@ifundefined{definecolor}
 {\@ifundefined{definecolor}
 {\@ifundefined{definecolor}
 {\@ifundefined{definecolor}
 {\usepackage{color}}{}
}{}
}{}
}{}
}{}
\@ifundefined{definecolor}{\@ifundefined{definecolor}
 {\@ifundefined{definecolor}
 {\@ifundefined{definecolor}
 {\@ifundefined{definecolor}
 {\@ifundefined{definecolor}
 {\usepackage{color}}{}
}{}
}{}
}{}
}{}
}{}
\@ifundefined{definecolor}{\@ifundefined{definecolor}
 {\@ifundefined{definecolor}
 {\@ifundefined{definecolor}
 {\@ifundefined{definecolor}
 {\@ifundefined{definecolor}
 {\@ifundefined{definecolor}
 {\usepackage{color}}{}
}{}
}{}
}{}
}{}
}{}
}{}
\@ifundefined{definecolor}{\@ifundefined{definecolor}
 {\@ifundefined{definecolor}
 {\@ifundefined{definecolor}
 {\@ifundefined{definecolor}
 {\@ifundefined{definecolor}
 {\@ifundefined{definecolor}
 {\@ifundefined{definecolor}
 {\usepackage{color}}{}
}{}
}{}
}{}
}{}
}{}
}{}
}{}
\@ifundefined{definecolor}{\@ifundefined{definecolor}
 {\@ifundefined{definecolor}
 {\@ifundefined{definecolor}
 {\@ifundefined{definecolor}
 {\@ifundefined{definecolor}
 {\@ifundefined{definecolor}
 {\@ifundefined{definecolor}
 {\@ifundefined{definecolor}
 {\usepackage{color}}{}
}{}
}{}
}{}
}{}
}{}
}{}
}{}
}{}
\@ifundefined{definecolor}{\@ifundefined{definecolor}
 {\@ifundefined{definecolor}
 {\@ifundefined{definecolor}
 {\@ifundefined{definecolor}
 {\@ifundefined{definecolor}
 {\@ifundefined{definecolor}
 {\@ifundefined{definecolor}
 {\@ifundefined{definecolor}
 {\@ifundefined{definecolor}
 {\usepackage{color}}{}
}{}
}{}
}{}
}{}
}{}
}{}
}{}
}{}
}{}
\@ifundefined{definecolor}{\@ifundefined{definecolor}
 {\@ifundefined{definecolor}
 {\@ifundefined{definecolor}
 {\@ifundefined{definecolor}
 {\@ifundefined{definecolor}
 {\@ifundefined{definecolor}
 {\@ifundefined{definecolor}
 {\@ifundefined{definecolor}
 {\@ifundefined{definecolor}
 {\@ifundefined{definecolor}
 {\usepackage{color}}{}
}{}
}{}
}{}
}{}
}{}
}{}
}{}
}{}
}{}
}{}
\@ifundefined{definecolor}{\@ifundefined{definecolor}
 {\@ifundefined{definecolor}
 {\@ifundefined{definecolor}
 {\@ifundefined{definecolor}
 {\@ifundefined{definecolor}
 {\@ifundefined{definecolor}
 {\@ifundefined{definecolor}
 {\@ifundefined{definecolor}
 {\@ifundefined{definecolor}
 {\@ifundefined{definecolor}
 {\@ifundefined{definecolor}
 {\usepackage{color}}{}
}{}
}{}
}{}
}{}
}{}
}{}
}{}
}{}
}{}
}{}
}{}
\@ifundefined{definecolor}{\@ifundefined{definecolor}
 {\@ifundefined{definecolor}
 {\@ifundefined{definecolor}
 {\@ifundefined{definecolor}
 {\@ifundefined{definecolor}
 {\@ifundefined{definecolor}
 {\@ifundefined{definecolor}
 {\@ifundefined{definecolor}
 {\@ifundefined{definecolor}
 {\@ifundefined{definecolor}
 {\@ifundefined{definecolor}
 {\usepackage{color}}{}
}{}
}{}
}{}
}{}
}{}
}{}
}{}
}{}
}{}
}{}
}{}
\@ifundefined{definecolor}{\@ifundefined{definecolor}
 {\@ifundefined{definecolor}
 {\@ifundefined{definecolor}
 {\@ifundefined{definecolor}
 {\@ifundefined{definecolor}
 {\@ifundefined{definecolor}
 {\@ifundefined{definecolor}
 {\@ifundefined{definecolor}
 {\@ifundefined{definecolor}
 {\@ifundefined{definecolor}
 {\@ifundefined{definecolor}
 {\@ifundefined{definecolor}
 {\@ifundefined{definecolor}
 {\usepackage{color}}{}
}{}
}{}
}{}
}{}
}{}
}{}
}{}
}{}
}{}
}{}
}{}
}{}
}{}
\@ifundefined{definecolor}{\@ifundefined{definecolor}
 {\@ifundefined{definecolor}
 {\@ifundefined{definecolor}
 {\@ifundefined{definecolor}
 {\@ifundefined{definecolor}
 {\@ifundefined{definecolor}
 {\@ifundefined{definecolor}
 {\@ifundefined{definecolor}
 {\@ifundefined{definecolor}
 {\@ifundefined{definecolor}
 {\@ifundefined{definecolor}
 {\@ifundefined{definecolor}
 {\@ifundefined{definecolor}
 {\@ifundefined{definecolor}
 {\usepackage{color}}{}
}{}
}{}
}{}
}{}
}{}
}{}
}{}
}{}
}{}
}{}
}{}
}{}
}{}
}{}
\usepackage{bm}

\@ifundefined{definecolor}{\@ifundefined{definecolor}
 {\@ifundefined{definecolor}
 {\@ifundefined{definecolor}
 {\@ifundefined{definecolor}
 {\@ifundefined{definecolor}
 {\@ifundefined{definecolor}
 {\@ifundefined{definecolor}
 {\@ifundefined{definecolor}
 {\@ifundefined{definecolor}
 {\@ifundefined{definecolor}
 {\@ifundefined{definecolor}
 {\@ifundefined{definecolor}
 {\@ifundefined{definecolor}
 {\@ifundefined{definecolor}
 {\@ifundefined{definecolor}
 {\usepackage{color}}{}
}{}
}{}
}{}
}{}
}{}
}{}
}{}
}{}
}{}
}{}
}{}
}{}
}{}
}{}
}{}

\usepackage{bm}

\@ifundefined{definecolor}{\@ifundefined{definecolor}
 {\@ifundefined{definecolor}
 {\@ifundefined{definecolor}
 {\@ifundefined{definecolor}
 {\@ifundefined{definecolor}
 {\@ifundefined{definecolor}
 {\@ifundefined{definecolor}
 {\@ifundefined{definecolor}
 {\@ifundefined{definecolor}
 {\@ifundefined{definecolor}
 {\@ifundefined{definecolor}
 {\@ifundefined{definecolor}
 {\@ifundefined{definecolor}
 {\@ifundefined{definecolor}
 {\@ifundefined{definecolor}
 {\@ifundefined{definecolor}
 {\usepackage{color}}{}
}{}
}{}
}{}
}{}
}{}
}{}
}{}
}{}
}{}
}{}
}{}
}{}
}{}
}{}
}{}
}{}

\usepackage{amsfonts}

\usepackage{epsfig}

\usepackage{latexsym}

\def\spa#1{\phantom{\fbox{\rule[-#1cm]{0cm}{0cm}}}}

\def\lesssim{\mathrel{\hbox{\rlap{\hbox{\lower4pt\hbox{$\sim$}}}\hbox{$<$}}}}
\def\gtrsim{\mathrel{\hbox{\rlap{\hbox{\lower4pt\hbox{$\sim$}}}\hbox{$>$}}}}

\usepackage{aecompl}

\usepackage{babel}

\usepackage{babel}

\usepackage{babel}

\usepackage{babel}

\usepackage{babel}

\usepackage{babel}

\usepackage{babel}

\usepackage{babel}

\usepackage{babel}

\usepackage{babel}

\usepackage{babel}

\usepackage{babel}

\usepackage{babel}

\usepackage{babel}

\usepackage{babel}

\usepackage{babel}

\makeatother

\begin{document}

\title{\textbf{{Gravito-electromagnetic analogies}}}

\author{L. Filipe O. Costa$^{1,2}$%
\thanks{lfpocosta@math.ist.utl.pt%
}, José Natário$^{2}$%
\thanks{jnatar@math.ist.utl.pt%
} \\
 \\
 {\em $^{1}$Centro de Física do Porto, Universidade do Porto}
\\
 {\em Rua do Campo Alegre, 687, 4169-007 Porto, Portugal}\\
 {\em $^{2}$Centro de Análise Matemática, Geometria e Sistemas
Dinâmicos,} \\
 {\em Instituto Superior Técnico, 1049-001, Lisboa, Portugal}\\
 }

\date{\today}
\maketitle
\begin{abstract}
We reexamine and further develop different gravito-electromagnetic
(GEM) analogies found in the literature, and clarify the connection
between them. Special emphasis is placed in two exact physical analogies:
the analogy based on inertial fields from the so-called ``1+3 formalism'',
and the analogy based on tidal tensors. Both are reformulated, extended
and generalized. We write in both formalisms the Maxwell and the full
exact Einstein field equations with sources, plus the algebraic Bianchi
identities, which are cast as the source-free equations for the gravitational
field. New results within each approach are unveiled. The well known
analogy between linearized gravity and electromagnetism in Lorentz
frames is obtained as a limiting case of the exact ones. The formal
analogies between the Maxwell and Weyl tensors are also discussed,
and, together with insight from the other approaches, used to physically
interpret gravitational radiation. The precise conditions under which
a similarity between gravity and electromagnetism occurs are discussed,
and we conclude by summarizing the main outcome of each approach.
1+3 Splitting · Quasi-Maxwell formalism · Gyroscope precession · Spin-curvature
force
\end{abstract}
\tableofcontents{}

\section{Introduction}

This work has two main goals: one is to establish the connection between
the several gravito-electromagnetic analogies existing in the literature,
summarizing the main results and insights offered by each of them;
the second is to further develop and extend some of these analogies.

In an earlier work by one of the authors \cite{CHPRD,CHPreprint},
a gravito-electromagnetic analogy based on tidal tensors was presented,
and its relationship with 1) the well known analogy between linearized
gravity and electromagnetism, 2) the mapping, via the Klein-Gordon
equation, between ultrastationary spacetimes and magnetic fields in
curved spacetimes, and 3) the formal analogies between the Weyl and
Maxwell tensors (their decomposition into electric and magnetic parts,
the quadratic scalar invariants they form, and the field equations
they obey) was discussed.

Building up on the work in \cite{CHPreprint}, another approach is
herein added to the discussion: the exact analogy based on the fields
of inertial forces, arising in the context of the 1+3 splitting of
spacetime. This approach, which is herein reformulated and suitably
generalized, is still not very well known, but very far reaching.
It is therefore important to understand how it relates with the other
known analogies, and in particular with the (also exact) approach
based on tidal tensors.

Each of the analogies discussed here are also further developed, and
some new results within each of them are presented. We start in Sec.~\ref{sec:Tidal tensor analogy}
by revisiting the approach based on tidal tensors introduced in \cite{CHPRD}
(and partly reviewed in \cite{CordaTidalTensors}), completing it
by extending the formalism to the full gravitational field equations
(cast herein as the Einstein field equations, plus the algebraic Bianchi
identities). More precisely, through suitable projector techniques,
we make a full 1+3 covariant splitting of the latter, obtaining a
set of six \emph{algebraic} equations: four of them involve only the
sources, the ``gravitoelectric'' ($\mathbb{E}_{\alpha\beta}$) and
``gravitomagnetic'' ($\mathbb{H}_{\alpha\beta}$) tidal tensors,
and are formally similar to the Maxwell equations written in this
formalism; plus an additional pair of equations that involve the purely
spatial curvature (encoded in the tensor $\mathbb{F}_{\alpha\beta}$)
and have no electromagnetic analogue. The formalism is then used to
contrast the gravitational and electromagnetic tidal effects. We also
add to the list of exact analogies manifest in this formalism the
one concerning relative precession of spinning particles.

In Sec.~\ref{sec:3+1}, we discuss another exact gravito-electromagnetic
analogy, the one drawing a parallelism between spatial inertial forces
-- described by the ``gravitoelectromagnetic'' (GEM) fields -- and
the electromagnetic fields. GEM fields are best known from linearized
theory, e.g.~\cite{Gravitation and Inertia,Gravitation and Spacetime,Harris1991,Tucker Clark,Ruggiero:2002hz,General Relativity,Wald,Carroll,Near Zero,Wald et al 2010,Mashhoon:2003ax}.
Less well known are the exact analogies based on inertial fields that
arise in the splitting of spacetime into time + space with respect
to two preferred congruences of observers: time-like Killing congruences
in stationary spacetimes, introduced by Landau-Lifshitz \cite{LandauLifshitz},
and further worked out by other authors \cite{Oliva,Natario,NatarioCosta,MenaNatario,ZonozBell,ZonozPRD},
which leads to the so-called ``quasi-Maxwell'' analogy; and hypersurface
orthogonal observers, which have motivated substantially different
approaches, e.g. \cite{Black Holes,SemerakInertial}, leading also
to analogies based on exact equations, albeit not as close as in the
former case. Lesser known still is the existence of an exact formulation
applying to \emph{arbitrary observer congruences} in \emph{arbitrary}
\emph{spacetimes} \cite{The many faces,GEM User Manual}, where a
framework is developed to encompass the inertial fields arising in
the different spacetime splittings in the literature. Herein we propose
a reformulation of the problem, also applying to arbitrary observers
in arbitrary spacetimes, but whose most distinctive feature is allowing
for an arbitrary rotation of the spatial frame, which was achieved
by defining a suitable connection ($\tilde{\nabla}$) on the bundle
of vectors orthogonal to a congruence of time-like curves. That allows
to describe the inertial forces of any orthonormal frame, and manifests
that the so-called ``gravitomagnetic field'' \textcolor{black}{($\vec{H}$)}
consists of a combination of \textcolor{black}{two effects of different
(independent, in a general formulation) origin: }the vorticity \textcolor{black}{$\vec{\omega}$}
of the observer congruence, plus the angular velocity \textcolor{black}{$\vec{\Omega}$}
of rotation along the congruence (relative to Fermi-Walker transport)
of the triad of spatial axes that each observer ``carries''. Such
formulation encompasses the gravitomagnetic field of different frames
studied in the literature, e.g. the one measured by the Killing (or
``static'') observers in a stationary spacetime, studied in e.g.
\cite{Natario,ZonozBell,ZonozPRD}, or the one arising in the so-called
``locally non-rotating frames'' (LNR) in e.g. \cite{SemerakInertial,Bardeen et al}.
Drawing a parallelism with what is done in Sec.~\ref{sec:Tidal tensor analogy},
we express the Maxwell equations and the full gravitational field
equations in this formalism. Again, a set of four equations are produced
which exhibit many similarities with their electromagnetic counterparts
(closer in the case of rigid observer congruences in stationary spacetimes,
with an almost one to one correspondence), plus two additional equations
which have no electromagnetic analogue. We also add to the list of
exact analogies the one between the electromagnetic force on a magnetic
dipole and the gravitational force on a gyroscope, when both are at
rest with respect to a rigid (arbitrarily accelerated and rotating)
frame in a stationary spacetime. In Sec. \ref{sub:Relation1+3_TTensors}
we establish the relationship between the inertial fields and the
tidal tensors of Sec.~\ref{sec:Tidal tensor analogy} --- a particularly
important result in the context of this work.

In Sec. \ref{sec:Ultrastationary} we discuss a special class of spacetimes
admitting global rigid geodesic congruences, the ``ultra-stationary''
spacetimes. They have interesting properties in the context of GEM,
which were discussed in \cite{CHPRD} in the framework of the tidal
tensor analogy; herein we revisit those spacetimes in the framework
of the GEM inertial fields of Sec. \ref{sec:3+1}, shedding some new
light on questions left open in \cite{CHPRD}.

In Sec.~\ref{sec:Linear-gravitoelectromagnetism} we explain the
relation between the exact approach based in the inertial GEM fields
of Sec.~\ref{sec:3+1}, and the popular gravito-electromagnetic analogy
based on linearized theory, e.g.~\cite{Gravitation and Inertia,Gravitation and Spacetime,Tucker Clark,Harris1991}.
The latter is obtained as a special limit of the exact equations of
the former. Taking this route gives a clearer account of the physical
meaning of the GEM fields, which are often somewhat naively derived
from the temporal components of the metric tensor (drawing a parallelism
with the electromagnetic potentials), without making apparent their
status as artifacts of the reference frame, and in particular their
relation with the kinematical quantities associated to the observer's
congruence. It is also a procedure for obtaining the field equations
that does not rely on choosing the harmonic gauge condition and its
inherent subtleties (which have been posing some difficulties in the
literature, see e.g.~\cite{Harris1991,Harmonic Gauge,Tucker Clark,CHPreprint}).

In Sec.~\ref{sec:Weyl analogy} we briefly review the \emph{formal}
analogies between the electric ($\mathcal{E}_{\alpha\beta}$) and
magnetic ($\mathcal{H}_{\alpha\beta}$) parts of the Weyl tensor,
and the electric ($E^{\alpha}$) and magnetic ($B^{\alpha}$) fields,
e.g. \cite{matte,bel,Tchrakian,CampbellMorganAjp,Maartens:1997fg,EMMbook,general relativity and cosmology,ellis:97,Bonnor:1995zf,cherubini:02,McIntosh et al 1994,WyllePRD,LozanovskiAarons:99}.
One of its interesting results is that in the case of vacuum, in the
linear regime, one obtains a set of equations formally similar to
Maxwell's equations in a Lorentz frame, only with the gravitational
tidal tensors ($\mathcal{E}_{ij}=\mathbb{E}_{ij}$, $\mathcal{H}_{ij}=\mathbb{H}_{ij}$)
in place of $E^{\alpha}$ and $B^{\alpha}$. Formally analogous wave
equations follow likewise, whose physical interpretation is discussed
based on what is learned from the approaches herein.

\subsection{Notation and conventions\label{sub:Notation-and-conventions}}
\begin{enumerate}
\item \begin{flushleft}
\emph{Signature and signs}. We use the signature $-+++$; $\epsilon_{\alpha\beta\gamma\delta}\equiv\sqrt{-g}[\alpha\beta\gamma\delta]$
denotes the Levi-Civita tensor, and we follow the orientation $[1230]=1$
(i.e., in flat spacetime $\epsilon_{1230}=1$). $\epsilon_{ijk}\equiv\epsilon_{ijk0}$
is the 3-D alternating tensor. $\star$ denotes the Hodge dual. 
\par\end{flushleft}
\item \begin{flushleft}
\emph{\label{enu:Time-and-space  Proj}Time and space projectors}.
$(\top^{u})_{\,\,\,\beta}^{\alpha}\equiv-u^{\alpha}u_{\beta}$, $(h^{u})_{\ \beta}^{\alpha}\equiv u^{\alpha}u_{\beta}+g_{\ \beta}^{\alpha}$
are, respectively, the projectors parallel and orthogonal to a unit
time-like vector $u^{\alpha}$; may be interpreted as the time and
space projectors in the local rest frame of an observer of 4-velocity
$u^{\alpha}$. $\langle\alpha\rangle$ denotes the index of a spatially
projected tensor: $A^{\langle\alpha\rangle\beta\ldots}\equiv(h^{u})_{\ \mu}^{\alpha}A^{\mu\beta\ldots}$. 
\par\end{flushleft}
\item \begin{flushleft}
$\rho_{c}=-j^{\alpha}u_{\alpha}$ and $j^{\alpha}$ are, respectively,
the charge density and current 4-vector; $\rho=T_{\alpha\beta}u^{\alpha}u^{\beta}$
and $J^{\alpha}=-T_{\ \beta}^{\alpha}u^{\beta}$ are the mass/energy
density and current (quantities measured by the observer of 4-velocity
$u^{\alpha}$); $T_{\alpha\beta}\equiv$ energy-momentum tensor. 
\par\end{flushleft}
\item \begin{flushleft}
$S^{\alpha}\equiv$ spin 4-vector; $\mu^{\alpha}\equiv$ magnetic
dipole moment; defined such that their components in the particle's
proper frame are $S^{\alpha}=(0,\vec{S})$, $\mu^{\alpha}=(0,\vec{\mu})$.
For their precise definition in terms of moments of $T^{\alpha\beta}$
and $j^{\alpha}$, see \cite{Gyros}. 
\par\end{flushleft}
\item \emph{Tensors as measured by an observer}. $(A^{u})_{\ \ }^{\alpha_{1}..\alpha_{n}}$
denotes the tensor $\mathbf{A}$ as measured by an observer $\mathcal{O}(u)$
of 4-velocity $u^{\alpha}$. For example, $(E^{u})^{\alpha}\equiv F_{\ \beta}^{\alpha}u^{\beta}$,
$(E^{u})_{\alpha\beta}\equiv F_{\alpha\gamma;\beta}u^{\gamma}$ and
$(\mathbb{E}^{u})_{\alpha\beta}\equiv R_{\alpha\mu\beta\nu}u^{\nu}u^{\mu}$
denote, respectively, the electric field, electric tidal tensor, and
gravitoelectric tidal tensor as measured by $\mathcal{O}(u)$. Analogous
forms apply to their magnetic/gravitomagnetic counterparts. \\
 For 3-vectors we use notation $\vec{A}(u)$; for example, $\vec{E}(u)$
denotes the electric 3-vector field as measured by $\mathcal{O}(u)$
(i.e., the space part of $(E^{u})^{\alpha}$, \emph{written in a frame}
where $u^{i}=0$). Often we drop the superscript (e.g.~$(E^{U})^{\alpha}\equiv E^{\alpha}$),
or the argument of the 3-vector: $\vec{E}(U)\equiv\vec{E}$, when
the meaning is clear. 
\item \emph{Electromagnetic field.} The Maxwell tensor $F^{\alpha\beta}$
and its Hodge dual $\star F^{\alpha\beta}\equiv\epsilon_{\mu\nu}^{\ \ \alpha\beta}F_{\mu\nu}/2$
decompose in terms of the electric $(E^{u})^{\alpha}\equiv F_{\ \beta}^{\alpha}u^{\beta}$
and magnetic $(B^{u})^{\alpha}\equiv\star F_{\ \beta}^{\alpha}u^{\beta}$
fields measured by an observer of 4-velocity $u^{\alpha}$ as 
\begin{equation}
F_{\alpha\beta}=2u_{[\alpha}(E^{u})_{\beta]}+\epsilon_{\alpha\beta\gamma\delta}u^{\delta}(B^{u})^{\gamma}\ \quad{\rm (a)}\qquad\star F_{\alpha\beta}=2u_{[\alpha}(B^{u})_{\beta]}-\epsilon_{\alpha\beta\gamma\sigma}u^{\sigma}(E^{u})^{\gamma}\ \quad{\rm (b)}\label{eq:Fdecomp}
\end{equation}

\end{enumerate}

\section{The gravito-electromagnetic analogy based on tidal tensors\label{sec:Tidal tensor analogy}}

The rationale behind the tidal tensor gravito-electromagnetic analogy
is to make a comparison between the two interactions based on physical
forces present in both theories. The electromagnetic Lorentz force
has no physical counterpart in gravity\textbf{,} as monopole point
test particles in a gravitational field move along geodesics, without
any force being exerted on them. In this sense, the analogy drawn
in Sec.~\ref{sub: 1+3 Geodesics} between Eqs.~(\ref{eq:Lorentz})
and (\ref{eq:Geo3+1}) is a comparison of a physical electromagnetic
force to an artifact of the reference frame. Tidal forces, by their
turn, are covariantly present in both theories, and their mathematical
description in terms of objects called ``tidal tensors'' is the
basis of this approach. Tidal forces manifest themselves in essentially
two basic effects: the relative acceleration of two nearby monopole
test particles, and in the net force exerted on dipoles. The notions
of multipole moments arise from a description of the \emph{test} bodies
in terms of the fields they \emph{would} produce. In electromagnetism
they are the multipole expansions of the 4-current density vector
$j^{\alpha}=(\rho_{c},\vec{j}$), rigorously established in \cite{Dixon1967},
and well known in textbooks as the moments of the charge and current
densities. In gravity they are are the moments of the energy momentum
tensor $T_{\alpha\beta}$, the so called~\cite{MathissonNeueMechanik}
``gravitational skeleton'', of which only the moments of the 4-current
density $J^{\alpha}=-T^{\alpha\beta}U_{\beta}$ have an electromagnetic
counterpart. Monopole particles in the context of electromagnetism
are those whose only non-vanishing moment is the total charge; dipole
particles are particles with non-vanishing electric and magnetic dipole
moments (i.e., respectively the dipole moments of $\rho_{c}$ and
$\vec{j}$); see \cite{Dixon1967} and companion paper \cite{Gyros}
for precise definitions of these moments. Monopole particles in gravity
are particles whose only non-vanishing moment of $T^{\alpha\beta}$
is the mass, and correspond to the usual notion of \emph{point test
particle}, which moves along geodesics. There is no gravitational
analogue of the intrinsic electric dipole, as there are no negative
masses; but there is an analogue of the magnetic dipole moment, which
is the ``intrinsic'' angular momentum (i.e.~the angular momentum
about the particle's center of mass), usually dubbed spin vector/tensor.
Sometimes we will also call it, for obvious reasons, the ``gravitomagnetic
dipole moment''. A particle possessing only pole-dipole gravitational
moments corresponds to the notion of an ideal gyroscope. We thus have
two physically analogous effects suited to compare gravitational and
electromagnetic tidal forces: worldline deviation of nearby monopole
test particles, and the force exerted on magnetic dipoles/gyroscopes.
An exact gravito-electromagnetic analogy, summarized in Table \ref{analogy},
emerges from this comparison.

Eqs.~(\ref{analogy}.1) are the worldline deviations for nearby test
particles with the same%
\footnote{\label{fn:DeviatonVelocity}We want to emphasize this point, which,
even today, is not clear in the literature. Eqs.~(\ref{analogy}.1)
apply to the \emph{instant} where the two particles have the same
(or infinitesimally close, in the gravitational case) tangent vector.
When the particles have arbitrary velocities, both in electromagnetism
and gravity, their relative acceleration is not given by a simple
contraction of a tidal tensor with a separation vector; the equations
include more terms, see \cite{CHPRD,CHPreprint,Generalized Geodesic}.
There is however a difference: whereas Eq. (\ref{analogy}.1a) requires
strictly $\delta\mathbf{U}=\mathbf{U}_{2}-\mathbf{U}_{1}=0$, see
\cite{BalakinHoltenKerner,CHPRD,CHPreprint}, Eq. (\ref{analogy}.1b)
allows for an infinitesimal $\delta U\propto\delta x$, as can be
seen from Eq. (6) of \cite{Generalized Geodesic}. That means that
(\ref{analogy}.1b) holds for \emph{infinitesimally close} curves
belonging to an arbitrary geodesic \emph{congruence} (it is in this
sense that in e.g. \cite{Gravitation,Dixon1970I} $\delta U$ is portrayed
as ``arbitrary'' --- it is understood to be infinitesimal therein,
as those treatments deal with congruences of curves).%
} tangent vector (and the same ratio charge/mass in the electromagnetic
case), separated by the infinitesimal vector $\delta x^{\alpha}$.
They tell us that the so-called (e.g. \cite{nareshdadhich:00}) electric
part of the Riemann tensor $\mathbb{E}_{\,\,\,\beta}^{\alpha}\equiv R_{\,\,\,\mu\beta\nu}^{\alpha}U^{\mu}U^{\nu}$
plays in the geodesic deviation equation (\ref{analogy}.1b) the same
physical role as the tensor $E_{\alpha\beta}\equiv F_{\alpha\gamma;\beta}U^{\gamma}$
in the electromagnetic worldline deviation (\ref{analogy}.1a): in
a gravitational field the ``relative acceleration'' between two
nearby test particles, with the \emph{same 4-velocity} $U^{\alpha}$,
is given by a contraction of $\mathbb{E}_{\alpha\beta}$ with the
separation vector $\delta x^{\beta}$; just like in an electromagnetic
field the ``relative acceleration'' between two nearby charged particles
(with the same $U^{\alpha}$ and ratio $q/m)$ is given by a contraction
of the electric tidal tensor $E_{\alpha\beta}$ with $\delta x^{\alpha}$.
$E_{\alpha\beta}$ measures the tidal effects produced by the electric
field $E^{\alpha}=F_{\ \gamma}^{\alpha}U^{\gamma}$ \emph{as measured
by the test particle} of 4-velocity $U^{\alpha}$. We can define it
as a covariant derivative of the electric field as measured in the
inertial frame momentarily comoving with the particle: $E_{\alpha\beta}={E_{\alpha;\beta}}|_{U=const}$.
Hence we dub it the ``electric tidal tensor'', and its gravitational
counterpart the ``gravitoelectric tidal tensor''.

Eqs (\ref{analogy}.2) are, respectively, the electromagnetic force
on a magnetic dipole \cite{Gyros}, and the Mathisson-Papapetrou equation
\cite{MathissonNeueMechanik,Papapetrou I} for the gravitational force
exerted on a gyroscope (supplemented by the Mathisson-Pirani spin
condition \cite{MathissonNeueMechanik,Pirani 1956}; see \cite{Gyros}
for more details). They tell us that the magnetic part of the Riemann
tensor $\mathbb{H}_{\,\,\,\beta}^{\alpha}\equiv\star R_{\,\,\,\mu\beta\nu}^{\alpha}U^{\mu}U^{\nu}$
plays in the gravitational force (\ref{analogy}.2b) the same physical
role as the tensor $B_{\alpha\beta}\equiv\star F_{\alpha\gamma;\beta}U^{\gamma}$
in the electromagnetic force (\ref{analogy}.2a): the gravitational
force exerted on a spinning particle of 4-velocity $U^{\alpha}$ is
exactly given by a contraction of $\mathbb{H}_{\alpha\beta}$ with
the spin vector $S^{\alpha}$ (the ``gravitomagnetic dipole moment''),
just like its electromagnetic counterpart is exactly given by a contraction
of the magnetic tidal tensor $B_{\alpha\beta}$ with the magnetic
dipole moment $\mu^{\alpha}$. $B_{\alpha\beta}$ measures the tidal
effects produced by the magnetic field $B^{\alpha}=\star F_{\ \gamma}^{\alpha}U^{\gamma}$
\emph{as measured by the particle} of 4-velocity $U^{\alpha}$; for
this reason we dub it the ``magnetic tidal tensor'', and its gravitational
analogue $\mathbb{H}_{\alpha\beta}$ the ``gravitomagnetic tidal
tensor''.

\subsection{Tidal tensor formulation of Maxwell and Einstein equations}

Taking time and space projections, Maxwell's and Einstein's equations
can be expressed in tidal tensor formalism; that makes explicit a
striking aspect of the analogy: Maxwell's equations (the source equations
plus the Bianchi identity) may be cast as a set of \emph{algebraic}
equations involving only tidal tensors and source terms (the charge
current 4-vector); and the gravitational field equations (Einstein's
source equations plus the algebraic Bianchi identity) as a set of
five independent equations, consisting of two parts: i) a subset of
four equations formally very similar to Maxwell's, that are likewise
algebraic equations involving only tidal tensors and sources (the
mass-energy current vector), and ii) a fifth equation involving the
space part of $T^{\alpha\beta}$ and a spatial rank 2 tensor which
has no electromagnetic analogue. This is what we are going to show
next. For that, we first introduce the time and space projectors with
respect to a unit time-like vector $U^{\alpha}$ (i.e., the projectors
parallel and orthogonal to $U^{\alpha}$): 
\begin{equation}
\top_{\ \beta}^{\alpha}\equiv(\top^{U})_{\ \beta}^{\alpha}=-U^{\alpha}U_{\beta};\qquad h_{\ \beta}^{\alpha}\equiv(h^{U})_{\ \beta}^{\alpha}=U^{\alpha}U_{\beta}+\delta_{\ \beta}^{\alpha}.\label{eq:Time_Space_Proj}
\end{equation}
A vector $A^{\alpha}$ can be split into its time and space projections
with respect to $U^{\alpha}$; and an arbitrary rank $n$ tensor can
be completely decomposed taking time and space projections in each
of its indices (e.g.~\cite{The many faces}): 
\begin{equation}
A^{\alpha}=\top_{\ \beta}^{\alpha}A^{\beta}+h_{\ \beta}^{\alpha}A^{\beta};\ \qquad A^{\alpha_{1}...\alpha_{n}}=\left(\top_{\ \beta_{1}}^{\alpha_{1}}+h_{\ \beta_{1}}^{\alpha_{1}}\right)...\left(\top_{\ \beta_{n}}^{\alpha_{n}}+h_{\ \beta_{n}}^{\alpha_{n}}\right)A^{\beta_{1}...\beta_{n}}.\label{eq:Time-Space-Decomp}
\end{equation}
Instead of using $h_{\ \sigma}^{\mu}$, one can also, if convenient,
spatially project an index of a tensor $A^{\sigma...}$ contracting
it with the spatial 3-form $\epsilon_{\alpha\beta\sigma\gamma}U^{\gamma}$;
for instance, for the case of vector $A^{\sigma}$, one obtains the
spatial 2-form $\epsilon_{\alpha\beta\sigma\gamma}U^{\gamma}A^{\sigma}=\star A_{\alpha\beta\gamma}U^{\gamma}$,
which contains precisely the same information as the spatial vector
$A^{\mu}h_{\ \mu}^{\sigma}\equiv A^{\langle\sigma\rangle}$ (the former
is the spatial dual of the latter). New contraction with $\epsilon_{\ \ \mu\nu}^{\alpha\beta}U^{\nu}$
yields $A^{\langle\sigma\rangle}$ again. Indeed we may write 
\[
h_{\sigma}^{\mu}=\frac{1}{2}\epsilon^{\alpha\beta\mu\nu}U_{\nu}\epsilon_{\alpha\beta\sigma\gamma}U^{\gamma}.
\]

Another very useful relation is the following. The space projection
$h_{\ \alpha}^{\mu}h_{\ \beta}^{\nu}F_{\mu\nu}\equiv F_{\langle\alpha\rangle\langle\beta\rangle}$
of a 2-form $F_{\alpha\beta}=F_{[\alpha\beta]}$ is equivalent to
the tensor $\epsilon_{\mu\nu\alpha\beta}F^{\alpha\beta}U^{\nu}=2\star\!\! F_{\mu\nu}U^{\nu}$
(i.e., spatially projecting $F_{\alpha\beta}$ is equivalent to time-projecting
its Hodge dual). We have: 
\begin{equation}
F_{\langle\alpha\rangle\langle\beta\rangle}=\frac{1}{2}\epsilon_{\mu\alpha\beta\lambda}U^{\lambda}\epsilon_{\ \nu\sigma\delta}^{\mu}U^{\nu}F^{\sigma\delta}=\epsilon_{\mu\alpha\beta\lambda}U^{\lambda}\star\! F_{\ \nu}^{\mu}U^{\nu}.\label{eq:SpaceProj2Form}
\end{equation}
Now let $F_{\gamma_{1}...\gamma_{n}\alpha\beta\delta_{1}...\delta_{m}}=F_{\gamma_{1}...\gamma_{n}[\alpha\beta]\delta_{1}...\delta_{m}}$,
be some tensor antisymmetric in the pair $\alpha,\beta$; an equality
similar to the one above applies: 
\begin{equation}
F_{\gamma_{1}...\gamma_{n}\langle\alpha\rangle\langle\beta\rangle\delta_{1}...\delta_{m}}=\frac{1}{2}\epsilon_{\mu\alpha\beta\lambda}U^{\lambda}\epsilon_{\ \nu\sigma\delta}^{\mu}U^{\nu}F_{\gamma_{1}...\gamma_{n}\ \ \delta_{1}...\delta_{m}}^{\ \ \ \ \ \ \sigma\delta}\,.\label{eq:SpaceProjAsym}
\end{equation}

\subsubsection{Maxwell's equations\label{sub:Maxwell's-EqsTidal}}

Maxwell's equations are given in tensor form by the pair of equations:
\begin{equation}
F_{\ \ \ ;\beta}^{\alpha\beta}=4\pi j^{\alpha}\quad{\rm (a)};\qquad\ \star F_{\ \ \ ;\beta}^{\alpha\beta}=0\quad{\rm (b)}.\label{eq:MaxwellFieldEqs}
\end{equation}
Here (\ref{eq:MaxwellFieldEqs}a) are the Maxwell source equations,
and (\ref{eq:MaxwellFieldEqs}b) are the source-free equations, equivalent
to $F_{[\alpha\beta;\gamma]}=0$, and commonly called the Bianchi
identity for $F_{\alpha\beta}$. $j^{\alpha}$ denotes the current
4-vector. Decomposing the tensors $F_{\alpha\beta;\gamma}$ and $\star F_{\alpha\beta;\gamma}$
into their time and space projections in the first two indices, using
Eq.~(\ref{eq:SpaceProj2Form}) to project spatially, we obtain their
decomposition in terms of tidal tensors, 
\begin{eqnarray}
F_{\alpha\beta;\gamma} & = & 2U_{[\alpha}E_{\beta]\gamma}+\epsilon_{\alpha\beta\mu\sigma}U^{\sigma}B_{\ \gamma}^{\mu}\ ;\label{Fcovdecomp}\\
\star F_{\alpha\beta;\gamma} & = & 2U_{[\alpha}B_{\beta]\gamma}-\epsilon_{\alpha\beta\mu\sigma}U^{\sigma}E_{\ \gamma}^{\mu}\ .\label{FcivStarDecomp}
\end{eqnarray}
Substituting these decompositions into Eqs. (\ref{eq:MaxwellFieldEqs})
and taking time and space projections, we obtain the set of four equations:
\begin{eqnarray}
E_{\,\,\,\alpha}^{\alpha} & = & 4\pi\rho_{c}\ ;\label{Et}\\
E_{[\alpha\beta]} & = & U_{[\alpha}E_{\beta]\gamma}U^{\gamma}+\frac{1}{2}\epsilon_{\alpha\beta\mu\sigma}U^{\sigma}B^{\mu\gamma}U_{\gamma}\ ;\label{Eanti}\\
B_{\,\,\,\alpha}^{\alpha} & = & 0\ ;\label{Bt}\\
B_{[\alpha\beta]} & = & U_{[\alpha}B_{\beta]\gamma}U^{\gamma}-\frac{1}{2}\epsilon_{\alpha\beta\mu\sigma}U^{\sigma}E^{\mu\gamma}U_{\gamma}-2\pi\epsilon_{\alpha\beta\sigma\gamma}j^{\sigma}U^{\gamma}\ .\label{Banti}
\end{eqnarray}
Here $\rho_{c}\equiv-j^{\alpha}U_{\alpha}$ is the charge density
as measured by an observer of 4-velocity $U^{\alpha}$. In more detail,
taking the time projection of (\ref{eq:MaxwellFieldEqs}a), we obtain
Eq.~(\ref{Et}); taking the space projection, by contracting with
the spatial 3-form $\epsilon_{\mu\nu\alpha\sigma}U^{\sigma}$, yields
Eq.~(\ref{Banti}). The same procedure applied to Eq. (\ref{eq:MaxwellFieldEqs}b)
yields Eqs.~(\ref{Bt}) and (\ref{Eanti}) as time and space projections,
respectively. 

Hence, in this formalism, Maxwell's equations are cast as the equations
for the traces and antisymmetric parts of the electromagnetic tidal
tensors; and they involve only \emph{tidal tensors and sources}. Substituting
(\ref{Fcovdecomp})-(\ref{FcivStarDecomp}) back into (\ref{Eanti})
and (\ref{Banti}), leads to the equivalent set Eqs. (\ref{analogy}.4a)-(\ref{analogy}.8a)
of Table \ref{analogy}. It is also useful to note that the pair of
Eqs.~(\ref{Eanti}) and (\ref{Banti}) can be condensed into the
equivalent pair 
\begin{equation}
\epsilon_{\ \ \alpha\delta}^{\beta\gamma}U^{\delta}E_{[\gamma\beta]}=-B_{\alpha\beta}U^{\beta};\quad{\rm (a)}\qquad\epsilon_{\ \ \alpha\delta}^{\beta\gamma}U^{\delta}B_{[\gamma\beta]}=E_{\alpha\beta}U^{\beta}+4\pi j_{\alpha}.\quad{\rm (b)}\label{eq:SpaceE[ab]B[ab]}
\end{equation}
In a Lorentz frame in flat spacetime, since $U_{\ ;\beta}^{\alpha}=U_{\ ,\beta}^{\alpha}=0$,
we have $E_{\gamma\beta}=E_{\gamma;\beta}$, $B_{\gamma\beta}=B_{\gamma;\beta}$;
and (using $U^{\alpha}=\delta_{0}^{\alpha}$) Eqs.~(\ref{eq:SpaceE[ab]B[ab]})
can be written in the familiar vector forms $\nabla\times\vec{E}=-\partial\vec{B}/\partial t$
and $\nabla\times\vec{B}=\partial\vec{E}/\partial t+4\pi\vec{j}$,
respectively. Likewise, Eqs.~(\ref{Et}) and (\ref{Bt}) reduce in
this frame to the familiar forms $\nabla\cdot\vec{E}=4\pi\rho_{c}$
and $\nabla\cdot\vec{B}=0$, respectively.

\subsubsection{Einstein's equations\label{sub:Einstein's-EqsTidal}}

Equations (\ref{eq:EinsteinField}a) below are the Einstein source
equations for the gravitational field; Eqs (\ref{eq:EinsteinField}b)
are the algebraic Bianchi identity, equivalent to $R_{[\alpha\beta\gamma]\delta}=0$:

\begin{equation}
R_{\ \alpha\gamma\beta}^{\gamma}\equiv R_{\alpha\beta}=8\pi\left(T_{\alpha\beta}^{\ }-\frac{1}{2}g_{\alpha\beta}^{\ }T_{\ \gamma}^{\gamma}\right)\quad{\rm (a)};\qquad\star R_{\ \ \ \gamma\beta}^{\gamma\alpha}=0\quad{\rm (b)}.\label{eq:EinsteinField}
\end{equation}
In order to express these equations in the tidal tensor formalism
we will decompose the Riemann tensor into its time and space projections
(in each of its indices) with respect to a unit time-like vector $U^{\alpha}$,
$R^{\alpha\beta\gamma\delta}=\left(\top_{\ \ \rho}^{\alpha}+h_{\ \rho}^{\alpha}\right)..\left(\top_{\ \sigma}^{\delta}+h_{\ \sigma}^{\delta}\right)R^{\rho..\sigma}$,
cf.~Eq.~(\ref{eq:Time-Space-Decomp}); we obtain%
\footnote{The characterization of the Riemann tensor by these three spatial
rank 2 tensors is known as the ``Bel decomposition'', even though
the explicit decomposition (\ref{Bel}) is not presented in any of
Bel's papers (e.g.~\cite{BelDecomp}). To the author's knowledge,
an equivalent expression (Eq.~(4.6) therein) can only be found at
\cite{AlfonsoSE}.%
} 
\begin{eqnarray}
R_{\ \ \gamma\delta}^{\alpha\beta} & = & 4\mathbb{E}_{\ \ [\gamma}^{[\alpha}U_{\delta]}U^{\beta]}+2\left\{ \epsilon_{\ \ \gamma\delta}^{\mu\chi}U_{\chi}\mathbb{H}_{\mu}^{\ [\beta}U^{\alpha]}+\epsilon^{\mu\alpha\beta\chi}U_{\chi}\mathbb{H}_{\mu[\delta}U_{\gamma]}\right\} \nonumber \\
 &  & +\epsilon^{\alpha\beta\phi\psi}U_{\psi}\epsilon_{\ \ \gamma\delta}^{\mu\nu}U_{\nu}\mathbb{F}_{\phi\mu},\label{Bel}
\end{eqnarray}
where we made use of the identity (\ref{eq:SpaceProj2Form}) to project
spatially an antisymmetric pair of indices, noting that $R_{\alpha\beta\gamma\delta}$
can be regarded as a double 2-form. This equation tells us that the
Riemann tensor decomposes, with respect to $U^{\alpha}$, into three
\emph{spatial} tensors: the gravitoelectric tidal tensor $\mathbb{E}_{\alpha\beta}$,
the gravitomagnetic tidal tensor $\mathbb{H}_{\alpha\beta}$, plus
a third tensor 
\[
\mathbb{F}_{\alpha\beta}\equiv\star R\star_{\alpha\gamma\beta\delta}U^{\gamma}U^{\delta}=\epsilon_{\ \ \alpha\gamma}^{\mu\nu}\epsilon_{\ \ \beta\delta}^{\lambda\tau}R_{\mu\nu\lambda\tau}U^{\gamma}U^{\delta}\,\,,
\]
introduced by Bel \cite{BelDecomp}, which encodes the purely spatial
curvature with respect to $U^{\alpha}$, and has no electromagnetic
analogue. In order to obtain Eq.~(\ref{Bel}), we made use of the
symmetries $R^{\alpha\beta\gamma\delta}=R^{[\alpha\beta][\gamma\delta]}$,
and in the case of the terms involving $\mathbb{H}_{\alpha\beta}$
(and \emph{only for these terms}) we also assumed the pair exchange
symmetry $R^{\alpha\beta\gamma\delta}=R^{\gamma\delta\alpha\beta}$.
$\mathbb{E}_{\alpha\beta}$ and $\mathbb{F}_{\alpha\beta}$ are symmetric
(and spatial), and therefore have 6 independent components each; $\mathbb{H}_{\alpha\beta}$
is traceless (and spatial), and so has 8 independent components. Therefore
these three tensors together encode the 20 independent components
of the Riemann tensor.

In what follows we will need also the Hodge dual, in the first two
indices, of the decomposition (\ref{Bel}):\textbf{ 
\begin{eqnarray}
\star R_{\ \ \gamma\delta}^{\alpha\beta} & = & 2\epsilon_{\ \ \lambda\tau}^{\alpha\beta}\mathbb{E}_{\ [\gamma}^{\lambda}U_{\delta]}U^{\tau}+4U^{[\alpha}\mathbb{H}_{\ [\delta}^{\beta]}U_{\gamma]}+\epsilon_{\ \ \lambda\tau}^{\alpha\beta}\epsilon_{\ \ \gamma\delta}^{\mu\nu}U_{\nu}\mathbb{H}_{\mu}^{\ \tau}U^{\lambda}\nonumber \\
 &  & -2U^{[\alpha}\mathbb{F}_{\ \ \mu}^{\beta]}\epsilon_{\ \ \gamma\delta}^{\mu\nu}U_{\nu}\ .\label{eq:Belstar}
\end{eqnarray}
} The Ricci tensor $R_{\ \delta}^{\beta}=R_{\ \ \ \alpha\delta}^{\alpha\beta}$
and the tensor $\star R_{\ \alpha\mu\beta}^{\mu}$ follow as: 
\begin{equation}
R_{\ \delta}^{\beta}=-\epsilon^{\alpha\beta\mu\nu}\mathbb{H}_{\mu\alpha}U_{\delta}U_{\nu}-\epsilon_{\alpha\delta\mu\nu}\mathbb{H}^{\mu\alpha}U^{\beta}U^{\nu}-\mathbb{F}_{\ \delta}^{\beta}-\mathbb{E}_{\ \delta}^{\beta}+\mathbb{E}_{\ \sigma}^{\sigma}U^{\beta}U_{\delta}+\mathbb{F}_{\ \sigma}^{\sigma}h_{\ \delta}^{\beta},\label{RicciDecomp}
\end{equation}

\begin{equation}
\star R_{\ \ \alpha\delta}^{\alpha\beta}=\epsilon_{\ \ \lambda\tau}^{\alpha\beta}\mathbb{E}_{\ \alpha}^{\lambda}U_{\delta}U^{\tau}-\delta_{\ \delta}^{\beta}\mathbb{H}_{\ \alpha}^{\alpha}+U^{\beta}\mathbb{F}_{\ \mu}^{\alpha}\epsilon_{\ \ \alpha\delta}^{\mu\nu}U_{\nu}.\label{eq:RicStarDecomp}
\end{equation}
Substituting (\ref{RicciDecomp}) into (\ref{eq:EinsteinField}a),
and (\ref{eq:RicStarDecomp}) into (\ref{eq:EinsteinField}b), we
obtain Einstein's equations and the algebraic Bianchi identities in
terms of the tensors $\mathbb{E}_{\alpha\beta}$, $\mathbb{H}_{\alpha\beta}$,
$\mathbb{F}_{\alpha\beta}$. Now let us make the time-space splitting
of these equations. Eq.~(\ref{eq:EinsteinField}a) is symmetric,
hence it only has 3 non-trivial projections: time-time, time-space,
and space-space. The time-time projection yields 
\begin{equation}
\mathbb{E}_{\ \alpha}^{\alpha}=4\pi\left(2\rho+T_{\ \alpha}^{\alpha}\right),\label{eq:EgravTrace}
\end{equation}
where $\rho\equiv T^{\alpha\beta}U_{\beta}U_{\alpha}$ denotes the
mass-energy density as measured by an observer of 4-velocity $U^{\alpha}$.
Contraction of (\ref{RicciDecomp}) with the time-space projector
$\top_{\ \beta}^{\theta}\epsilon_{\ \sigma\tau\gamma}^{\delta}U^{\gamma}$
yields, using (\ref{eq:EinsteinField}a): 
\begin{equation}
\mathbb{H}_{[\sigma\tau]}=-4\pi\epsilon_{\lambda\sigma\tau\gamma}J^{\lambda}U^{\gamma}\,,\label{eq:H[ab]}
\end{equation}
where $J^{\alpha}\equiv-T^{\alpha\beta}U_{\beta}$ is the mass/energy
current as measured by an observer of 4-velocity $U^{\alpha}$. The
space-space projection yields: 
\begin{equation}
\mathbb{F}_{\ \theta}^{\lambda}+\mathbb{E}_{\ \theta}^{\lambda}-\mathbb{F}_{\ \sigma}^{\sigma}h_{\ \theta}^{\lambda}=8\pi\left[h_{\ \theta}^{\lambda}\frac{1}{2}T_{\,\,\,\alpha}^{\alpha}-T_{\langle\theta\rangle}^{\ \langle\lambda\rangle}\right].\label{eq:SpaceSpaceEinsteinTidal}
\end{equation}
where $T_{\langle\theta\rangle}^{\ \langle\lambda\rangle}\equiv h_{\ \delta}^{\lambda}h_{\ \theta}^{\beta}T_{\beta}^{\ \delta}$.

Since the tensor $\star R_{\ \ \ \gamma\beta}^{\gamma\alpha}$ is
not symmetric, Eq.~(\ref{eq:EinsteinField}b) seemingly splits into
four parts: a time-time, time-space, space-time, and space-space projections.
However, the time-time and space-space projections yield the same
equation. Substituting decomposition (\ref{eq:RicStarDecomp}) into
Eq.~(\ref{eq:EinsteinField}b), and taking the time-time (or space-space),
time-space, and space-time projections, yields, respectively: 
\begin{equation}
\mathbb{H}_{\ \alpha}^{\alpha}=0;\quad({\rm a})\qquad\mathbb{F}_{[\alpha\beta]}=0;\quad({\rm b})\qquad\mathbb{E}_{[\alpha\beta]}=0\quad({\rm c)}.\label{eq:BianchiIdTidalTensors}
\end{equation}

Note however that Eqs. (\ref{eq:EgravTrace})-(\ref{eq:BianchiIdTidalTensors})
are not a set of six \emph{independent} equations (only five), as
Eqs. (\ref{eq:BianchiIdTidalTensors}b), (\ref{eq:BianchiIdTidalTensors}c)
and (\ref{eq:SpaceSpaceEinsteinTidal}) are not independent; using
the latter, together with (\ref{eq:BianchiIdTidalTensors}b)/(\ref{eq:BianchiIdTidalTensors}c),
one can obtain the remaining one, (\ref{eq:BianchiIdTidalTensors}c)/(\ref{eq:BianchiIdTidalTensors}b).

The gravitational field equations are summarized and contrasted with
their electromagnetic counterparts in Table \ref{analogy}. Eqs.~(\ref{analogy}.4b)-(\ref{analogy}.5b),
(\ref{analogy}.7b)-(\ref{analogy}.8b) are very similar in form to
Maxwell Eqs.~(\ref{analogy}.4a)-(\ref{analogy}.8a); they are their
\emph{physical} gravitational analogues, since both are the traces
and antisymmetric parts of the tensors $\{E_{\alpha\beta},\ B_{\alpha\beta}\}\leftrightarrow\{\mathbb{E}_{\alpha\beta},\ \mathbb{H}_{\alpha\beta}\}$,
which we know, from equations (\ref{analogy}.1) and (\ref{analogy}.2),
to play analogous physical roles in the two theories. Note this interesting
aspect of the analogy: if one replaces, in Eqs.~(\ref{Et})-(\ref{Banti}),
the electromagnetic tidal tensors ($E_{\alpha\beta}$ and $B_{\alpha\beta}$)
by the gravitational ones ($\mathbb{E}_{\alpha\beta}$ and $\mathbb{H}_{\alpha\beta}$),
and the charges by masses (i.e., density $\rho_{c}$ and current $j^{\alpha}$
of charge by density $\rho$ and current $J^{\alpha}$ of mass), one
\emph{almost} obtains Eqs.~(\ref{analogy}.4b)-(\ref{analogy}.5b),
(\ref{analogy}.7b)-(\ref{analogy}.8b), apart from a factor of 2
in the source term in (\ref{analogy}.5b) and the difference in the
source of (\ref{analogy}.4b). This happens because, since $\mathbb{E}_{\alpha\beta}$
and $\mathbb{H}_{\alpha\beta}$ are spatial tensors, all the contractions
with $U^{\alpha}$ present in Eqs.~(\ref{Eanti}) and (\ref{Banti})
vanish. In the case of vacuum, the four gravitational equations which
are analogous to Maxwell's are thus \emph{exactly }obtained from the
latter by simply replacing $\{E_{\alpha\beta},B_{\alpha\beta}\}\rightarrow\{\mathbb{E}_{\alpha\beta},\mathbb{H}_{\alpha\beta}\}$.

Eqs.~(\ref{eq:BianchiIdTidalTensors}b) and (\ref{eq:SpaceSpaceEinsteinTidal}),
involving $\mathbb{F}_{\alpha\beta}$, have no electromagnetic analogue.
Eq.~(\ref{eq:SpaceSpaceEinsteinTidal}) involves also, as a source,
the space-space part of the energy momentum tensor, $T^{\langle\alpha\rangle\langle\beta\rangle}$,
which, unlike the energy current 4-vector $J^{\alpha}=-T^{\alpha\beta}U_{\beta}$
(analogous to the charge current 4-vector $j^{\alpha}$), has no electromagnetic
counterpart. It is worth discussing this equation in some detail.
It has a fundamental difference%
\footnote{We thank João Penedones for drawing our attention to this point.%
} with respect to the other gravitational field equations in Table
\ref{analogy}, and with their electromagnetic analogues: the latter
are algebraic equations involving only the traces and antisymmetric
parts of the tidal tensors (or of $\mathbb{F}_{\alpha\beta}$), plus
the source terms; they impose no condition on the symmetric parts.
In electromagnetism, this is what allows the field to be dynamical,
and waves to exist (their tidal tensors are described, in an inertial
frame, by Eqs.~(\ref{eq:HigherOrderEMEqs1})-(\ref{eq:HigherOrderEMEqs2})
below); were there additional independent algebraic equations for
the traceless symmetric part of the tidal tensors, and these fields
would be fixed. But Eq.~(\ref{eq:SpaceSpaceEinsteinTidal}), by contrast,
is an equation for the symmetric parts of the tensors $\mathbb{E}_{\alpha\beta}$
and $\mathbb{F}_{\alpha\beta}$. It can be split into two parts. Taking
the trace, and using (\ref{eq:EgravTrace}), one obtains the source
equation for $\mathbb{F}_{\alpha\beta}$: 
\begin{equation}
\mathbb{F}_{\ \sigma}^{\sigma}=8\pi\rho\ ;\label{eq:FSource}
\end{equation}
substituting back into (\ref{eq:SpaceSpaceEinsteinTidal}) we get:
\begin{equation}
\mathbb{F}_{\ \beta}^{\alpha}+\mathbb{E}_{\ \beta}^{\alpha}=8\pi\left[h_{\ \beta}^{\alpha}\left(\frac{1}{2}T_{\ \gamma}^{\gamma}+\rho\right)-T_{\ \langle\beta\rangle}^{\langle\alpha\rangle}\right].\label{eq:FtensorEq2}
\end{equation}
This equation tells us that the tensor $\mathbb{F}_{\ \beta}^{\alpha}$
is not an extra (comparing with electrodynamics) \emph{independent}
object; given the sources and the gravitoelectric tidal tensor $\mathbb{E}_{\alpha\beta}$,
$\mathbb{F}_{\alpha\beta}$ is completely determined by (\ref{eq:FtensorEq2}).

In vacuum ($T^{\alpha\beta}=0$, $j^{\alpha}=0$), the Riemann tensor
becomes the Weyl tensor: $R_{\alpha\beta\gamma\delta}=C_{\alpha\beta\gamma\delta}$;
due to the self duality property of the latter: $C_{\alpha\beta\gamma\delta}=-\star C\star_{\alpha\beta\gamma\delta}$,
it follows that $\mathbb{F}_{\alpha\beta}=-\mathbb{E}_{\alpha\beta}$.

The equations in Table \ref{analogy} have the status of constraints
for the \emph{tidal} fields. They are especially suited to compare
the tidal dynamics (i.e., Eqs.~(\ref{analogy}.1) and (\ref{analogy}.2))
of the two interactions, which is discussed in the next section. But
they do not tell us about the dynamics of the fields \emph{themselves}.
To obtain dynamical field equations, one possible route is to take
one step back and express the tidal tensors in terms of gauge fields
(such as the GEM inertial fields $\vec{G}$, $\vec{H}$ and the shear
$K_{(\alpha\beta)}$ of the 1+3 formalism of Sec.~\ref{sec:3+1};
the general expressions of Einstein equations in terms of these fields
is given Sec.~\ref{sub:Einstein-Eqs-1+3} below; but is also possible
to write the equations for the dynamics of the tidal tensors (the
\emph{physical} fields); that is done not through Einstein equations
(\ref{eq:EinsteinField}), but through the differential Bianchi identity
$R_{\sigma\tau[\mu\nu;\alpha]}=0$, together with decomposition (\ref{Bel}),
and using (\ref{eq:EinsteinField}) to substitute $R^{\alpha\beta}$
by the source terms. The resulting equations, for the case of vacuum
(where $\{\mathcal{E}_{\alpha\beta},\ \mathcal{H}_{\alpha\beta}\}=\{\mathbb{E}_{\alpha\beta},\ \mathbb{H}_{\alpha\beta}\}$),
are Eqs.~(\ref{MarteensG1})-(\ref{MaartensG2}) of Sec.~\ref{sec:Weyl analogy}
below. One may write as well dynamical equations for the electromagnetic
tidal tensors, which for the case of vacuum, and an inertial frame,
are Eqs.~(\ref{eq:HigherOrderEMEqs1})-(\ref{eq:HigherOrderEMEqs2})
of Sec.~\ref{sub:Matte's-equations-vs}; however in the electromagnetic
case the fundamental physical fields are the vectors $E^{\alpha},\ B^{\alpha}$,
whose covariant field equations are Eqs. (\ref{eq:Gaus3+1})-(\ref{eq:CurlEMaartens})
(the tidal field equations (\ref{eq:HigherOrderEMEqs1})-(\ref{eq:HigherOrderEMEqs2})
follow trivially from these).

\subsection{Gravity vs Electromagnetism\label{sub:Gravity-vs-Electromagnetism}}

In the tidal tensor formalism, cf.~Table \ref{analogy}, the gravitational
field is described by five (independent) algebraic equations, four
of which analogous to the Maxwell equations, plus an additional equation,
involving the tensor $\mathbb{F}_{\alpha\beta}$, which has no parallel
in electromagnetism. Conversely, in Maxwell equations there are terms
with no gravitational counterpart; these correspond to the antisymmetric
parts / time projections (with respect to the observer congruence)
of the electromagnetic tidal tensors.

\emph{The tensor} $\mathbb{F}_{\alpha\beta}$ --- whereas Maxwell's
equations can be fully expressed in terms of tidal tensors and sources,
the same is only true, in general, for the temporal part of Einstein's
equations. The Space-Space part, Eq.~(\ref{eq:SpaceSpaceEinsteinTidal}),
involves the tensor $\mathbb{F}_{\alpha\beta}$, which has no electromagnetic
analogue. This tensor, however, is not an additional independent object,
as it is completely determined via (\ref{eq:SpaceSpaceEinsteinTidal})
given the sources and $\mathbb{E}_{\alpha\beta}$. In vacuum $\mathbb{F}_{\alpha\beta}=-\mathbb{E}_{\alpha\beta}$.

\emph{Sources} --- The source of the gravitational field is the rank
two energy momentum tensor $T^{\alpha\beta}$, whereas the source
of the electromagnetic field is the current 4-vector $j^{\alpha}$.
Using the projectors (\ref{eq:Time_Space_Proj}) one can split $T^{\alpha\beta}=\rho U^{\alpha}U^{\beta}+2U^{(\alpha}h_{\ \mu}^{\beta)}J^{\mu}+T^{\langle\alpha\rangle\langle\beta\rangle}$,
and $j^{\alpha}=\rho_{c}U^{\alpha}+h_{\ \mu}^{\alpha}j^{\mu}$. Eqs.
(\ref{analogy}.4) show that the source of $E_{\alpha\beta}$ is $\rho_{c}$,
and its gravitational analogue, as the source of $\mathbb{E}_{\alpha\beta}$,
is $2\rho+T_{\ \alpha}^{\alpha}$ ($\rho+3p$ for a perfect fluid).
The magnetic/gravitomagnetic tidal tensors are analogously sourced
by the charge/mass-energy currents $j^{\langle\mu\rangle}$/$J^{\langle\mu\rangle}$,
as shown by Eqs.~(\ref{analogy}.5). Note that, when the Maxwell
tensor is covariantly constant \emph{along }the observer's worldline,
$\star F_{\alpha\beta;\gamma}U^{\gamma}$ vanishes and equations (\ref{analogy}.5a)
and (\ref{analogy}.5b) match up to a factor of 2, identifying $j^{\langle\mu\rangle}\leftrightarrow J^{\langle\mu\rangle}$.
Eq.~(\ref{eq:FSource}) shows that $\rho$ is the source of $\mathbb{F}_{\alpha\beta}$.
Eq.~(\ref{analogy}.6), sourced by the space-space part $T^{\langle\alpha\rangle\langle\beta\rangle}$,
as well as the contribution $T_{\ \alpha}^{\alpha}$ for (\ref{analogy}.4b),
manifest the well known fact that in gravity, by contrast with electromagnetism,
pressure and stresses act as sources of the field.

\emph{Symmetries and time projections of tidal tensors} --- The gravitational
and electromagnetic tidal tensors do not generically exhibit the same
symmetries; moreover, the former tidal tensors\pagebreak{}
\begin{table}[h]
\caption{{\small{{\label{analogy}The gravito-electromagnetic analogy based
on tidal tensors. }}}}

\centering{}{\small{{\setlength{\arrayrulewidth}{0.8pt}}}}%
\begin{tabular}{>{\centering}p{39.6ex}c>{\centering}p{41.9ex}c}
\hline 
\multicolumn{2}{c}{{\small{{\raisebox{3.2ex}{}\raisebox{0.5ex}{Electromagnetism}}}}} & \multicolumn{2}{c}{{\small{{\raisebox{3.2ex}{}\raisebox{0.5ex}{Gravity}}}}}\tabularnewline
\hline 
{\small{{\raisebox{3ex}{}Worldline deviation:}}}  &  & {\small{{\raisebox{3ex}{}Geodesic deviation:}}}  & \tabularnewline
{\small{{\raisebox{6ex}{}\raisebox{2ex}{${\displaystyle \frac{D^{2}\delta x^{\alpha}}{d\tau^{2}}=\frac{q}{m}E_{\,\,\,\beta}^{\alpha}\delta x^{\beta}},\,\,\, E_{\,\,\,\beta}^{\alpha}\equiv F_{\ \mu;\beta}^{\alpha}U^{\mu}$}}}}  & {\small{{\raisebox{2ex}{(\ref{analogy}.1a)}}}}  & {\small{{\raisebox{5.5ex}{}\raisebox{2ex}{${\displaystyle \frac{D^{2}\delta x^{\alpha}}{d\tau^{2}}=-\mathbb{E}_{\,\,\,\beta}^{\alpha}\delta x^{\beta}},\,\,\,\mathbb{E}_{\,\,\,\beta}^{\alpha}\equiv R_{\,\,\,\mu\beta\nu}^{\alpha}U^{\mu}U^{\nu}$}}}}  & {\small{{\raisebox{2ex}{(\ref{analogy}.1b)}}}}\tabularnewline
\hline 
{\small{{\raisebox{3ex}{}Force on magnetic dipole:}}}  &  & {\small{{\raisebox{3ex}{}Force on gyroscope:}}}  & \tabularnewline
{\small{{\raisebox{6ex}{}\raisebox{2ex}{${\displaystyle F_{EM}^{\beta}=B_{\alpha}^{\,\,\,\beta}\mu^{\alpha}},\,\,\, B_{\,\,\,\beta}^{\alpha}\equiv\star F_{\ \mu;\beta}^{\alpha}U^{\mu}$}}}}  & {\small{{\raisebox{2ex}{(\ref{analogy}.2a)}}}}  & {\small{{\raisebox{6ex}{}\raisebox{2ex}{~~${\displaystyle F_{G}^{\beta}=-\mathbb{H}_{\alpha}^{\,\,\,\beta}S^{\alpha}},\,\,\,\mathbb{H}_{\,\,\,\beta}^{\alpha}\equiv\star R_{\,\,\,\mu\beta\nu}^{\alpha}U^{\mu}U^{\nu}$}}}}  & {\small{{\raisebox{2ex}{(\ref{analogy}.2b)}}}}\tabularnewline
\hline 
{\small{{\raisebox{3ex}{}Differential precession of magnetic dipoles:}}}  &  & {\small{{Differential precession of gyroscopes:}}}  & \tabularnewline
{\small{{\raisebox{6ex}{}\raisebox{2ex}{$\delta\Omega_{{\rm EM}}^{i}=-\sigma B_{\ \beta}^{i}\delta x^{\beta}$}}}}  & {\small{{\raisebox{2ex}{(\ref{analogy}.3a)}}}}  & {\small{{\raisebox{6ex}{}\raisebox{2ex}{$\delta\Omega_{{\rm G}}^{i}=\mathbb{H}_{\ \beta}^{i}\delta x^{\beta}$}}}}  & {\small{{\raisebox{2ex}{(\ref{analogy}.3b)}}}}\tabularnewline
\hline 
{\small{{\raisebox{3ex}{}Maxwell Source Equations}}}  &  & {\small{{\raisebox{3ex}{}Einstein Equations}}}  & \tabularnewline
{\small{{\raisebox{3ex}{}$F_{\,\,\,\,;\beta}^{\alpha\beta}=4\pi j^{\alpha}$}}}  &  & {\small{{$R_{\mu\nu}=8\pi\left(T_{\mu\nu}-\frac{1}{2}g_{\mu\nu}T_{\,\,\,\alpha}^{\alpha}\right)$}}}  & \tabularnewline
{\small{{\raisebox{4ex}{}$\bullet$ Time Projection:}}}  &  & {\small{{$\bullet$ Time-Time Projection:}}}  & \tabularnewline
{\small{{\raisebox{3ex}{}$E_{\,\,\,\alpha}^{\alpha}=4\pi\rho_{c}$}}}  & {\small{{(\ref{analogy}.4a)}}}  & {\small{{\raisebox{3ex}{}$\mathbb{E}_{\,\,\,\alpha}^{\alpha}=4\pi\left(2\rho+T_{\,\,\alpha}^{\alpha}\right)$}}}  & {\small{{(\ref{analogy}.4b)}}}\tabularnewline
{\small{{\raisebox{3.8ex}{}$\bullet$ Space Projection:}}}  &  & {\small{{$\bullet$ Time-Space Projection:}}}  & \tabularnewline
{\small{{\raisebox{3ex}{}$B_{[\alpha\beta]}=\frac{1}{2}\star F_{\alpha\beta;\gamma}U^{\gamma}-2\pi\epsilon_{\alpha\beta\sigma\gamma}j^{\sigma}U^{\gamma}$~~}}}  & {\small{{(\ref{analogy}.5a)}}}  & {\small{{$\mathbb{H}_{[\alpha\beta]}=-4\pi\epsilon_{\alpha\beta\sigma\gamma}J^{\sigma}U^{\gamma}$}}}  & {\small{{(\ref{analogy}.5b)}}}\tabularnewline
 &  & {\small{{\raisebox{3.8ex}{}$\bullet$ Space-Space Projection:}}}  & \tabularnewline
{\small{{\raisebox{6.2ex}{}\raisebox{2.5ex}{$No\,\, electromagnetic\,\, analogue$}}}}  &  & {\small{{\raisebox{2.5ex}{$\mathbb{F}_{\ \beta}^{\alpha}+\mathbb{E}_{\ \beta}^{\alpha}-\mathbb{F}_{\ \sigma}^{\sigma}h_{\ \beta}^{\alpha}=8\pi\left[\frac{1}{2}T_{\ \gamma}^{\gamma}h_{\ \beta}^{\alpha}-T_{\ \langle\beta\rangle}^{\langle\alpha\rangle}\right]$}~~}}}  & {\small{{\raisebox{2ex}{(\ref{analogy}.6)}}}}\tabularnewline
\hline 
{\small{{\raisebox{3ex}{}Bianchi Identity}}}  &  & {\small{{\raisebox{3ex}{}Algebraic Bianchi Identity}}}  & \tabularnewline
{\small{{\raisebox{3ex}{}$\ \star F_{\ \ \ ;\beta}^{\alpha\beta}=0\quad(\Leftrightarrow F_{[\alpha\beta;\gamma]}=0\ )$}}}  &  & {\small{{\raisebox{3ex}{}$\star R_{\ \ \ \gamma\beta}^{\gamma\alpha}=0\quad(\Leftrightarrow R_{[\alpha\beta\gamma]\delta}=0)$}}}  & \tabularnewline
{\small{{\raisebox{4ex}{}$\bullet$ Time Projection:}}}  &  & {\small{{$\bullet$ Time-Time (or Space-Space) Proj.:}}}  & \tabularnewline
{\small{{\raisebox{3ex}{}$B_{\,\,\,\alpha}^{\alpha}=0$}}}  & {\small{{(\ref{analogy}.7a)}}}  & {\small{{$\mathbb{H}_{\,\,\,\alpha}^{\alpha}=0$}}}  & {\small{{(\ref{analogy}.7b)}}}\tabularnewline
{\small{{\raisebox{3.8ex}{}$\bullet$ Space Projection:}}}  &  & {\small{{$\bullet$ Space-Time Projection:}}}  & \tabularnewline
{\small{{\raisebox{3.5ex}{}$E_{[\alpha\beta]}=\frac{1}{2}F_{\alpha\beta;\gamma}U^{\gamma}$}}}  & {\small{{(\ref{analogy}.8a)}}}  & {\small{{\raisebox{3.5ex}{}$\mathbb{E}_{[\alpha\beta]}=0$}}}  & {\small{{(\ref{analogy}.8b)}}}\tabularnewline
 &  & {\small{{\raisebox{3.8ex}{}$\bullet$ Time-Space Projection:}}}  & \tabularnewline
{\small{{\raisebox{2.5ex}{$No\,\, electromagnetic\,\, analogue$}}}}  &  & {\small{{\raisebox{2ex}{}\raisebox{1.2ex}{$\mathbb{F}_{[\alpha\beta]}=0$}}}}  & \tabularnewline
\hline 
\end{tabular}{\small{{{} }}} 
\end{table}
 are spatial, whereas the latter have a time projection (with respect
to the observer measuring them), signaling fundamental differences
between the two interactions. In the general case of fields that vary
along the observer's worldline (that is the case of an intrinsically
non-stationary field, or an observer moving in a stationary non-uniform
field), $E_{\alpha\beta}$ possesses an antisymmetric part; $\mathbb{E}_{\alpha\beta}$,
by contrast, is always symmetric. $E_{[\alpha\beta]}$ encodes Faraday's
law of induction: as discussed above, $E_{\alpha\beta}$ is a covariant
derivative of the electric field as measured in the momentarily comoving
reference frame (MCRF); thus Eq.~(\ref{eq:SpaceE[ab]B[ab]}a) is
a covariant way of writing the Maxwell-Faraday equation $\nabla\times\vec{E}=-\partial\vec{B}/\partial t$.
Therefore, the statement encoded in the equation $\mathbb{E}_{[\alpha\beta]}=0$
is that \textit{\emph{there are no analogous induction effects in
the }}\textit{physical}\textit{\emph{ (i.e., tidal) gravitational
forces}} (in the language of GEM vector fields of Sec.~\ref{sec:3+1},
we can say that the curl of the gravitoelectric field $\vec{G}$ does
not manifest itself in the tidal forces, unlike its electromagnetic
counterpart; see Sec.~\ref{sub:Relation1+3_TTensors} for explicit
demonstration). To see a physical consequence, let $\delta x^{\alpha}$
in Eq.~(\ref{analogy}.1a) --- the separation vector between a pair
of particles with the same $q/m$ and the same 4-velocity $U^{\alpha}$
--- be spatial with respect to $U^{\alpha}$ $(\delta x^{\alpha}U_{\alpha}=0$);
and note that the spatially projected antisymmetric part of $E_{\mu\nu}$
can be written in terms of the dual spatial vector $\alpha^{\mu}$:
$E_{[\langle\mu\rangle\langle\nu\rangle]}=\epsilon_{\mu\nu\gamma\delta}\alpha^{\gamma}U^{\delta}$.
Then the spatial components (\ref{analogy}.1a) can be written as
(using $E_{\langle\mu\rangle\langle\nu\rangle}=E_{(\langle\mu\rangle\langle\nu\rangle)}+E_{[\langle\mu\rangle\langle\nu\rangle]}$):
\begin{equation}
\frac{D^{2}\delta x_{\langle\mu\rangle}}{d\tau^{2}}=\frac{q}{m}\left[E_{(\langle\mu\rangle\langle\nu\rangle)}\delta x^{\nu}+\epsilon_{\mu\nu\gamma\delta}\alpha^{\gamma}U^{\delta}\delta x^{\nu}\right]\quad\Leftrightarrow\quad\frac{D^{2}\delta\vec{x}}{d\tau^{2}}=\frac{q}{m}\left[\overleftrightarrow{E}\cdot\delta\vec{x}+\delta\vec{x}\times\vec{\alpha}\right],\label{eq:TidalEM}
\end{equation}
the second equation holding in the frame $U^{i}=0$, where we used
the dyadic notation $\overleftrightarrow{E}$ of e.g.~\cite{CampbelMacekMorgan}.
From the form of the second equation we see that $q\vec{\alpha}/m$
is \emph{minus} an angular acceleration. Using equation (\ref{eq:SpaceE[ab]B[ab]}a),
we see that $\alpha^{\mu}=B_{\ \beta}^{\mu}U^{\beta}$; and in an
inertial frame $\vec{\alpha}=\partial\vec{B}/\partial t=-\nabla\times\vec{E}$.
In the gravitational case, since $\mathbb{E}_{\mu\nu}=\mathbb{E}_{(\mu\nu)}=\mathbb{E}_{\langle\mu\rangle\langle\nu\rangle}$,
we have 
\begin{equation}
\frac{D^{2}\delta x_{\langle\mu\rangle}}{d\tau^{2}}=\frac{D^{2}\delta x_{\mu}}{d\tau^{2}}=-\mathbb{E}_{(\mu\nu)}\delta x^{\nu}\quad\Leftrightarrow\quad\frac{D^{2}\delta\vec{x}}{d\tau^{2}}=-\overleftrightarrow{\mathbb{E}}\cdot\delta\vec{x}\,\,.\label{eq:TidalGR}
\end{equation}
That is, given a set of neighboring charged test particles, the electromagnetic
field ``shears'' the set via $E_{(\mu\nu)}$, and induces an accelerated
rotation%
\footnote{By rotation we mean here absolute rotation, i.e, measured with respect
to a comoving Fermi-Walker transported frame. As one can check from
the connection coefficients (\ref{eq:Connectioni0j}) below, in such
frame ($\Omega_{\alpha\beta}=0$) we have $d^{2}\delta x^{\hat{i}}/d\tau^{2}=D^{2}\delta x^{\hat{i}}/d\tau^{2}$.
See also in this respect \cite{BiniStrains}.%
} via the laws of electromagnetic induction encoded in $E_{[\mu\nu]}$.
The gravitational field, by contrast, only shears%
\footnote{If the two particles were connected by a ``rigid'' rod then the
symmetric part of the electric tidal tensor would also, in general,
torque the rod; hence in such system we would have a rotation even
in the gravitational case, see \cite{Black Holes} pp.~154-155. The
same is true for a quasi-rigid extended body; however, even in this
case the effects due to the symmetric part are very different from
the ones arising from electromagnetic induction: first, the former
do not require the fields to vary along the particle's worldline,
they exist even if the body is at rest in a stationary field; second,
they vanish if the body is spherical, which does not happen with the
torque generated by the induced electric field, see \cite{Gyros}.%
} the set, since $\mathbb{E}_{[\mu\nu]}=0$.

Further physical evidence for the absence of a gravitational analogue
for Faraday's law of induction in the \emph{physical} forces and torques
is given in the companion paper \cite{Gyros}: consider a spinning
spherical charged body in an electromagnetic field; and choose the
MCRF; if the magnetic field is not constant in this frame, by virtue
of equation $\nabla\times\vec{E}=-\partial\vec{B}/\partial t$, a
torque will in general be exerted on the body by the induced electric
field, changing its angular momentum and kinetic energy of rotation.
By contrast, \emph{no gravitational torque} is exerted on a spinning
``spherical'' body (i.e., a particle whose multipole moments in
a local orthonormal frame match the ones of a spherical body in flat
spacetime) placed in an \emph{arbitrary} gravitational field; its
angular momentum and kinetic energy of rotation are \emph{constant}.

As discussed in the previous section, the symmetry of $\mathbb{E}_{\alpha\beta}$
follows from the algebraic Bianchi identity $R_{\ [\beta\gamma\delta]}^{\alpha}=0$;
this identity states that $R_{\ \beta\gamma\delta}^{\alpha}$ is the
curvature tensor of a connection with vanishing torsion (the Levi-Civita
connection of the space-time manifold). So one can say that the absence
of electromagnetic-like induction effects is the statement that the
physical gravitational forces are described by the curvature tensor
of a connection without torsion.

There is also an antisymmetric contribution $\star F_{\alpha\beta;\gamma}U^{\gamma}$
to $B_{\alpha\beta}$; in vacuum, Eq.~(\ref{analogy}.5a) is a covariant
form of $\nabla\times\vec{B}=\partial\vec{E}/\partial t$; hence,
the fact that, in vacuum, $\mathbb{H}_{[\alpha\beta]}=0$, means that
there is no gravitational analogue to the antisymmetric part $B_{[\alpha\beta]}$
(i.e., the curl of $\vec{B}$) induced by the time varying field $\vec{E}$.
Some physical consequences of this fact are explored in \cite{Gyros}:
Eq.~(\ref{analogy}.5a) implies, via (\ref{analogy}.2a), that whenever
a magnetic dipole moves in a non-homogeneous field, it measures a
non-vanishing $B_{[\alpha\beta]}$ (thus also $B_{\alpha\beta}\ne0$),
and therefore (except for very special orientations of the dipole
moment $\mu^{\alpha}$) a force will be exerted on it; in the gravitational
case, by contrast, the gravitational force on a gyroscope is not constrained
to be non-vanishing when it moves in a non-homogeneous field; it is
found that it may actually move along geodesics, as is the case of
radial motion in Schwarzschild spacetime%
\footnote{This example is particularly interesting in this discussion. In the
electromagnetic analogous problem, a magnetic dipole in (initially)
radial motion in the Coulomb field of a point charge experiences a
force; that force, as shown in \cite{Gyros}, comes entirely from
the antisymmetric part of the magnetic tidal tensor, $B_{\alpha\beta}=B_{[\alpha\beta]}$;
it is thus a natural realization of the arguments above that $\mathbb{H}_{\alpha\beta}=0$
in the analogous gravitational problem.%
}, or circular geodesics in Kerr-dS spacetime.

The spatial character of the gravitational tidal tensors, contrasting
with their electromagnetic counterparts, is another difference in
the tensorial structure related to the laws of electromagnetic induction:
as can be seen from Eqs.~(\ref{Eanti}) and (\ref{Banti}), the antisymmetric
parts of $E_{\alpha\beta}$ and $B_{\alpha\beta}$ (in vacuum, for
the latter) consist of time projections of these tidal tensors. Physically,
these time projections are manifest for instance in the fact that
the electromagnetic force on a magnetic dipole has a non-vanishing
projection along the particle's 4-velocity $U^{\alpha}$, which is
the rate of work done on it by the induced electric field \cite{CHPRD,Gyros},
and is reflected in a variation of the particle's proper mass. The
projection, along $U^{\alpha}$, of the gravitational force (\ref{analogy}.2b),
in turn, vanishes, and the gyroscope's proper mass is constant.

\subsection{The analogy for differential precession\label{sub:Diff-Precession}}

Eqs.~(\ref{analogy}.2) in Table \ref{analogy} give $B_{\alpha\beta}$
and $\mathbb{H}_{\alpha\beta}$ a physical interpretation as the tensors
which, when contracted with a magnetic/gravitomagnetic dipole vector,
yield the force exerted on magnetic dipoles/gyroscopes. We will now
show that these tensors can also be interpreted as tensors of ``relative'',
or ``differential'', precession for these test particles; i.e.,
tensors that, when contracted with a separation vector $\delta x^{\beta}$,
yield the angular velocity of precession of a spinning particle at
a point $\mathcal{P}_{2}$ relative to a system of axes anchored to
spinning particles, with \emph{the same 4-velocity }(and the same
gyromagnetic ratio $\sigma$, if an electromagnetic field is present),
at the infinitesimally close point $\mathcal{P}_{1}$. This is analogous
to the electric tidal tensors $E_{\alpha\beta}$ and $\mathbb{E}_{\alpha\beta}$,
which, when contracted with $\delta x^{\beta}$, yield the relative
acceleration of two infinitesimally close test particles with the
same 4-velocity (and the same ratio $q/m$ in the electromagnetic
case).

For clarity we will treat the gravitational and electromagnetic interactions
separately. We will start by the gravitational problem. Our goal is
to compute the precession of a gyroscope at some point $\mathcal{P}_{2}$
relative to a frame attached to guiding gyroscopes at the neighboring
point $\mathcal{P}_{1}$. Let $U^{\alpha}$ be the 4-velocity of the
gyroscope, and $a^{\alpha}=DU^{\alpha}/d\tau$ its acceleration. According
to the Mathisson-Papapetrou equations \cite{MathissonNeueMechanik,Papapetrou I},
no torque is exerted on a gyroscope (taken as a \emph{pole-dipole}
particle) in a gravitational field; if one moreover assumes the Mathisson-Pirani
\cite{MathissonNeueMechanik,Pirani 1956} spin condition $S^{\alpha\beta}U_{\beta}=0$,
it follows from these equations that its spin vector undergoes Fermi-Walker
transport, 
\begin{equation}
\frac{DS^{\alpha}}{d\tau}=S_{\nu}a^{\nu}U^{\alpha}\ ,\label{eq:Fermi-Walker}
\end{equation}
(for more details, see ~\cite{Gyros}; in the comoving frame, the
spatial part reads $D\vec{S}/d\tau=0$). Thus the frame we are looking
for is a tetrad Fermi-Walker transported along the worldline $L$
of the set of gyroscopes 1 (passing trough the location $\mathcal{P}_{1}$).
There is a locally rectangular coordinate system associated to such
tetrad, the so-called%
\footnote{Following Synge \cite{Synge}, by Fermi coordinates we mean the locally
rectangular coordinate system associated to a tetrad Fermi-Walker
transported along a worldline (the curve being the origin of the frame,
and its tangent the time axis). Note the existence of different conventions
in the literature: the so-called ``Fermi normal coordinates'' of
e.g. \cite{ManasseMisner,Gravitation} are a special case of our definition,
for the case that the worldline is geodesic. The ``Fermi coordinates''
of \cite{Nesterov}, in turn, are a generalization of our definition,
for the case that the tetrad is not Fermi-Walker transported.%
} ``Fermi coordinates''; let $\mathbf{e}_{\alpha}$ denote its basis
vectors and $\Gamma_{\beta\gamma}^{\alpha}$ its Christoffel symbols,
$\Gamma_{\beta\gamma}^{\alpha}\mathbf{e}_{\alpha}=\nabla_{\mathbf{e}_{\beta}}\mathbf{e}_{\gamma}$.
The vectors $\mathbf{e}_{\alpha}$ are Fermi-Walker transported along
$L$, so $\left\langle \nabla_{\mathbf{e}_{0}}\mathbf{e}_{i},\mathbf{e}_{j}\right\rangle |_{\mathcal{P}_{1}}=0\Rightarrow\Gamma_{0i}^{j}(\mathcal{P}_{1})=0$.
Hence, a gyroscope at $\mathcal{P}_{1}$, momentarily at rest in this
frame, by Eq. (\ref{eq:Fermi-Walker}) obviously does not precess
relative to it, $d\vec{S}/d\tau|_{\mathcal{P}_{1}}=\dot{\vec{S}}|_{\mathcal{P}_{1}}=0$.
Here the dot denotes ordinary derivative along $\mathbf{e}_{0}$,
$\dot{A}^{\alpha}\equiv\partial_{0}A^{\alpha}$. However, outside
$L$, the basis vectors $\mathbf{e}_{\alpha}$ are no longer Fermi-Walker
transported, $\left\langle \nabla_{\mathbf{e}_{0}}\mathbf{e}_{i},\mathbf{e}_{j}\right\rangle |_{\mathcal{P}_{2}}\ne0\Rightarrow\Gamma_{0i}^{j}(\mathcal{P}_{2})\ne0$.
That means that gyroscope 2, at a point $\mathcal{P}_{2}$ (outside
$L$), will be be seen to precess relative to the frame $\mathbf{e}_{\alpha}$:
$d\vec{S}_{2}/d\tau_{2}|_{\mathcal{P}_{2}}\ne0$. If the gyroscope
is at rest in this frame ($U_{2}^{i}=0$), we have 
\begin{equation}
\left.\frac{dS_{2}^{i}}{d\tau_{2}}\right|_{\mathcal{P}_{2}}=-\Gamma_{0j}^{i}(\mathcal{P}_{2})S_{2}^{j}U_{2}^{0}\ .\label{eq:RelPrecGrav00}
\end{equation}
The Christoffel symbol $\Gamma_{0j}^{i}(\mathcal{P}_{2})$, to first
order in $\delta x$, can be obtained from e.g. Eqs. (20) of \cite{Nesterov}
(making $\Omega_{ij}=0$ therein); it reads $\Gamma_{0j}^{i}(\mathcal{P}_{2})=R_{\ jk0}^{i}(\mathcal{P}_{1})\delta x^{k}$.
From Eq.~(\ref{eq:SpaceProj2Form}) above, we note that 
\[
R_{\langle\alpha\rangle\langle\beta\rangle\gamma\tau\ }=\epsilon_{\ \ \sigma\delta}^{\mu\nu}\epsilon_{\mu\alpha\beta\lambda}U^{\lambda}U_{\nu}R_{\ \ \gamma\tau\ }^{\sigma\delta}=\epsilon_{\mu\alpha\beta\lambda}U^{\lambda}U_{\nu}\star\! R_{\ \ \gamma\tau\ }^{\mu\nu}\;,
\]
which, in the Fermi frame $\mathbf{e}_{\alpha}$ (orthonormal at $\mathcal{P}_{1}$),
reads: $R_{ij\gamma\tau\ }=\epsilon_{ijk}\star\! R_{\ 0\gamma\tau\ }^{k}$.
We thus have 
\[
\Gamma_{0j}^{i}(\mathcal{P}_{2})=\epsilon_{\ jk}^{i}\star\! R_{\ 0l0}^{k}(\mathcal{P}_{1})\delta x^{l}=\epsilon_{\ jk}^{i}\mathbb{H}_{\ l}^{k}\delta x^{l}\ .
\]
$\Gamma_{0j}^{i}(\mathcal{P}_{2})$ is an antisymmetric matrix, which
we can write as $\Gamma_{0j}^{i}(\mathcal{P}_{2})=\epsilon_{\ jk}^{i}\delta\Omega_{{\rm G}}^{k}$,
where 
\begin{equation}
\delta\Omega_{{\rm G}}^{i}\equiv\mathbb{H}_{\ l}^{i}\delta x^{l}\ .\label{eq:RelPrecGrav0}
\end{equation}
Substituting into (\ref{eq:RelPrecGrav00}), and noting that $U_{2}^{0}=(-g_{00})^{-1/2}=1+\mathcal{O}(\delta x^{2})$,
we have, to first order in $\delta x$, 
\begin{equation}
\left.\frac{d\vec{S}_{2}}{d\tau_{2}}\right|_{\mathcal{P}_{2}}=\dot{\vec{S}}_{2}|_{\mathcal{P}_{2}}=\delta\vec{\Omega}_{{\rm G}}\times\vec{S}_{2}\ ,\label{eq:RelPrecGrav}
\end{equation}
where in the first equality we noted that %
$d\vec{S}_{2}/d\tau_{2}|_{\mathcal{P}_{2}}=\dot{\vec{S}}_{2}|_{\mathcal{P}_{2}}+\mathcal{O}(\delta x^{3})$.
Thus, $\delta\vec{\Omega}_{{\rm G}}$ is the angular velocity of precession
of gyroscopes at $\mathcal{P}_{2}$ with respect to the Fermi frame
$\mathbf{e}_{i}$, locked to the guiding gyroscopes at $\mathcal{P}_{1}$.
Obviously, this is just \emph{minus} the angular velocity of rotation
of the basis vectors $\mathbf{e}_{i}$ relative to Fermi-Walker transport
\emph{at} $\mathcal{P}_{2}$. This result was first obtained in a
recent work \cite{DiffPrecession} through a different procedure;
we believe the derivation above is more clear, and shows that one
of the assumptions made in \cite{DiffPrecession} to obtain $\delta\vec{\Omega}_{{\rm G}}$
--- that the gyroscopes at $\mathcal{P}_{1}$ and $\mathcal{P}_{2}$
have the same acceleration --- is not necessary; in order for (\ref{eq:RelPrecGrav})
to hold, they only need to be momentarily comoving%
\footnote{The relative velocity of objects at different points is not a well
defined notion in curved spacetime, since there is no natural way
of comparing vectors at different points, see e.g. \cite{BolosIntrinsic,BolosLightlike}.
The notion of relative velocity implied above is dubbed in \cite{BolosIntrinsic}
``Fermi relative velocity''.%
}, in the sense that they have zero 3-velocity in the Fermi coordinate
system of $\mathcal{P}_{1}$. It is also worth noting that if the
basis worldline $L$ through $\mathcal{P}_{1}$ is geodesic, then
(\ref{eq:RelPrecGrav}) still holds when the gyroscopes at $\mathcal{P}_{2}$
have an infinitesimal velocity $v\propto\delta x$ in such frame (as
is straightforward to check by a similar computation for the case
$U_{2}^{i}\ne0$). Hence Eq. (\ref{eq:RelPrecGrav}) applies to gyroscopes
carried by \emph{infinitesimally close} observers belonging to an
arbitrary \emph{geodesic congruence} (in a certain analogy with the
geodesic deviation equation (\ref{analogy}.1b) of Table \ref{analogy}).
The differential dragging effect in terms of the eigenvectors of $\mathbb{H}_{\alpha\beta}$
and their associated field lines, as well as their visualization in
different spacetimes, are discussed in detail in \cite{DiffPrecession}.

Let us turn now to the analogous electromagnetic problem. Consider,
in flat spacetime, a triad of orthogonal magnetic dipoles (magnetic
moment $\mu^{\alpha}=\sigma S^{\alpha}$), with the same gyromagnetic
ratio $\sigma$ (so that they all precess with the same frequency),
moving along a worldline $L$ of tangent $\mathbf{U}$ passing through
$\mathcal{P}_{1}$. If the Mathisson-Pirani condition holds, the spin
vector of a magnetic dipole evolves along $L$ as (e.g.~\cite{Gyros}):

\begin{equation}
\frac{DS^{\mu}}{d\tau}=S_{\nu}a^{\nu}U^{\mu}+\sigma\epsilon_{\ \alpha\beta\nu}^{\mu}U^{\nu}S^{\alpha}B^{\beta}\;,\label{eq:SpinVector}
\end{equation}
where $B^{\alpha}\equiv\star F^{\alpha\beta}U_{\beta}$. The second
term marks an obvious difference with the case of the gyroscope, as
it means that magnetic dipoles \emph{do precess} (with angular velocity
$\Omega^{\alpha}=-\sigma B^{\alpha}$) with respect to the comoving
Fermi-Walker transported frame. We shall compare this precession for
dipoles at infinitesimally close points, and relate with the gravitational
analogue. In the Fermi frame $\mathbf{e}_{\alpha}$ with origin at
$L$, the space part of Eq. (\ref{eq:SpinVector}) reads (at $\mathcal{P}_{1}$),
\begin{equation}
\left.\frac{d\vec{S}}{d\tau}\right|_{\mathcal{P}_{1}}=\vec{\Omega}(\mathcal{P}_{1})\times\vec{S}\;;\qquad\vec{\Omega}(\mathcal{P}_{1})\equiv-\sigma\vec{B}(\mathcal{P}_{1})\;.\label{eq:PrecEM1}
\end{equation}

Now consider a magnetic dipole at the neighboring point $\mathcal{P}_{2}$,
\emph{at rest in the frame} $\mathbf{e}_{\alpha}$ (i.e. $U_{2}^{i}=0$,
so that we are using the same notion of comoving as in the gravitational
case); we have 
\begin{equation}
\left.\frac{dS_{2}^{i}}{d\tau_{2}}\right|_{\mathcal{P}_{2}}=-\Gamma_{0j}^{i}(\mathcal{P}_{2})S_{2}^{j}+(\vec{\Omega}(\mathcal{P}_{2})\times\vec{S}_{2})^{i}=(\vec{\Omega}(\mathcal{P}_{2})\times\vec{S}_{2})^{i}\;;\qquad\vec{\Omega}(\mathcal{P}_{2})\equiv-\sigma\vec{B}(\mathcal{P}_{2})\ ,\label{eq:PrecEM2}
\end{equation}
where we noted that, in flat spacetime, $\Gamma_{0j}^{i}=0$ everywhere
in a Fermi frame (cf. Eqs (20) of \cite{Nesterov}). Being at rest
in the frame $\mathbf{e}_{\alpha}$ implies, in flat spacetime%
\footnote{This can be easily seen from the fact that the triad $\mathbf{e}_{i}$
coincides with the basis vectors of a momentarily comoving inertial
frame; thus $U_{2}^{i}=0$ implies that, in the inertial frame, $U_{2}^{\alpha}=(1,\vec{0})=U^{\alpha}$.
This can also be seen from Eqs. (22)-(23) of \cite{BolosIntrinsic},
to which we refer for a detailed discussion of the Fermi relative
velocity in flat spacetime.%
}, $\mathbf{U}_{2}=\mathbf{U}$ (note that parallelism between vectors
at different points is well defined herein). Thus, $dS_{2}^{i}/d\tau_{2}\equiv(S_{2})_{\ ,\alpha}^{i}U_{2}^{\alpha}=dS_{2}^{i}/d\tau$,
and $B^{\alpha}(\mathcal{P}_{2})\equiv\star F_{\ \beta}^{\alpha}(\mathcal{P}_{2})U_{2}^{\beta}=\star F_{\ \beta}^{\alpha}(\mathcal{P}_{2})U^{\beta}$.
Performing a Taylor expansion of $F_{\ \beta}^{\alpha}$ about $\mathcal{P}_{1}$
(and using \emph{for this operation} a rectangular coordinate system,
which one can always do in flat spacetime, so that $F_{\ \beta,\gamma}^{\alpha}=F_{\ \beta;\gamma}^{\alpha}$),
we may write $F_{\ \beta}^{\alpha}(\mathcal{P}_{2})=F_{\ \beta}^{\alpha}(\mathcal{P}_{2})+F_{\ \beta;\gamma}^{\alpha}(\mathcal{P}_{1})\delta x^{\gamma}+\mathcal{O}(\delta x^{2})$.
Therefore, 
\begin{equation}
B^{\alpha}(\mathcal{P}_{2})=B^{\alpha}(\mathcal{P}_{1})+B_{\ \gamma}^{\alpha}(\mathcal{P}_{1})\delta x^{\gamma}+\mathcal{O}(\delta x^{2})\ ,\label{eq:Bexpansion}
\end{equation}
where $B_{\alpha\beta}=\star F_{\alpha\gamma;\beta}U^{\gamma}$ is
the magnetic tidal tensor as defined in Eq. (\ref{analogy}.2a) of
Table \ref{analogy}. Eqs. (\ref{eq:PrecEM1}) and (\ref{eq:PrecEM2})
are precessions measured with respect to the same frame $\mathbf{e}_{\alpha}$;
taking the difference $\delta\vec{\Omega}_{{\rm EM}}$, we obtain
\begin{equation}
\delta\Omega_{{\rm EM}}^{i}=\Omega^{i}(\mathcal{P}_{2})-\Omega^{i}(\mathcal{P}_{1})=-\sigma B_{\ \gamma}^{i}\delta x^{\gamma}\ ,\label{eq:RelPrecEM}
\end{equation}
which is analogous to Eq. (\ref{eq:RelPrecGrav0}), only with $-\sigma B_{\alpha\beta}$
in the place of $\mathbb{H}_{\alpha\beta}$.

It should be mentioned, however, that Eq. (\ref{eq:RelPrecEM}) does
\emph{not}, \emph{in general}, yield the precession of dipole 2 with
respect to a frame whose axes are \emph{anchored} to the spin vectors
of guiding magnetic dipoles at $\mathcal{P}_{1}$ (which would be
perhaps the most natural analogue of the gravitational problem considered
above). Let us denote by $(\mathbf{e}_{{\rm dip}})_{\alpha}$ the
basis vectors of the coordinate system adapted to such frame (originating
at $L$, where it is rectangular, with $(\mathbf{e}_{{\rm dip}})_{0}=\mathbf{U}$;
this is a generalized version of the Fermi-coordinates of $L$, for
the case that the spatial triad is not Fermi-Walker transported).
The spin evolution equation for dipole 2 reads, in this frame, 
\begin{equation}
\left.\frac{d\vec{S}_{2}}{d\tau}\right|_{\mathcal{P}_{2}}=\delta\vec{\Omega}_{{\rm EM}}\times\vec{S}_{2}+(\vec{a}\times\vec{U}_{2})\times\vec{S}_{2}\ ,\label{eq:RelPrecEMdip}
\end{equation}
where we used the connection coefficients given in Eqs. (20) of \cite{Nesterov}
(in particular, $\Gamma_{0j}^{i}(\mathcal{P}_{2})=\Omega_{\ j}^{i}=-\sigma\epsilon_{\ kj}^{i}B^{k}(\mathcal{P}_{1})$),
and noted that dipole 2 (since it is at rest in the Fermi frame $\mathbf{e}_{\alpha}$)
moves in the frame $(\mathbf{e}_{{\rm dip}})_{\alpha}$ with spatial
velocity $U_{2}^{i}\approx-\vec{\Omega}(\mathcal{P}_{1})\times\vec{\delta}x$.
Thus, only when $L$ is geodesic one has in such frame $d\vec{S}_{2}/d\tau|_{\mathcal{P}_{2}}=\delta\vec{\Omega}_{{\rm EM}}\times\vec{S}_{2}$
(as for the acceleration of dipole 2, it can be arbitrary). It should
also be noted that, by contrast with the gravitational Eqs. (\ref{eq:RelPrecGrav0})-(\ref{eq:RelPrecGrav}),
Eqs. (\ref{eq:RelPrecEM})-(\ref{eq:RelPrecEMdip}), do not hold when
the dipoles possess an infinitesimal relative velocity $\delta U\propto\delta x$
(even if the basis worldline $L$ is geodesic, as an extra term $\star F_{\ \beta}^{\alpha}\delta U^{\beta}$
would show up in (\ref{eq:Bexpansion}); $\delta\mathbf{U}=\mathbf{U}_{2}-\mathbf{U}$);
they must be \emph{strictly} comoving. This is analogous to the situation
with the worldline deviation equations (\ref{analogy}.1) of Table
\ref{analogy}, where the gravitational equation allows the particles
to have an infinitesimal deviation velocity, whereas the electromagnetic
one does not (cf. footnote \ref{fn:DeviatonVelocity}).

\section{\label{sec:3+1}Gravito-electromagnetic analogy based on inertial
fields from the 1+3 splitting of spacetime}

This approach has a different philosophy from the tidal tensor analogy
of Sec.~\ref{sec:Tidal tensor analogy}. \textcolor{black}{Therein
we aimed to compare physical, covariant forces of both theories; which
was accomplished through the tidal forces. Herein the analogy drawn
is between the electromagnetic fields $E^{\alpha},\ B^{\alpha}$ and
spatial inertial fields} $G^{\alpha}$, $H^{\alpha}$\textcolor{black}{{}
(i.e., fields of inertial forces, or ``acceleration'' fields), usually
dubbed ``gravitoelectromagnetic'' (GEM) fields, that mimic $E^{\alpha}$
and $B^{\alpha}$ in gravitational dynamics. Inertial forces are fictitious
forces, attached to a specific reference frame, and in this sense
one can regard this analogy as a parallelism between physical forces
from one theory, and reference frame effects from the other.}

\textcolor{black}{The GEM 3-vector fields are best known in the context
of linearized theory for stationary spacetimes, e.g.~\cite{Gravitation and Inertia,Ruggiero:2002hz},
where they are (somewhat naively) formulated as derivatives of the
temporal components of the linearized metric tensor (the GEM potentials,
in analogy with the EM potentials). More general approaches are possible
if one observes that these are fields associated not to the local
properties of a particular spacetime, but, as stated above, to the
kinematical quantities of the reference frame. In particular, the
GEM fields of the usual linearized approaches are but, up to some
factors, the acceleration and vorticity of the congruence of zero
3-velocity observers ($u^{\alpha}\simeq\delta_{0}^{\alpha}$) in the
chosen background. Taking this perspective, the GEM fields may actually
be cast in an} \textcolor{black}{\emph{exact}} \textcolor{black}{form,
applying to arbitrary reference frames in arbitrary fields,} through
a general 1+3 splitting of spacetime. \textcolor{black}{In this section
we present such an exact and general formulation.} We take an arbitrary
orthonormal reference frame, which can be thought as a continuous
field of orthonormal tetrads, or, alternatively, as consisting of
a congruence of observers, each of them carrying an orthonormal tetrad
whose time axis is the observer's 4-velocity; the spatial triads,
spanning the local rest space of the observers, are generically left
arbitrary (namely their rotation with respect to Fermi-Walker transport).
The inertial fields associated to this frame are, in this framework,
encoded in the mixed time-space part of the connection coefficients:
the acceleration $a^{\alpha}$ and vorticity $\omega^{\alpha}$ of
the observer congruence, plus the rotation frequency $\Omega^{\alpha}$
of the spatial triads with respect to Fermi-Walker transport. The
connection coefficients encode also the shear/expansion $K_{(\alpha\beta)}$
of the congruence. A ``gravitoelectric'' field is defined in this
framework as $G^{\alpha}\equiv-a^{\alpha}$, and a gravitomagnetic
field as $H^{\alpha}\equiv\Omega^{\alpha}+\omega^{\alpha}$; the motivation
for these definitions being the geodesic equation, whose space part,
in such frame, resembles the Lorentz force, with $G^{\alpha}$ in
the role of an electric field, $H^{\alpha}$ in the role of a magnetic
field, plus a third term with no electromagnetic analogue, involving
$K_{(\alpha\beta)}$.

The treatment herein \textcolor{black}{is} to a large extent equivalent\textcolor{black}{{}
to what is dubbed in~\cite{The many faces,GEM User Manual}} the
``congruence point of view''; the main difference \textcolor{black}{(apart
from the differences in the formalism) }is that we use a more general
definition of gravitomagnetic field $H^{\alpha}$, allowing for an
arbitrary rotation of the spatial frame (i.e., $\Omega^{\alpha}$
is left arbitrary), so that it adjusts to any frame. We also try to
use a simplified notation, as the one in \textcolor{black}{\cite{The many faces,GEM User Manual}},
albeit very precise and rigorous, is not easy to follow. For that
we work with orthonormal frames in most of our analysis, which are
especially suited for our purposes because the connection coefficients
associated to them are very simply related with the inertial fields.
The price to pay is not having manifestly covariant equations at each
step, by contrast with the formalism in \textcolor{black}{\cite{The many faces,GEM User Manual}}
(although the results are easily written in covariant form).

\subsection{The reference frame\label{sub:The-reference-frame}}

\begin{figure}
\includegraphics[width=0.45\textwidth]{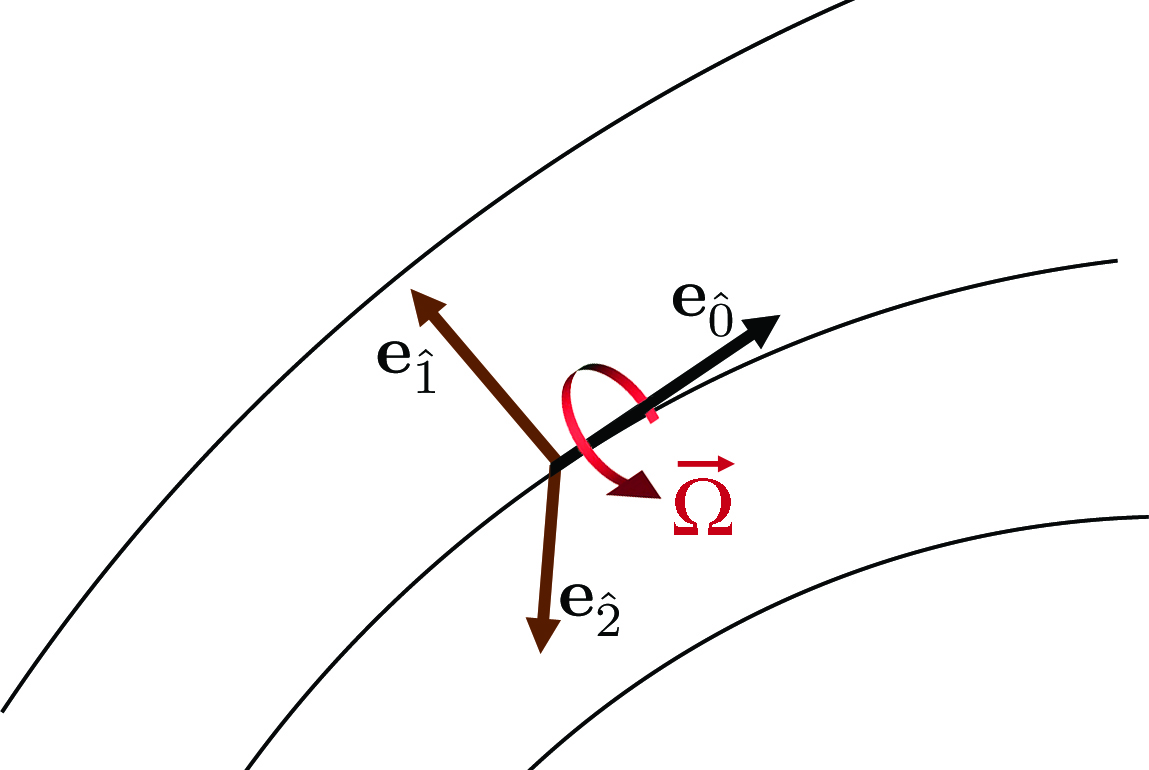}

\caption{\label{fig:RefFrame}The reference frame: a congruence of of time-like
curves --- the observers' worldlines --- of of tangent $\mathbf{u}$;
each observer carries an orthonormal tetrad $\mathbf{e}_{\hat{\alpha}}$
such that $\mathbf{e}_{\hat{0}}=\mathbf{u}$ and the spatial triad
$\mathbf{e}_{\hat{i}}$ spans the observer's local rest space. The
triad $\mathbf{e}_{\hat{i}}$ rotates, relative to Fermi-Walker transport,
with some prescribed angular velocity $\vec{\Omega}$.}
\end{figure}
To an arbitrary observer moving along a worldline of tangent vector
$u^{\alpha}$, one naturally associates an adapted \emph{local} orthonormal
frame (e.g.~\cite{The many faces}), which is a tetrad $\mathbf{e}_{\hat{\alpha}}$
whose time axis is the observer's 4-velocity, $\mathbf{e}_{\hat{0}}=\mathbf{u}$,
and whose spatial triad $\mathbf{e}_{\hat{i}}$ spans the local rest
space of the observer. The latter is for now undefined up to an arbitrary
rotation. The evolution of the tetrad along the observer's worldline
is generically described by the equation: 
\begin{equation}
\nabla_{\mathbf{u}}\mathbf{e}_{\hat{\beta}}=\Omega_{\,\,\hat{\beta}}^{\hat{\alpha}}\mathbf{e}_{\hat{\alpha}};\quad\Omega^{\alpha\beta}=2u^{[\alpha}a^{\beta]}+\epsilon_{\ \ \nu\mu}^{\alpha\beta}\Omega^{\mu}u^{\nu}\label{eq:TransportTetrad}
\end{equation}
where $\Omega^{\alpha\beta}$ is the (anti-symmetric) infinitesimal
generator of Lorentz transformations, whose spatial part $\Omega_{\hat{i}\hat{j}}=\epsilon_{\hat{i}\hat{k}\hat{j}}\Omega^{\hat{k}}$
describes the arbitrary angular velocity $\vec{\Omega}$ of rotation
of the spatial triad $\mathbf{e}_{\hat{i}}$ relative to a Fermi-Walker
transported triad. $a^{\alpha}\equiv\nabla_{\mathbf{u}}u^{\alpha}$
is the observers' acceleration. Alternatively, from the definition
of the connection coefficients, 
\[
\nabla_{\mathbf{e}_{\hat{\beta}}}\mathbf{e}_{\hat{\gamma}}=\Gamma_{\hat{\beta}\hat{\gamma}}^{\hat{\alpha}}\mathbf{e}_{\hat{\alpha}}\ ,
\]
we can think of the components of $\Omega^{\alpha\beta}$ as some
of these coefficients: 
\begin{eqnarray}
\Omega_{\,\,\hat{0}}^{\hat{i}} & = & \Gamma_{\hat{0}\hat{0}}^{\hat{i}}\ =\ \ \Gamma_{\hat{0}\hat{i}}^{\hat{0}}\ =\ a^{\hat{i}}\ ;\label{eq:Connectioni00}\\
\Omega_{\,\,\hat{j}}^{\hat{i}} & = & \Gamma_{\hat{0}\hat{j}}^{\hat{i}}\ =\ \epsilon_{\hat{i}\hat{k}\hat{j}}\Omega^{\hat{k}}\ .\label{eq:Connectioni0j}
\end{eqnarray}

Unlike the situation in flat spacetime (and Lorentz coordinates),
where one can take the tetrad adapted to a given inertial observer
as a global frame, in the general case such tetrad is a valid frame
only locally, in an infinitesimal neighborhood of the observer. In
order to define a reference frame over an extended region of spacetime,
one needs a congruence of observers, that is, one needs to extend
$u^{\alpha}$ to a field of unit timelike vectors tangent to a congruence
of time-like curves. A connecting vector $X^{\alpha}$ between two
neighboring observers in the congruence satisfies 
\begin{equation}
\left[{\bf u},{\bf X}\right]={\bf 0}\Leftrightarrow u^{\beta}\nabla_{\beta}X^{\alpha}-X^{\beta}\nabla_{\beta}u^{\alpha}=0.
\end{equation}
The evolution of the connecting vector along the worldline of an observer
in the congruence is then given by the linear equation 
\begin{equation}
\nabla_{{\bf u}}X^{\alpha}=\left(\nabla^{\beta}u^{\alpha}\right)X_{\beta}.
\end{equation}
The component of the connecting vector orthogonal to the congruence,
\begin{equation}
Y^{\alpha}=(h^{u})_{\ \beta}^{\alpha}X^{\beta}=X^{\alpha}+\left(u_{\beta}X^{\beta}\right)u^{\alpha},
\end{equation}
satisfies 
\begin{equation}
\nabla_{{\bf u}}Y^{\alpha}=K^{\alpha\beta}Y_{\beta}+\left(a_{\beta}Y^{\beta}\right)u^{\alpha},\label{eq:NablaY}
\end{equation}
where $K^{\alpha\beta}$ denotes the spatially projected covariant
derivative of $u^{\alpha}$ 
\begin{equation}
K^{\alpha\beta}\equiv(h^{u})_{\ \lambda}^{\alpha}(h^{u})_{\ \tau}^{\beta}u^{\lambda;\tau}=\nabla^{\beta}u^{\alpha}+a^{\alpha}u^{\beta}\label{eq:Kab}
\end{equation}
The decomposition of this tensor into its trace, symmetric trace-free
and anti-symmetric parts yields the expansion 
\[
\theta=u_{\ ;\alpha}^{\alpha}
\]
the shear 
\begin{equation}
\sigma_{\alpha\beta}=K_{(\alpha\beta)}-\frac{1}{3}\theta g_{\alpha\beta}-\frac{1}{3}\theta u_{\alpha}u_{\beta}\label{eq:Expansion}
\end{equation}
and the vorticity 
\begin{equation}
\omega_{\alpha\beta}=K_{[\alpha\beta]}\label{eq:Vorticity}
\end{equation}
of the congruence. It is useful to introduce the vorticity vector
\begin{equation}
\omega^{\alpha}=\frac{1}{2}\epsilon^{\alpha\beta\gamma\delta}u_{\gamma;\beta}u_{\delta}=-\frac{1}{2}\epsilon^{\alpha\beta\gamma\delta}\omega_{\alpha\beta}u_{\delta}.\label{eq:VorticityVector}
\end{equation}
According to definition above, $\omega^{\alpha}$ yields \emph{half}
the curl of $u^{\alpha}$; this is in agreement with the convention
in e.g.~\cite{Natario,The many faces}, but differs by a minus sign
from the definition in e.g.~\cite{Maartens:1997fg,EMMbook,HawkingEllis}.
Note however that for the vorticity tensor $\omega_{\alpha\beta}$
we are using the most usual definition given in \cite{Maartens:1997fg,EMMbook,HawkingEllis},
differing from a minus sign from the one in \cite{The many faces}
(consequently, $\omega^{\alpha}$ given by Eq.~(\ref{eq:VorticityVector})
is \emph{minus} the dual of $\omega_{\alpha\beta}$). The non-vanishing
tetrad components of $K_{\alpha\beta}$ are 
\begin{equation}
K_{\hat{i}\hat{j}}=\sigma_{\hat{i}\hat{j}}+\frac{1}{3}\theta\delta_{\hat{i}\hat{j}}+\omega_{\hat{i}\hat{j}}.\label{eq:Kij}
\end{equation}
These components determine the following connection coefficients:
\begin{equation}
K_{\hat{i}\hat{j}}=\nabla_{\hat{j}}u_{\hat{i}}=\Gamma_{\hat{j}\hat{i}}^{\hat{0}}=\Gamma_{\hat{j}\hat{0}}^{\hat{i}}.\label{eq:Connection0ij}
\end{equation}
The remaining temporal connection coefficients (other than the ones
given in Eqs.~(\ref{eq:Connectioni00})-(\ref{eq:Connectioni0j}),
(\ref{eq:Connection0ij}) above) are trivially zero: 
\[
\Gamma_{\hat{\alpha}\hat{0}}^{\hat{0}}=-\mathbf{e}_{\hat{0}}\cdot\nabla_{\mathbf{e}_{\hat{\alpha}}}\mathbf{e}_{\hat{0}}=-\frac{1}{2}\nabla_{\mathbf{e}_{\hat{\alpha}}}(\mathbf{e}_{\hat{0}}\cdot\mathbf{e}_{\hat{0}})=0.
\]

Each observer in the congruence carries its own adapted tetrad, c.f.
Fig. \ref{fig:RefFrame}, and to define the reference frame one must
provide the law of evolution for the spatial triads orthogonal to
$u^{\alpha}$. A natural choice would be Fermi-Walker transport, $\vec{\Omega}=0$
(the triad does not rotate relative to local guiding gyroscopes);
another natural choice, of great usefulness in this framework, is
to lock the rotation of the spatial triads to the vorticity of the
congruence, $\vec{\Omega}=\vec{\omega}$. We will dub such frame ``congruence
adapted frame''%
\footnote{Note however that in some literature, e.g.~\cite{BiniStrains}, the
term ``congruence adapted'' is employed with a different meaning,
designating any tetrad field whose time axis is tangent to the congruence,
without any restriction on the transport law for the spatial triad
(namely without requiring the triads to co-rotate with the congruence).
Hence ``adapted'' therein means what, in our convention, we would
call adapted to each individual observer.%
}; \textcolor{black}{as argued in \cite{MassaII,MassaZordan} (see
also \cite{MassaI}), this is the most natural generalization of the
non-relativistic concept of reference frame}; and the corresponding
transport law $\vec{\Omega}=\vec{\omega}$ has been dubbed ``co-rotating
Fermi-Walker transport'' \cite{The many faces,GEM User Manual}.
This choice is more intuitive in the special case of a shear-free
congruence, where, as we will show next, the axes of the frame thereby
defined point towards \emph{fixed} neighboring observers. Indeed,
if $X^{\alpha}$ is a connecting vector between two neighboring observers
of the congruence and $Y^{\alpha}$ is its component orthogonal to
the congruence, we have 
\begin{equation}
\nabla_{\mathbf{u}}Y^{\hat{i}}=\dot{Y}^{\hat{i}}+\Gamma_{\hat{0}\hat{0}}^{\hat{i}}Y^{\hat{0}}+\Gamma_{\hat{0}\hat{j}}^{\hat{i}}Y^{\hat{j}}=\dot{Y}^{\hat{i}}+\Omega_{\,\,\hat{j}}^{\hat{i}}X^{\hat{j}}\ ,\label{eq:X_{i;a}U^a}
\end{equation}
where the dot denotes the ordinary derivative along $\mathbf{u}$:
$\dot{A}_{\hat{\alpha}}\equiv A_{\hat{\alpha},\hat{\beta}}u^{\hat{\beta}}$.
Since, from (\ref{eq:NablaY}), $\nabla_{\mathbf{u}}Y^{\hat{i}}=K_{\ \hat{j}}^{\hat{i}}Y^{\hat{j}}$,
we conclude that 
\begin{equation}
\dot{Y}_{\hat{i}}=\left(\sigma_{\hat{i}\hat{j}}+\frac{1}{3}\theta\delta_{\hat{i}\hat{j}}+\omega_{\hat{i}\hat{j}}-\Omega_{\hat{i}\hat{j}}\right)Y^{\hat{j}}.\label{eq:ConnectingVector}
\end{equation}
This tells us that for a shear-free congruence ($\sigma_{\hat{i}\hat{j}}=0$),
if we lock the rotation $\vec{\Omega}$ of the tetrad to the vorticity
$\vec{\omega}$ of the congruence, $\Omega_{\hat{i}\hat{j}}=\omega_{\hat{i}\hat{j}}$,
the connecting vector's direction is fixed on the tetrad (and if in
addition $\theta=0$, i.e., the congruence is rigid, the connecting
vectors have constant components on the tetrad). A familiar example
is the rigidly rotating frame in flat spacetime; in the non-relativistic
limit, the vorticity of the congruence formed by the rigidly rotating
observers is constant, and equals the angular velocity of the frame;
in this case, by choosing $\vec{\Omega}=\vec{\omega}$, one is demanding
that the spatial triads $\mathbf{e}_{\hat{i}}$ carried by the observers
co-rotate with the angular velocity of the congruence; hence it is
clear that the axes $\mathbf{e}_{\hat{i}}$ always point to the same
neighboring observers. For relativistic rotation, the vorticity $\vec{\omega}$
is not constant and no longer equals the (constant) angular velocity
of the rotating observers; but it is still the condition $\vec{\Omega}=\vec{\omega}$
that ensures that the tetrads are rigidly anchored to the observer
congruence. Another example is the family of the so-called ``static''
observers in Kerr spacetime, which is very important in this context,
because it is this construction which allows us to determine the rotation
of the frame of the ``distant stars'' with respect to a local gyroscope,
as we shall see in Sec.~\ref{sub:3+1 Gyroscope-precession}.

\subsection{Geodesics. Inertial forces --- ``gravitoelectromagnetic fields''\label{sub: 1+3 Geodesics}}

The spatial part of the geodesic equation for a test particle of 4-velocity
$U^{\alpha}$, $\nabla_{\mathbf{U}}U^{\alpha}\equiv DU^{\alpha}/d\tau=0$,
reads, in the frame $e_{\hat{\alpha}}$: 
\[
\frac{dU^{\hat{i}}}{d\tau}+\Gamma_{\hat{0}\hat{0}}^{\hat{i}}(U^{\hat{0}})^{2}+\left(\Gamma_{\hat{0}\hat{j}}^{\hat{i}}+\Gamma_{\hat{j}\hat{0}}^{\hat{i}}\right)U^{\hat{0}}U^{\hat{j}}+\Gamma_{\hat{j}\hat{k}}^{\hat{i}}U^{\hat{k}}U^{\hat{j}}=0\;.
\]
Substituting \eqref{eq:Connectioni00}, \eqref{eq:Connectioni0j}
and \eqref{eq:Connection0ij}, we have 
\begin{equation}
\frac{\tilde{D}\vec{U}}{d\tau}=U^{\hat{0}}\left[U^{\hat{0}}\vec{G}+\vec{U}\times\vec{H}-\sigma_{\ \hat{j}}^{\hat{i}}U^{\hat{j}}\mathbf{e}_{\hat{i}}-\frac{1}{3}\theta\vec{U}\right]\equiv\vec{F}_{{\rm GEM}}\label{eq:Geo3+1}
\end{equation}
where 
\begin{equation}
\frac{\tilde{D}U^{\hat{i}}}{d\tau}=\frac{dU^{\hat{i}}}{d\tau}+\Gamma_{\hat{j}\hat{k}}^{\hat{i}}U^{\hat{k}}U^{\hat{j}}\ .\label{eq:FGEM_Def}
\end{equation}
Here $\vec{G}=-\vec{a}$ is the \emph{``gravitoelectric field}'',
and $\vec{H}=\vec{\omega}+\vec{\Omega}$ is the \emph{``gravitomagnetic
field}''. These designations are due to the analogy with the roles
that the electric and magnetic fields play in the electromagnetic
Lorentz force, which reads in the tetrad 
\begin{equation}
\left(\nabla_{\mathbf{U}}\mathbf{U}\right)^{\hat{i}}=\frac{q}{m}\left[U^{\hat{0}}E^{\hat{i}}+(\vec{U}\times\vec{B})^{\hat{i}}\right]\ ,\label{eq:Lorentz}
\end{equation}
with $\vec{E}\equiv\vec{E}(u)$ and $\vec{B}\equiv\vec{B}(u)$ denoting
the electric and magnetic fields as measured by the observers $u^{\alpha}$.
It is useful to write the GEM fields in a manifestly covariant form:
\begin{equation}
(G^{u})^{\alpha}=-\nabla_{\mathbf{u}}u^{\alpha}\equiv-u_{\ ;\beta}^{\alpha}u^{\beta}\;;\qquad(H^{u})^{\alpha}=\omega^{\alpha}+\Omega^{\alpha}\;.\label{eq:GEM Fields Cov}
\end{equation}
The gravitomagnetic field $(H^{u})^{\alpha}$ thus consists of two
parts of different origins: the angular velocity $\Omega^{\alpha}$
of rotation of the spatial triads relative to Fermi-Walker transport
(i.e., to local guiding gyroscopes), plus the vorticity $\omega^{\alpha}$
of the congruence of observers $u^{\alpha}$. If we lock the rotation
of the triads to the vorticity of the congruence, $\Omega^{\alpha}=\omega^{\alpha}$,
the gravitomagnetic field becomes simply twice the vorticity: $(H^{u})^{\alpha}=2\omega^{\alpha}$.

The last two terms of (\ref{eq:Geo3+1}) have no electromagnetic counterpart;
they consist of the shear/expansion tensor $K_{(\alpha\beta)}$, which
is sometimes called the {\em second fundamental form} of the distribution
of hyperplanes orthogonal to ${\bf u}$. If this distribution is integrable
(that is, if there is no vorticity) then $K_{(\alpha\beta)}$ is just
the extrinsic curvature of the time slices orthogonal to ${\bf u}$.
These terms correspond to the time derivative of the spatial metric
$(h^{u})_{\alpha\beta}$, that locally measures the spatial distances
between neighboring observers; this can be seen noting that $2K_{(\alpha\beta)}=(h^{u})_{\ \alpha}^{\gamma}(h^{u})_{\ \beta}^{\delta}\mathcal{L}_{\mathbf{u}}(h^{u})_{\gamma\delta}=u^{0}\partial_{0}(h^{u})_{\alpha\beta}$,
the last equality holding in a coordinate system where $\mathbf{u}$
is proportional to $\partial/\partial t$.

\subsubsection{The derivative operator $\tilde{D}/d\tau$ and inertial forces\label{sub:The-derivative-operator}}

$\vec{F}_{{\rm GEM}}=\tilde{D}\vec{U}/d\tau$ describes the inertial
accelerations (forces) associated to an \emph{arbitrary} orthonormal
frame; we shall now justify this statement, and the splitting of the
connection made in Eq.~(\ref{eq:Geo3+1}). We start by noticing that
$\tilde{D}\vec{U}/d\tau$ is a spatial vector which is the derivative
of another spatial vector (the spatial velocity $U^{\langle\alpha\rangle}=(h^{u})_{\ \beta}^{\alpha}U^{\beta}$,
or $\vec{U}$ in the tetrad $\mathbf{e}_{\hat{\alpha}}$); mathematically,
this is determined by a connection on the vector bundle of all spatial
vectors. There is a Riemannian metric naturally defined on this vector
bundle, the spatial metric $(h^{u})_{\alpha\beta}$, and the most
obvious connection preserving it is the orthogonal projection $\nabla^{\perp}$
of the ordinary spacetime covariant derivative, 
\begin{equation}
\nabla_{\alpha}^{\perp}X^{\beta}\equiv(h^{u})_{\ \gamma}^{\beta}{\nabla}_{\alpha}X^{\gamma},\label{eq:NablaPerpCov}
\end{equation}
which in terms of the tetrad components is written 
\begin{equation}
\nabla_{\hat{\alpha}}^{\perp}X^{\hat{i}}={\nabla}_{\hat{\alpha}}X^{\hat{i}}=X_{\ ,\hat{\alpha}}^{\hat{i}}+\Gamma_{\hat{\alpha}\hat{j}}^{\hat{i}}X^{\hat{j}}.\label{eq:NablaPerpTetrad}
\end{equation}
We shall call $\nabla^{\perp}$ the \emph{Fermi-Walker connection},
since the parallel transport that it defines along the congruence
is exactly the Fermi-Walker transport; this is because the spatially
projected covariant derivative of a \emph{spatial} vector $X^{\alpha}$
equals its Fermi-Walker derivative: 
\begin{equation}
\nabla_{\mathbf{u}}^{\perp}X^{\alpha}=(h^{u})_{\beta}^{\alpha}\nabla_{\mathbf{u}}X^{\beta}=\nabla_{\mathbf{u}}X^{\alpha}-u^{\alpha}X_{\beta}\nabla_{\mathbf{u}}u^{\beta}\;\label{eq:DdtDFdt}
\end{equation}
Along any curve with tangent vector ${\bf U}$ we have 
\begin{align}
\frac{D^{\perp}X^{\hat{i}}}{d\tau} & \equiv\nabla_{\mathbf{U}}^{\perp}X^{\hat{i}}=\frac{dX^{\hat{i}}}{d\tau}+\Gamma_{\hat{0}\hat{j}}^{\hat{i}}U^{\hat{0}}X^{\hat{j}}+\Gamma_{\hat{j}\hat{k}}^{\hat{i}}U^{\hat{j}}X^{\hat{k}}\nonumber \\
 & =\frac{dX^{\hat{i}}}{d\tau}+\Omega_{\,\,\hat{j}}^{\hat{i}}U^{\hat{0}}X^{\hat{j}}+\Gamma_{\hat{j}\hat{k}}^{\hat{i}}U^{\hat{j}}X^{\hat{k}}\;,\label{eq:Dperp}
\end{align}
and so along the congruence, 
\begin{equation}
\nabla_{\mathbf{u}}^{\perp}X^{\hat{i}}=\dot{X}^{\hat{i}}+\Omega_{\,\,\hat{j}}^{\hat{i}}X^{\hat{j}}.\label{eq:DperpCong}
\end{equation}
Notice that the Fermi-Walker connection preserves the spatial metric:
if $\vec{X}$ and $\vec{Y}$ are spatial vector fields then we have
\begin{align*}
\frac{d}{d\tau}(\vec{X}\cdot\vec{Y}) & =\frac{d}{d\tau}(\delta_{\hat{i}\hat{j}}X^{\hat{i}}Y^{\hat{j}})=\delta_{\hat{i}\hat{j}}\left(\frac{d{X}^{\hat{i}}}{d\tau}Y^{\hat{j}}+X^{\hat{i}}\frac{d{Y}^{\hat{j}}}{d\tau}\right)=\delta_{\hat{i}\hat{j}}\left(\frac{D^{\perp}X^{\hat{i}}}{d\tau}Y^{\hat{j}}+X^{\hat{i}}\frac{D^{\perp}Y^{\hat{j}}}{d\tau}\right)\\
 & =\frac{D^{\perp}\vec{X}}{d\tau}\cdot\vec{Y}+\vec{X}\cdot\frac{D^{\perp}\vec{Y}}{d\tau},
\end{align*}
where we used $\Omega_{\,\,\hat{j}}^{\hat{i}}=-\Omega_{\,\,\hat{i}}^{\hat{j}}$
and $\Gamma_{\hat{j}\hat{k}}^{\hat{i}}=-\Gamma_{\hat{j}\hat{i}}^{\hat{k}}$.

Eq.~(\ref{eq:Dperp}) yields the variation, along a curve of tangent
$\mathbf{U}$, of a spatial vector $X^{\alpha}$, with respect to
a triad of spatial axes undergoing Fermi-Walker transport \emph{along
the congruence}. But our goal is to measure ``accelerations'' (i.e,
the variation of the spatial velocity $U^{\langle\alpha\rangle}\equiv(h^{u})_{\ \beta}^{\alpha}U^{\beta}$)
with respect to some chosen orthonormal frame, whose triad of spatial
vectors $\mathbf{e}_{\hat{i}}$ rotate along the congruence (according
to the Fermi-Walker connection, cf.~Eq.~(\ref{eq:TransportTetrad}))
with an angular velocity $\vec{\Omega}$ that one may arbitrarily
specify. We need thus to define a connection for which the $\mathbf{e}_{\hat{i}}$
are constant along $\mathbf{u}$, whilst still equaling the projection
($\nabla^{\perp}$) of the spacetime covariant derivative $\nabla$
along the directions orthogonal to $\mathbf{u}$ (so that it still
corrects, via the coefficients $\Gamma_{\hat{j}\hat{k}}^{\hat{i}}$,
for the variation%
\footnote{e.g. the trivial variation from point to point of the triads associated
with a non-rectangular coordinate system in flat spacetime. These
do not encode inertial forces, nor do they necessarily vanish in an
inertial frame. %
} of the $\mathbf{e}_{\hat{i}}$ in the directions orthogonal to the
congruence, which are not related with inertial forces). This is achieved
by the connection 
\begin{equation}
\tilde{\nabla}_{\alpha}X^{\beta}=\nabla_{\alpha}^{\perp}X^{\beta}+u_{\alpha}(h^{u})_{\ \gamma}^{\beta}\Omega_{\ \delta}^{\gamma}X^{\delta}=(h^{u})_{\ \gamma}^{\beta}{\nabla}_{\alpha}X^{\gamma}+u_{\alpha}(h^{u})_{\ \gamma}^{\beta}\Omega_{\ \delta}^{\gamma}X^{\delta},\label{eq:NablaTildaCov}
\end{equation}
or, in tetrad components, 
\begin{equation}
\tilde{\nabla}_{\hat{\alpha}}X^{\hat{i}}=\nabla_{\hat{\alpha}}^{\perp}X^{\hat{i}}-\delta_{\hat{\alpha}}^{\hat{0}}\Omega_{\,\,\hat{j}}^{\hat{i}}X^{\hat{j}}=X_{\ ,\hat{\alpha}}^{\hat{i}}+\Gamma_{\hat{\alpha}\hat{j}}^{\hat{i}}X^{\hat{j}}-\delta_{\hat{\alpha}}^{\hat{0}}\Omega_{\,\,\hat{j}}^{\hat{i}}X^{\hat{j}},\label{eq:NablaTildaTetrad}
\end{equation}
so that 
\begin{equation}
\frac{\tilde{D}X^{\hat{i}}}{d\tau}=\nabla_{\mathbf{U}}^{\perp}X^{\hat{i}}-\Omega_{\,\,\hat{j}}^{\hat{i}}U^{\hat{0}}X^{\hat{j}}=\frac{dX^{\hat{i}}}{d\tau}+\Gamma_{\hat{j}\hat{k}}^{\hat{i}}U^{\hat{j}}X^{\hat{k}}.\label{eq:DtildeX}
\end{equation}
Similarly to what was done for $\nabla^{\perp}$, it is easy to see
that $\tilde{\nabla}$ preserves the spatial metric: 
\[
\frac{d}{d\tau}(\vec{X}\cdot\vec{Y})=\frac{\tilde{D}\vec{X}}{d\tau}\cdot\vec{Y}+\vec{X}\cdot\frac{\tilde{D}\vec{Y}}{d\tau}.
\]
Hence $\tilde{D}/d\tau$ is a \emph{covariant} derivative along a
curve, \emph{preserving the spatial metric} $(h^{u})_{\alpha\beta}$,
for spatial vectors. The inertial forces ($F_{{\rm GEM}}^{\alpha}$)
of a given frame are given by the derivative $\tilde{D}/d\tau$ acting
on the spatial velocity $U^{\langle\alpha\rangle}$ of a particle
undergoing geodesic motion. In covariant form, we have 
\begin{equation}
F_{{\rm GEM}}^{\alpha}\equiv\frac{\tilde{D}U^{\langle\alpha\rangle}}{d\tau}=\frac{D^{\perp}U^{\langle\alpha\rangle}}{d\tau}+\gamma\epsilon_{\ \beta\gamma\delta}^{\alpha}u^{\delta}U^{\beta}\Omega^{\gamma}\ =-\gamma\frac{D^{\perp}u^{\alpha}}{d\tau}+\gamma\epsilon_{\ \beta\gamma\delta}^{\alpha}u^{\delta}U^{\beta}\Omega^{\gamma}.\label{eq:FGEMCov}
\end{equation}
where $\gamma\equiv-U_{\alpha}u^{\alpha}$. In the last equality,
we decomposed $U^{\alpha}$ into its projections parallel and orthogonal
to the congruence, $U^{\alpha}=\gamma u^{\alpha}+U^{\langle\alpha\rangle}$,
and used the geodesic equation, $DU^{\alpha}/d\tau=0$, to note that
$D^{\perp}U^{\langle\alpha\rangle}/d\tau=-\gamma D^{\perp}u^{\alpha}/d\tau$.
Eq.~(\ref{eq:FGEMCov}) manifests that $F_{{\rm GEM}}^{\alpha}$
consists of two terms of distinct origin: the first term which depends
only on the variation of the observer velocity $u^{\alpha}$ along
the test particle's worldline, and the second term which is \emph{independent}
of the observer congruence, and arises from the transport law for
the spatial triads along $u^{\alpha}$. These two contributions are
illustrated, for simple examples in flat spacetime, in Appendix \ref{sub:Simple-examples-in_flat}.

Using $Du^{\alpha}/d\tau=u^{\alpha;\beta}U_{\beta}$, and decomposing
$u_{\alpha;\beta}$ in the congruence kinematics, cf.~Eqs.~(\ref{eq:Kab}),
(\ref{eq:Vorticity}), (\ref{eq:VorticityVector}), \textbf{ 
\begin{equation}
u_{\alpha;\beta}=-a(u)_{\alpha}u_{\beta}-\epsilon_{\alpha\beta\gamma\delta}\omega^{\gamma}u^{\delta}+K_{(\alpha\beta)}\ ,\label{eq:CongKinematics}
\end{equation}
} we get, substituting into (\ref{eq:FGEMCov})%
\footnote{This corresponds to a generalized version, for arbitrary \emph{orthonormal}
frames, of Eqs.~(6.13) or (6.18) of \cite{The many faces}, which
in their scheme would follow from a ``derivative'' of the type (5.3),
but allowing for an arbitrary $\Omega^{\alpha}$, rather than the
two choices $\Omega^{\alpha}=0$ and $\Omega^{\alpha}=\omega^{\alpha}$
(``fw'' and ``cfw'' in their notation, respectively), cf. Eq.
(2.16). On the other hand, their Lie transport option (``lie'')
in (5.3), which does not preserve orthonormality of the axes, is not
encompassed in our derivative (\ref{eq:FGEMCov}). %
}, 
\begin{equation}
F_{{\rm GEM}}^{\alpha}=\gamma\left[\gamma G^{\alpha}+\epsilon_{\ \beta\gamma\delta}^{\alpha}u^{\delta}U^{\beta}(\omega^{\gamma}+\Omega^{\gamma})-K^{(\alpha\beta)}U_{\beta}\right]\ ,\label{eq:FGEMCov2}
\end{equation}
which is Eq.~(\ref{eq:Geo3+1}) in covariant form.

The derivative (\ref{eq:FGEM_Def}) has a geometrical interpretation
that is more familiar when $\Omega^{\alpha}=\omega^{\alpha}$ and
the restriction of $\tilde{\nabla}$ to the spatial directions can
be interpreted as the Levi-Civita connection of some 3-D Riemannian
manifold. However, as we will see, this happens only in special cases.
The quotient of the spacetime by the congruence is diffeomorphic to
any time slice $\Sigma$, locally given by $t=f(x^{i})$ in a coordinate
system where $\mathbf{u}$ is proportional to $\partial/\partial t$.
The restriction to $\Sigma$ of the spatial projection of the 4-D
metric, $(h^{u})_{\alpha\beta}|_{\Sigma}\equiv h_{\alpha\beta}|_{\Sigma}$,
yields the spatial distances between neighboring observers along the
slice; choosing it to be the Riemannian metric on $\Sigma$ leads,
in general, to a slice dependent metric, $h_{\alpha\beta}|_{\Sigma}\equiv h_{\alpha\beta}(f(x^{i}),x^{i})$.
Let $\mathbf{E}{}_{\hat{i}}=\mathbf{e}_{\hat{i}}+A_{\hat{i}}\mathbf{e}_{\hat{0}}$
be tangent to $\Sigma$; they form (by construction) an orthonormal
basis with respect to $h_{\alpha\beta}|_{\Sigma}$. Let $\Gamma(\Sigma){}_{\hat{i}\hat{j}}^{\hat{k}}$
be the Levi-Civita connection coefficients of $h_{\alpha\beta}|_{\Sigma}$
in this basis. Using the vanishing of the torsion, $[\mathbf{E}_{\hat{j}},\mathbf{E}_{\hat{k}}]=2\Gamma(\Sigma)_{[\hat{j}\hat{k]}}^{\hat{i}}\mathbf{e}_{\hat{i}}$,
and anti-symmetry in $\hat{i},\hat{k}$, one can show after some algebra
that 
\begin{equation}
\Gamma(\Sigma)_{\hat{j}\hat{k}}^{\hat{i}}=\Gamma_{\hat{j}\hat{k}}^{\hat{i}}+A_{\hat{k}}K_{(\hat{i}\hat{j})}-A_{\hat{i}}K_{(\hat{j}\hat{k})}\label{eq:ConnectionSlice}
\end{equation}
(for an equivalent expression in terms of a coordinate basis, see
Eq. (10.10) of \cite{The many faces}). This tells us that the coefficients
$\Gamma_{\hat{j}\hat{k}}^{\hat{i}}$ match $\Gamma(\Sigma)_{\hat{j}\hat{k}}^{\hat{i}}$
in two notable cases: (i) when $\omega^{\alpha}=0$ and/or (ii) $K_{(\alpha\beta)}=0$.
In case (i) the congruence is hypersurface orthogonal; let $\Sigma_{t}$
be one such particular hypersurface; it follows that $A_{i}=0\Rightarrow\Gamma(\Sigma)_{\hat{j}k}^{\hat{i}}=\Gamma_{\hat{j}\hat{k}}^{\hat{i}}$.
This is a natural result if we note that $h_{\alpha\beta}|_{\Sigma}$
is in this case the \emph{induced metric} on $\Sigma_{t}$, whose
Levi-Civita connection is well known (e.g. \cite{General Relativity},
Lemma 10.2.1) to be the projection of the spacetime connection $\nabla$
on $\Sigma_{t}$. In case (ii) the congruence is not hypersurface
orthogonal; but on the other hand it is rigid (i.e., the distance
between neighboring observers is \emph{constant} along $\mathbf{u}$),
and so $h_{\alpha\beta}|_{\Sigma}=h_{\alpha\beta}$ is independent
of the time slice (see e.g. \cite{Oliva} p. 221; indeed in this case
one can identify the quotient with a Riemannian manifold $(\Sigma,\bm{h}$)).
Therefore at each point of the quotient one can choose a slice which
is orthogonal to the congruence \emph{at that point}, and an argument
similar to the above applies.

Let us now see the geometrical meaning of $\vec{F}_{{\rm GEM}}=\tilde{D}\vec{U}/d\tau$
in these two cases. Consider, in case (ii), the 3-D curve obtained
by projecting the particle's worldline $z^{\alpha}(\tau)$ on the
(time independent) space manifold $\Sigma$. Let it still be parametrized
by $\tau$; then $\vec{U}$ is its tangent vector, and $\tilde{D}\vec{U}/d\tau$
is just the usual covariant derivative, \emph{with respect the spatial
metric}, of $\vec{U}$ --- that is, the 3-D acceleration of this curve.
Note that expression (\ref{eq:FGEM_Def}) is indeed identical to the
usual definition of 3-D acceleration for curved spaces (or non-rectangular
coordinate systems in Euclidean space), e.g.~Eq.~(6.9) of \cite{Kay}.
It is easy to see that this corresponds to the usual notion of inertial
force from classical mechanics. Take a familiar example, a rigidly
rotating frame in \emph{flat} spacetime (as discussed in Appendix
\ref{sub:Simple-examples-in_flat}); we are familiar with the inertial
forces arising in such frame, from e.g.~a merry-go-round. They are
in this case a gravitoelectric field $\vec{G}=\vec{\omega}\times(\vec{r}\times\vec{\omega})$,
due to the acceleration of the rigidly rotating observers, that causes
a centrifugal force, plus a gravitomagnetic field $\vec{H}=2\vec{\omega}$,
half of it originating from the observers' vorticity $\vec{\omega}$,
and the other half from the angular velocity of rotation $\vec{\Omega}=\vec{\omega}$
(relative to Fermi-Walker transport) of the spatial triads they carry.
$\vec{H}$ causes the Coriolis (or gravitomagnetic) force $\vec{U}\times\vec{H}=2\vec{U}\times\vec{\omega}$.
These centrifugal and Coriolis forces become, in the non-relativistic
limit (so that the vorticity $\vec{\omega}$ equals the angular velocity
of the rotating frame) the well known expressions from non-relativistic
mechanics e.g.~Eq.~(4.91) of \cite{Goldstein}; the derivative operator
$\tilde{D}/d\tau$ becomes the one in Eq. (4.86) therein, and Eq.
(\ref{eq:DtildeX}) becomes Eq. (4.82) therein, identifying $\nabla_{\mathbf{U}}^{\perp}X^{\hat{i}}$
and $\tilde{D}X^{\hat{i}}/d\tau$, respectively, with their ``time
derivatives observed in the space (i.e.~non-rotating) and rotating
frames''.

Let us turn now to the sub-case $\omega^{\alpha}=0$. If $K_{(\alpha\beta)}=0$
(static congruence), $\vec{F}_{{\rm GEM}}$ is just the 3-D acceleration
of the projection of the particle's worldline on the time independent
hypersurfaces $\Sigma$. If $K_{(\alpha\beta)}\ne0$, the geometric
interpretation of $\vec{F}_{{\rm GEM}}$ is a bit trickier because
$\Sigma_{t}$ changes with time. In this case, we can make a point-wise
interpretation: \emph{at each point}, it is the 3-D acceleration of
the projected curve on that particular $\Sigma_{t}$ \emph{at that
point}.

In the more general case $\omega^{\alpha}\ne0$ and $K_{(\alpha\beta)}\ne0$,
and/or if $\omega^{\alpha}\ne\Omega^{\alpha}$ (case of the so-called
``locally non-rotating frames'', see below), it is not possible
to interpret (\ref{eq:FGEM_Def}) as the acceleration of a projected
curve in some space manifold; but it still yields what one would call
the inertial forces of the given frame, which is exemplified in the
case of flat spacetime in Appendix~\ref{sub:Simple-examples-in_flat}.

\emph{Usefulness of the general equation.} --- An equation like (\ref{eq:Geo3+1}),
yielding the inertial forces in an arbitrary orthonormal frame, in
particular allowing for an arbitrary rotation $\Omega^{\alpha}$ of
the spatial triads along $u^{\alpha}$, \emph{independent} of $\omega^{\alpha}$,
is of interest in many applications. Although the congruence adapted
frame, $\vec{\Omega}=\vec{\omega}=\vec{H}/2$, might seem the most
natural frame associated to a given family of observers, other frames
are used in the literature, and the associated gravitomagnetic effects
(encompassed in our general definition of $\vec{H}$, Eq. (\ref{eq:GEM Fields Cov}))
discussed therein. That includes the case of the reference frames
sometimes employed in the context of black hole physics and astrophysics
\cite{SemerakInertial,Bardeen,Bardeen et al,SemerakStationaryFrames,SemerakForcesGyro,Gravitation}:
the tetrads carried by hypersurface orthogonal observers, whose spatial
axis are taken to be fixed to the background symmetries; for instance,
in the Kerr spacetime, the congruence are the zero angular momentum
observers (ZAMOS), and the spatial triads are fixed to the Boyer-Lindquist
spatial coordinate basis vectors. This tetrad field has been dubbed
in some literature ``locally non-rotating frames'' (LNR)%
\footnote{Somewhat erroneously, as the tetrads \emph{do rotate} with respect
to the local compass of inertia, since they are not Fermi-Walker transported
in general \cite{Gravitation,SemerakStationaryFrames,SemerakForcesGyro}.%
} \cite{Bardeen et al,Bardeen,SemerakStationaryFrames,SemerakInertial}
or ``proper reference frames of the fiducial observers'' \cite{Black Holes}.
It is regarded as important for black hole physics because it is a
reference frame that is defined everywhere (unlike for instance the
star fixed static observers, see Sec.~\ref{sub:3+1 Gyroscope-precession}
below, that do not exist past the ergosphere). Eq.~(\ref{eq:Geo3+1})
allows the description of the inertial forces in these frames, where
$\vec{\omega}=0$, and $\vec{H}=\vec{\Omega}$; that is, all the gravitomagnetic
accelerations come from%
\footnote{In \cite{SemerakInertial} the term involving $\Omega^{\alpha}$ in
Eq. (\ref{eq:FGEMCov2}) above is cast not as part of a gravitomagnetic,
but of a ``Coriolis'' acceleration (``$a_{{\rm C}}^{\alpha}$'').
Therein, what is cast as ``gravitomagnetic'' (``$a_{{\rm d}}^{\alpha}$''),
are the terms involving $K_{(\alpha\beta)}$ and $\omega^{\alpha}$.%
} $\vec{\Omega}$ (for explicit expressions of $\vec{\Omega}$, see
e.g. Eq. (33.24) of \cite{Gravitation}, or Eq. (75) of \cite{SemerakStationaryFrames}).
Frames corresponding to a congruence with vorticity, but where the
spatial triads are chosen to be Fermi-Walker transported, $\vec{\Omega}=0$,
have also been considered; in such frames $\vec{H}=\vec{\omega}$
(dubbed the ``Fermi-Walker gravitomagnetic field'' \cite{The many faces,GEM User Manual}).

Finally, it is worth noting that the GEM ``Lorentz'' forces from
the more popular linearized theory, e.g. \cite{Gravitation and Inertia,Gravitation and Spacetime,Ruggiero:2002hz,Harris1991,Carroll},
or post-Newtonian approaches, e.g. \cite{Nordvedt1973,DSX,Kaplan},
are special cases of Eq.~(\ref{eq:Geo3+1}) (e.g.~linearizing it,
one obtains Eq.~(2.5) of~\cite{PaperIAU}; further specializing
to stationary fields, one obtains e.g.~(6.1.26) of~\cite{Gravitation and Inertia}).

\subsubsection{Stationary fields --- ``quasi-Maxwell'' formalism\label{sub:quasi-Maxwell}}

If one considers a stationary spacetime, \emph{and} a frame where
it is explicitly time-independent (i.e., a congruence of observers
$u^{\alpha}$ tangent to a time-like Killing vector field, which necessarily
means that the congruence is rigid \cite{MasonPooe}), the last two
terms of Eq.~(\ref{eq:Geo3+1}) vanish and the geodesic equation
becomes formally similar to the Lorentz force (\ref{eq:Lorentz}):
\begin{equation}
\frac{\tilde{D}\vec{U}}{d\tau}=U^{\hat{0}}\left(U^{\hat{0}}\vec{G}+\vec{U}\times\vec{H}\right).\label{geoQM}
\end{equation}

The line element of a \emph{stationary spacetime} is generically described
by:

\begin{equation}
ds^{2}=-e^{2\Phi}(dt-\mathcal{A}_{i}dx^{i})^{2}+h_{ij}dx^{i}dx^{j}\label{eq:Stationary}
\end{equation}
with $\Phi$, $\vec{\mathcal{A}}$, $h_{ij}$ time-independent. Here
$h_{ij}=(h^{u})_{ij}$ is a Riemannian metric, not flat in general,
that measures the time-constant distance between stationary observers,
as measured by the Einstein light signaling procedure \cite{LandauLifshitz}.
As discussed above, Eq.~(\ref{geoQM}) is the acceleration of the
3-D curves obtained by projecting test-particle's geodesics in the
3-D manifold of metric $h_{ij}$. The GEM fields measured by the static
observers (i.e.~the observers of zero 3-velocity in the coordinate
system of \eqref{eq:Stationary}) are related with the metric potentials
by~\cite{Natario}: 
\begin{equation}
\vec{G}=-\tilde{\nabla}\Phi;\quad\vec{H}=e^{\Phi}\tilde{\nabla}\times\vec{\mathcal{A}},\label{eq:GEMFieldsQM}
\end{equation}
with $\tilde{\nabla}$ denoting the covariant differentiation operator
\emph{with respect to the} \emph{spatial metric} $h_{ij}$. The formulation
(\ref{eq:GEMFieldsQM}) of GEM fields applying to stationary spacetimes
is the most usual one; it was introduced in \cite{LandauLifshitz},
and further worked out in e.g.~\cite{NatarioCosta,Natario,Black Holes,ZonozBell,Oliva},
and is sometimes called the ``quasi-Maxwell formalism''.

\subsection{Gyroscope precession\label{sub:3+1 Gyroscope-precession}}

One of the main results of this approach is that, within this formalism,
the equation describing the evolution of the spin-vector of a gyroscope
in a gravitational field takes a form exactly analogous to the precession
of a magnetic dipole under the action of a magnetic field \emph{when
expressed in a local orthonormal tetrad comoving with the test particle}.

As we have seen in Sec. \ref{sub:Diff-Precession}, if the Mathisson-Pirani
condition holds, the spin vector of a torque-free gyroscope is Fermi-Walker
transported, cf. Eq. (\ref{eq:Fermi-Walker}). Let $U^{\alpha}$ be
the 4-velocity of the gyroscope; in a comoving orthonormal tetrad
$e_{\hat{\alpha}}$, $U^{\hat{\alpha}}=\delta_{\hat{0}}^{\hat{\alpha}}$
and $S^{\hat{0}}=0$; therefore, Eq.~(\ref{eq:Fermi-Walker}) reduces
in such frame to: 
\begin{equation}
\frac{DS^{\hat{i}}}{d\tau}=0\Leftrightarrow\frac{dS^{\hat{i}}}{d\tau}=-\Gamma_{\hat{0}\hat{k}}^{\hat{i}}S^{\hat{k}}=\left(\vec{S}\times\vec{\Omega}\right)^{\hat{i}}.\label{eq:Spin3+1Omega}
\end{equation}
This is the natural result. If we choose a Fermi-Walker transported
frame, obtained by setting in Eq. (\ref{eq:TransportTetrad}) $\Omega^{\alpha}=0$,
$u^{\alpha}=U^{\alpha}$ (mathematically this is defined as a frame
with no ``absolute'' spatial rotation), then gyroscopes, which are
understood as objects that ``oppose'' to changes in direction (and
interpreted as determining the local ``compass of inertia'' \cite{Gravitation and Inertia}),
have their axes fixed with respect to such frame: $d\vec{S}/d\tau=0$.
Otherwise gyroscopes are seen to ``precesses'' with an angular velocity
$-\vec{\Omega}$, that is in fact just \emph{minus} the angular velocity
of rotation \emph{of the chosen frame} relative to a Fermi-Walker
transported frame. Now, for a congruence adapted frame, $\vec{\Omega}=\vec{\omega}$,
Eq. (\ref{eq:Spin3+1Omega}) becomes: 
\begin{equation}
\frac{d\vec{S}}{d\tau}=\frac{1}{2}\vec{S}\times\vec{H}.\label{eq:Spin3+1v2}
\end{equation}
Thus, the ``precession'' of a gyroscope is given, in terms of the
gravitomagnetic field $\vec{H}$, by an expression identical (up to
a factor of 2) to the precession of a magnetic dipole under the action
of a magnetic field $\vec{B}$, cf.~Eq.~(\ref{eq:SpinVector}):
\begin{equation}
\frac{D\vec{S}}{d\tau}=\vec{\mu}\times\vec{B}.\label{eq:DipolePrec}
\end{equation}
This holds for arbitrary fields, hence in this case the one to one
correspondence with electromagnetism is more general than the one
for the geodesics described above (between Eqs. (\ref{geoQM}) and
(\ref{eq:Lorentz})), which required the fields to be stationary \emph{and}
the observers to be stationary (i.e., their worldlines be tangent
to a Killing vector field); herein by contrast the only conditions
are the observer to be comoving with the gyroscope, and using an orthonormal
tetrad. Also, the earlier result obtained for weak fields in~\cite{PaperIAU}
(that the analogy holds even if the fields are time dependent) is
just a special case of this principle.

However important differences should be noted: whereas in the electromagnetic
case it is the same field $\vec{B}$ that is at the origin of both
the magnetic force $q(\vec{U}\times\vec{B})$ in Eq.~(\ref{eq:Lorentz})
and the torque $\vec{\mu}\times\vec{B}$ on the magnetic dipole, in
the case of the gravitomagnetic force $\vec{U}\times\vec{H}$ it has,
in the general formulation, a different origin from gyroscope ``precession'',
since the former arises not only from the rotation $\vec{\Omega}$
of the frame relative to a local Fermi-Walker transported tetrad,
but also from the vorticity $\vec{\omega}$ of the congruence. In
this sense, one can say that the Lense-Thirring effect detected in
the LAGEOS satellite data~\cite{Ciufolini Lageos} (and presently
under experimental scrutiny by the ongoing LARES mission \cite{LARES}),
measuring $\vec{H}$ from test particle's deflection, is of a different
mathematical origin from the one which was under scrutiny by the Gravity
Probe B mission~\cite{GPB}, measuring $\vec{\Omega}$ from gyroscope
precession, the two being made to match by measuring both effects
relative to the ``frame of the distant stars'' (discussed below),
for which $\vec{\Omega}=\vec{\omega}=\vec{H}/2$. This type of frame
(i.e. congruence adapted) is the most usual in the literature; in
this case the fields causing the gravitomagnetic force and the precession
of a gyroscope differ only by a relative factor of 2. But it is important
to not overlook their distinct origin, as in the literature GEM fields
of frames which are not congruence adapted are considered as well;
for instance the ``Fermi-Walker gravitomagnetic field'' defined
in \cite{The many faces}, which is the $\vec{H}$ of a frame corresponding
to a congruence with vorticity, but where the spatial triads are chosen
to be Fermi-Walker transported: $\vec{\Omega}=0$. Thus there is a
non-vanishing $\vec{H}=\vec{\omega}$ in this frame, whereas at the
same time gyroscopes do not precess relative to it.

Another obvious difference between Eqs. (\ref{eq:DipolePrec}) and
(\ref{eq:Spin3+1v2}) is the presence of a covariant derivative in
the former, whereas in the latter we have a simple derivative, signaling
that $\vec{B}$ is a \emph{physical} field, and $\vec{H}$ a mere
artifact of the reference frame (which can be anything, depending
on the frame one chooses), that can always be made to vanish (in the
congruence adapted case, $\vec{H}=2\vec{\omega}$, by choosing a vorticity-free
observer congruence).

\emph{Frame dragging. ---} The fact that $\vec{H}$ is a reference
frame artifact does not mean it is necessarily meaningless; indeed
it has no \emph{local} physical significance, but it can tell us about
frame dragging, which is a \emph{non-local} physical effect. That
is the case when one chooses the so-called ``frame of the distant
stars'', a notion that applies to asymptotically flat spacetimes.
In stationary spacetimes, such frame is setup as follows: consider
a rigid congruence of stationary observers such that at infinity it
coincides with the asymptotic inertial rest frame of the source ---
the axes of the latter define the directions fixed relative to the
distant stars. These observers are interpreted as being ``at rest''
with respect to the distant stars (and also at rest with respect to
the asymptotic inertial frame of the source); since the congruence
is rigid, it may be thought as a grid of points rigidly fixed to the
distant stars. For this reason we dub them ``static observers''%
\footnote{In the case of Kerr spacetime, these are the observers whose worldlines
are tangent to the temporal Killing vector field $\xi=\partial/\partial t$,
i.e., the observers of zero 3-velocity in Boyer-Lindquist coordinates.
This agrees with the convention in \cite{GEM User Manual,BiniStaticObs,Gravitation,Bardeen}.
We note however that the denomination ``static observers'' is employed
in some literature (e.g.~\cite{WyllemanBeke,DahiaSilva}) with a
different meaning, where it designates hypersurface orthogonal time-like
Killing vector fields (which are rigid, \emph{vorticity-free} congruences,
existing only in \emph{static spacetimes}). %
}. This congruence fixes the time axis of the local tetrads of the
frame. Now if we demand the rotation $\vec{\Omega}$ of the spatial
triads (relative to Fermi-Walker transport) to equal the vorticity
$\vec{\omega}$ of the congruence, we see from Eq.~(\ref{eq:ConnectingVector})
that the spatial components of the connecting vectors between different
observers are constant in the tetrad: 
\[
\dot{Y}^{\hat{i}}=0\ ;
\]
in other words, each local spatial triad $e_{\hat{i}}$ is locked
to this grid, and therefore has directions fixed to the distant stars.
Hence, despite having no \emph{local} meaning, the gravitomagnetic
field $\vec{H}=2\vec{\Omega}=2\vec{\omega}$ describes in this case
a consequence of the frame dragging effect: the fact that a torque
free gyroscope at finite distance from a rotating source precesses
with respect to an inertial frame at infinity. This is a physical
effect, that clearly distinguishes, for instance, the Kerr from the
Schwarzschild spacetimes, but is non-local (i.e., it cannot be detected
in any local measurement; only by locking to the distant stars by
means of telescopes). It should be noted, however, that the relative
precession of two neighboring (comoving) systems of gyroscopes is
locally measurable, and encoded in the curvature tensor (more precisely,
in the gravitomagnetic tidal tensor $\mathbb{H}_{\alpha\beta}$, as
discussed in Sec.~\ref{sub:Diff-Precession}).

\subsection{Field equations\label{sub:Field-equations1+3}}

The Einstein field equations and the algebraic Bianchi identity, Eqs.~(\ref{eq:EinsteinField}),
can be generically written in this exact GEM formalism --- i.e., in
terms of $\vec{G}$, $\vec{\Omega}$, $\vec{\omega}$ and $K_{(\alpha\beta)}$.
These equations will be compared in this section with the analogous
electromagnetic situation: Maxwell's equations in an arbitrarily accelerated,
rotating and shearing frame. The latter will be of use also in Sec.~\ref{sec:Weyl analogy}.
As a special case, we will also consider \emph{stationary spacetimes}
(and rigid, congruence adapted frames therein), where we recover the
quasi-Maxwell formalism of e.g.~\cite{Natario,ZonozPRD,LandauLifshitz,MenaNatario,ZonozBell,Oliva}.
In this case, the similarity with the electromagnetic analogue ---
Maxwell's equations in a rigid, but arbitrarily accelerated and rotating
frame --- becomes closer, as we shall see.

Before proceeding, let us write the following relations which will
be useful. Let $A^{\alpha}$ be a \emph{spatial} vector; in the tetrad
we have: 
\begin{eqnarray}
\tilde{\nabla}_{\hat{i}}A^{\hat{j}} & = & A_{\ ,\hat{i}}^{\hat{j}}+\Gamma_{\hat{k}\hat{i}}^{\hat{j}}A^{\hat{k}}\;;\label{eq:TetradCov}\\
\nabla_{\mathbf{u}}A^{\hat{i}} & = & \dot{A}^{\hat{i}}+\Gamma_{\hat{0}\hat{j}}^{\hat{i}}A^{\hat{j}}\ =\ \dot{A}^{\hat{i}}+\Omega_{\ \hat{j}}^{\hat{i}}A^{\hat{j}}\;;\label{eq:DTetrad}\\
A_{\ ;\beta}^{\beta} & = & \left(A_{\ ,\hat{i}}^{\hat{i}}+A^{\hat{i}}\Gamma_{\hat{j}\hat{i}}^{\hat{j}}\right)+A^{\hat{i}}\Gamma_{\hat{0}\hat{i}}^{\hat{0}}\ =\ \tilde{\nabla}\cdot\vec{A}+\vec{A}\cdot\vec{a}\;,\label{eq:TRaceQM}
\end{eqnarray}
where we used (\ref{eq:NablaTildaTetrad}) and the connection coefficients
(\ref{eq:Connectioni00})-(\ref{eq:Connectioni0j}), and the dot denotes
the ordinary derivative along the observer worldline, $\dot{A}^{\hat{\alpha}}\equiv A_{\ ,\hat{\beta}}^{\hat{\alpha}}u^{\hat{\beta}}$.
$\tilde{\nabla}$ is the connection defined in Eqs. (\ref{eq:NablaTildaCov})-(\ref{eq:NablaTildaTetrad});
since in expressions (\ref{eq:TetradCov}) and (\ref{eq:TRaceQM})
the derivatives are along the spatial directions, one could as well
have used the Fermi-Walker connection $\nabla^{\perp}$, Eqs. (\ref{eq:NablaPerpCov})-(\ref{eq:NablaPerpTetrad}),
they are the same along these directions.

\subsubsection{\textit{\emph{Maxwell equations for the electromagnetic fields measured
by an arbitrary congruence of observers\label{sub:Maxwell-equations-1+3}}}}

Using decomposition (\ref{eq:Fdecomp}), we write Maxwell's Eqs.~(\ref{eq:MaxwellFieldEqs})
in terms of the electric and magnetic fields $(E^{u})^{\alpha}=F_{\ \beta}^{\alpha}u^{\beta}$
and $(B^{u})^{\alpha}=\star F_{\ \beta}^{\alpha}u^{\beta}$ measured
by the congruence of observers of 4-velocity $u^{\alpha}$. All the
fields below are measured with respect to this congruence, so we may
drop the superscripts $u$: $(E^{u})^{\alpha}\equiv E^{\alpha}$,
$(B^{u})^{\alpha}\equiv B^{\alpha}$. The time projection of Eq.~(\ref{eq:MaxwellFieldEqs}a)
with respect to $u^{\alpha}$ (see point \ref{enu:Time-and-space  Proj}
of Sec. \ref{sub:Notation-and-conventions}) reads: \textbf{ 
\begin{equation}
E_{\ ;\beta}^{\beta}=4\pi\rho_{c}+E^{\alpha}a_{\alpha}+2\omega_{\alpha}B^{\alpha}\;.\label{eq:GaussLawCov}
\end{equation}
} Using (\ref{eq:TRaceQM}), we have in the tetrad: 
\begin{equation}
\tilde{\nabla}\cdot\vec{E}=4\pi\rho_{c}+2\vec{\omega}\cdot\vec{B}\;.\label{eq:DivE1+3v0}
\end{equation}
Analogously, for the time projection of (\ref{eq:MaxwellFieldEqs}b)
we get 
\begin{eqnarray}
B_{\ ;\beta}^{\beta} & = & B^{\alpha}a_{\alpha}-2\omega^{\mu}E_{\mu}\;,\label{eq:DivBCov}
\end{eqnarray}
which in the tetrad becomes 
\begin{equation}
\tilde{\nabla}\cdot\vec{B}=-2\vec{\omega}\cdot\vec{E}\;.\label{eq:DivB1+3v0}
\end{equation}
The space projection of Eq.~(\ref{eq:MaxwellFieldEqs}a) reads: 
\begin{equation}
\epsilon^{\alpha\gamma\beta}B_{\beta;\gamma}=\nabla_{\mathbf{u}}^{\perp}E^{\alpha}-K^{(\alpha\beta)}E_{\beta}+\theta E^{\alpha}-\epsilon_{\ \beta\gamma}^{\alpha}\omega^{\beta}E^{\gamma}+\epsilon_{\ \beta\gamma}^{\alpha}B^{\beta}a^{\gamma}+4\pi j^{\langle\alpha\rangle}\;,\label{eq:CurlB1}
\end{equation}
where the index notation $\langle\mu\rangle$ stands for the spatially
projected part of a vector, $V_{\langle\mu\rangle}\equiv h_{\mu}^{\ \nu}V_{\nu}$,
and $\epsilon^{\mu\beta\sigma}\equiv\epsilon^{\mu\beta\sigma\alpha}u_{\alpha}$.
The tetrad components of (\ref{eq:CurlB1}) in the frame defined in
Sec.~\ref{sub:The-reference-frame} read: 
\begin{equation}
(\tilde{\nabla}\times\vec{B})^{\hat{i}}=\nabla_{\mathbf{u}}E^{\hat{i}}-K_{\ }^{(\hat{i}\hat{j})}E_{\hat{j}}+\theta E^{\hat{i}}-(\vec{\omega}\times\vec{E})^{\hat{i}}+(\vec{G}\times\vec{B})^{\hat{i}}+4\pi j^{\hat{i}}\;.\label{eq:CurlB2}
\end{equation}
Using (\ref{eq:DTetrad}), this becomes 
\begin{equation}
(\tilde{\nabla}\times\vec{B})^{\hat{i}}=\dot{E}^{\hat{i}}+(\vec{G}\times\vec{B})^{\hat{i}}+\left[(\vec{\Omega}-\vec{\omega})\times\vec{E}\right]^{\hat{i}}-K_{\ }^{(\hat{i}\hat{j})}E_{\hat{j}}+\theta E^{\hat{i}}+4\pi j^{\hat{i}}\;.\label{eq:CurlB2v2}
\end{equation}

The space projection of (\ref{eq:MaxwellFieldEqs}b) is 
\begin{equation}
\epsilon^{\alpha\gamma\beta}E_{\beta;\gamma}=-\nabla_{\mathbf{u}}^{\perp}B^{\alpha}+K^{(\alpha\beta)}B_{\beta}-\theta B^{\alpha}+\epsilon_{\ \beta\gamma}^{\alpha}\omega^{\beta}B^{\gamma}+\epsilon_{\ \ }^{\alpha\mu\sigma}E_{\mu}a_{\sigma}\;,\label{eq:CurlE1}
\end{equation}
which, analogously, reads in the tetrad, 
\begin{equation}
(\tilde{\nabla}\times\vec{E})^{\hat{i}}=-\dot{B}^{\hat{i}}+(\vec{G}\times\vec{E})^{\hat{i}}+\left[(\vec{\omega}-\vec{\Omega})\times\vec{B}\right]^{\hat{i}}+K^{(\hat{i}\hat{j})}B_{\hat{j}}-\theta B^{\hat{i}}\;.\label{eq:CurlE2v0}
\end{equation}
In the congruence adapted frame ($\vec{\omega}=\vec{\Omega}=\vec{H}/2$),
Eqs.~(\ref{eq:DivE1+3v0}), (\ref{eq:DivB1+3v0}), (\ref{eq:CurlB2v2})
and (\ref{eq:CurlE2v0}) above become, 
\begin{eqnarray}
\tilde{\nabla}\cdot\vec{E} & = & 4\pi\rho_{c}+\vec{H}\cdot\vec{B}\;;\label{eq:DivE1+3}\\
\tilde{\nabla}\times\vec{B} & = & \dot{\vec{E}}+\vec{G}\times\vec{B}+4\pi\vec{j}-K_{\ }^{(\hat{i}\hat{j})}E_{\hat{j}}\vec{e}_{\hat{i}}+\theta\vec{E}\;;\label{eq:CurlB3}\\
\tilde{\nabla}\cdot\vec{B} & = & -\vec{H}\cdot\vec{E}\;;\label{eq:DivB1+3}\\
\tilde{\nabla}\times\vec{E} & = & -\dot{\vec{B}}+\vec{G}\times\vec{E}+K^{(\hat{i}\hat{j})}B_{\hat{j}}\vec{e}_{\hat{i}}-\theta\vec{B}\;.\label{eq:CurlE2}
\end{eqnarray}

In the special case of a rigid frame ($K_{\ }^{(\hat{i}\hat{j})}=\theta=0$)
and time-independent fields ($\dot{\vec{E}}=\dot{\vec{B}}=0$), this
yields Eqs.~(\ref{Tab:analogy1+3}.4a)-(3.8a) of Table \ref{Tab:analogy1+3}.

\subsubsection{Einstein equations\label{sub:Einstein-Eqs-1+3}}

We start by computing the tetrad components of the Riemann tensor
in the frame of Sec.~\ref{sub:The-reference-frame}: 
\begin{eqnarray}
R_{\hat{0}\hat{i}\hat{0}\hat{j}} & = & -\tilde{\nabla}_{\hat{i}}G_{\hat{j}}+G_{\hat{i}}G_{\hat{j}}-\dot{K}_{\hat{j}\hat{i}}+K_{\hat{l}\hat{i}}\Omega_{\ \hat{j}}^{\hat{l}}+\Omega_{\ \hat{i}}^{\hat{l}}K_{\hat{j}\hat{l}}-K_{\ \hat{i}}^{\hat{l}}K_{\hat{j}\hat{l}}\;;\label{eq:R0ij0}\\
R_{\hat{0}\hat{i}\hat{j}\hat{k}} & = & \tilde{\nabla}_{\hat{k}}K_{\hat{i}\hat{j}}-\tilde{\nabla}_{\hat{j}}K_{\hat{i}\hat{k}}+2G_{\hat{i}}\omega_{\hat{j}\hat{k}}\;;\label{eq:Rjkl0}\\
R_{\hat{i}\hat{j}\hat{k}\hat{l}} & = & \tilde{R}_{\hat{i}\hat{j}\hat{k}\hat{l}}-K_{\hat{l}\hat{i}}K_{\hat{k}\hat{j}}+K_{\hat{l}\hat{j}}K_{\hat{k}\hat{i}}+2\omega_{\hat{i}\hat{j}}\Omega_{\hat{k}\hat{l}}\;.\label{eq:Rijkl}
\end{eqnarray}
In the expressions above we kept $\vec{\Omega}$ independent of $\vec{\omega}$,
so that they apply to an arbitrary orthonormal tetrad field. Here
\begin{equation}
\tilde{R}_{\ \hat{j}\hat{k}\hat{l}}^{\hat{i}}\equiv\Gamma_{\hat{l}\hat{j},\hat{k}}^{\hat{i}}-\Gamma_{\hat{k}\hat{j},\hat{l}}^{\hat{i}}+\Gamma_{\hat{k}\hat{m}}^{\hat{i}}\Gamma_{\hat{l}\hat{j}}^{\hat{m}}-\Gamma_{\hat{l}\hat{m}}^{\hat{i}}\Gamma_{\hat{k}\hat{j}}^{\hat{m}}-C_{\hat{k}\hat{l}}^{\hat{m}}\Gamma_{\hat{m}\hat{j}}^{\hat{i}}\ \label{eq:RiemannSpaceDistr}
\end{equation}
(where $C_{\hat{k}\hat{l}}^{\hat{m}}=\Gamma_{\hat{k}\hat{l}}^{\hat{m}}-\Gamma_{\hat{l}\hat{k}}^{\hat{m}}$)
is the restriction to the spatial directions of the curvature of the
connection $\tilde{\nabla}$, given by 
\begin{equation}
\tilde{R}(\vec{X},\vec{Y})\vec{Z}=\tilde{\nabla}_{\vec{X}}\tilde{\nabla}_{\vec{Y}}\vec{Z}-\tilde{\nabla}_{\vec{Y}}\tilde{\nabla}_{\vec{X}}\vec{Z}-\tilde{\nabla}_{[\vec{X},\vec{Y}]}\vec{Z}\label{eq:RtildeDef}
\end{equation}
for any spatial vector fields $\vec{X},\vec{Y},\vec{Z}$. It is related
by 
\begin{equation}
\tilde{R}_{\hat{i}\hat{j}\hat{k}\hat{l}}=R_{\hat{i}\hat{j}\hat{k}\hat{l}}^{\perp}-2\omega_{\hat{i}\hat{j}}\Omega_{\hat{k}\hat{l}}\label{eq:Rtilde_Rperp}
\end{equation}
to the curvature tensor of the distribution of hyperplanes orthogonal
to the congruence, $R_{\hat{i}\hat{j}\hat{k}\hat{l}}^{\perp}$, that
is, the restriction to the spatial directions of the curvature of
the Fermi-Walker connection $\nabla^{\perp}$ on the vector bundle
of spatial vectors, given by a definition similar to (\ref{eq:RtildeDef}),
only replacing $\tilde{\nabla}\rightarrow\nabla^{\perp}$.

In the special cases (i)-(ii) discussed in Sec. \ref{sub:The-derivative-operator}
--- congruence adapted frames $\vec{\Omega}=\vec{\omega}$, and (i)
$\omega^{\alpha}=0$ or (ii) $K_{(\alpha\beta)}=0$ --- where the
spatial restriction of $\tilde{\nabla}$ is the Levi-Civita connection
of the spatial metric $(h^{u})_{\alpha\beta}$, $\tilde{R}_{\hat{i}\hat{j}\hat{k}\hat{l}}$
has a simple interpretation, it is the curvature tensor of such metric.
In case (i), the observers are hypersurface orthogonal (i.e., the
distribution is integrable), and $R_{\hat{i}\hat{j}\hat{k}\hat{l}}^{\perp}=\tilde{R}_{\hat{i}\hat{j}\hat{k}\hat{l}}$
is the curvature of the hypersurfaces. In case (ii) $\tilde{R}_{\hat{i}\hat{j}\hat{k}\hat{l}}\ne R_{\hat{i}\hat{j}\hat{k}\hat{l}}^{\perp}$,
and it is $\tilde{R}_{\hat{i}\hat{j}\hat{k}\hat{l}}$ (not $R_{\hat{i}\hat{j}\hat{k}\hat{l}}^{\perp}$)
that yields the curvature of the space manifold $(\Sigma,\mathbf{h})$%
. In the general case however $\tilde{R}_{\hat{i}\hat{j}\hat{k}\hat{l}}$
cannot be identified with the Levi-Civita connection of some 3D sub-manifold%
\footnote{\label{fn:TildeR_not_Manifold}This is manifest in the algebraic Bianchi
identities. The generalization of Eq. (\ref{eq:Bianchi1+3TimeSpace})
for non-congruence adapted frames is $\star\tilde{R}_{\ ji}^{j}=2\epsilon_{ikj}\omega^{j}\Omega^{k}-2K_{(ik)}\omega^{k}$;
the first term is not zero in general when $\vec{\Omega}\ne\vec{\omega}$
and/or both $K_{(ij)}$ and $\omega^{k}$ are different from zero.%
}.

We shall now compute the tetrad components of the Ricci tensor, but
specializing to congruence adapted frames: $\Omega_{ij}=\omega_{ij}=K_{[ij]}=-\epsilon_{ijk}H^{k}/2$,
so that the Ricci tensor comes in terms of the three GEM fields: $\vec{G}$,
$\vec{H}$ and $K_{(ij)}$. These read 
\begin{eqnarray}
R_{\hat{0}\hat{0}} & = & -\tilde{\nabla}\cdot\vec{G}+{\vec{G}}^{2}+\frac{1}{2}{\vec{H}}^{2}-\dot{\theta}-K^{(\hat{i}\hat{j})}K_{(\hat{i}\hat{j})}\;;\label{eq:R00}\\
R_{\hat{0}\hat{i}} & = & \tilde{\nabla}^{\hat{j}}K_{(\hat{j}\hat{i})}-\theta_{,\hat{i}}+\frac{1}{2}(\tilde{\nabla}\times\vec{H})_{\hat{i}}-(\vec{G}\times\vec{H})_{\hat{i}}\;;\label{eq:R0i}\\
R_{\hat{i}\hat{j}} & = & \tilde{R}_{\hat{i}\hat{j}}+\tilde{\nabla}_{\hat{i}}G_{\hat{j}}-G_{\hat{i}}G_{\hat{j}}+\dot{K}_{(\hat{i}\hat{j})}+K_{(\hat{i}\hat{j})}\theta\nonumber \\
 &  & +\frac{1}{2}\left[\dot{H}_{\hat{i}\hat{j}}+H_{\hat{i}\hat{j}}\theta+{\vec{H}}^{2}\delta_{\hat{i}\hat{j}}-H_{\hat{i}}H_{\hat{j}}+K_{(\hat{i}\hat{l})}H_{\ \hat{j}}^{\hat{l}}-H_{\hat{i}}^{\ \hat{l}}K_{(\hat{l}\hat{j})}\right]\;,\label{eq:Rij}
\end{eqnarray}
where $H_{ij}=\epsilon_{ijk}H^{k}$ is the dual of $\vec{H}$, and
$\tilde{R}_{\hat{i}\hat{j}}\equiv\tilde{R}_{\ \hat{i}\hat{l}\hat{j}}^{\hat{l}}$
is the Ricci tensor associated to $\tilde{R}_{\hat{i}\hat{j}\hat{k}\hat{l}}$;
this tensor is \emph{not} symmetric in the general case of a congruence
possessing both vorticity and shear. Using $T^{\hat{0}\hat{0}}=\rho$
and $T^{\hat{0}\hat{i}}=J^{\hat{i}}$, the time-time, time-space,
and space-space components of the Einstein field equations with sources,
Eq.~(\ref{eq:EinsteinField}a), read, respectively: 
\begin{eqnarray}
\tilde{\nabla}\cdot\vec{G} & = & -4\pi(2\rho+T_{\ \alpha}^{\alpha})+{\vec{G}}^{2}+\frac{1}{2}{\vec{H}}^{2}-\dot{\theta}-K^{(\hat{i}\hat{j})}K_{(\hat{i}\hat{j})}\;;\label{eq:DivG1+3}\\
\tilde{\nabla}\times\vec{H} & = & -16\pi\vec{J}+2\vec{G}\times\vec{H}+2\tilde{\nabla}\theta-2\tilde{\nabla}_{\hat{j}}K^{(\hat{j}\hat{i})}\vec{e}_{\hat{i}}\;;\label{eq:CurlH}\\
8\pi\left(T_{\hat{i}\hat{j}}-\frac{1}{2}\delta_{\hat{i}\hat{j}}T_{\ \alpha}^{\alpha}\right) & = & \tilde{R}_{\hat{i}\hat{j}}+\tilde{\nabla}_{\hat{i}}G_{\hat{j}}-G_{\hat{i}}G_{\hat{j}}+\dot{K}_{(\hat{i}\hat{j})}+K_{(\hat{i}\hat{j})}\theta\nonumber \\
 &  & +\frac{1}{2}\left[\dot{H}_{\hat{i}\hat{j}}+H_{\hat{i}\hat{j}}\theta+{\vec{H}}^{2}\delta_{\hat{i}\hat{j}}-H_{\hat{i}}H_{\hat{j}}+K_{(\hat{i}\hat{l})}H_{\ \hat{j}}^{\hat{l}}-H_{\hat{i}}^{\ \hat{l}}K_{(\hat{l}\hat{j})}\right]\;.\label{eq:SpaceSpace1+3}
\end{eqnarray}
Eqs.~(\ref{eq:DivG1+3})-(\ref{eq:CurlH}) are the gravitational
analogues of the electromagnetic equations (\ref{eq:DivE1+3}) and
(\ref{eq:CurlB3}), respectively; Eq.~(\ref{eq:SpaceSpace1+3}) has
no electromagnetic counterpart.

As for the algebraic Bianchi identities (\ref{eq:EinsteinField}b),
using (\ref{eq:R0ij0})-(\ref{eq:Rijkl}), the time-time (equal to
space-space, as discussed in Sec.~\ref{sub:Einstein's-EqsTidal}),
time-space and space-time components become, respectively: 
\begin{eqnarray}
\tilde{\nabla}\cdot\vec{H} & = & -\vec{G}\cdot\vec{H}\;;\label{eq:Bianchi1+3TimeTime}\\
\tilde{\nabla}\times\vec{G} & = & -\dot{\vec{H}}-\vec{H}\theta+H_{\hat{j}}K^{(\hat{i}\hat{j})}\vec{e}_{\hat{i}}\;;\label{eq:Bianchi1+3SpaceTime}\\
K_{(\hat{i}\hat{j})}H^{\hat{j}} & = & -\star\tilde{R}_{\ \hat{j}\hat{i}}^{\hat{j}}\;.\label{eq:Bianchi1+3TimeSpace}
\end{eqnarray}
Eqs.~(\ref{eq:Bianchi1+3TimeTime})-(\ref{eq:Bianchi1+3SpaceTime})
are the gravitational analogues of the time and space projections
of the electromagnetic Bianchi identities, Eqs.~(\ref{eq:DivB1+3})-(\ref{eq:CurlE2}),
respectively%
\footnote{Eqs.~(\ref{eq:Bianchi1+3TimeTime})-(\ref{eq:Bianchi1+3SpaceTime})
are equivalent to Eqs.~(7.3) of \cite{The many faces}; therein they
are obtained through a different procedure, not by projecting the
identity $\star R_{\ \ \ \gamma\beta}^{\gamma\alpha}=0\Leftrightarrow R_{[\alpha\beta\gamma]\delta}=0$,
but instead from the splitting of the identity $d^{2}\mathbf{u}=0\Leftrightarrow u_{[\alpha;\beta\gamma]}=0$.
Noting that $u_{[\alpha;\beta\gamma]}=-R_{[\alpha\beta\gamma]\lambda}u^{\lambda}$,
we see that the latter is indeed encoded in the time-time and space-time
parts (with respect to $u^{\alpha}$) of the former.%
}; Eq.~(\ref{eq:Bianchi1+3TimeSpace}) has no electromagnetic analogue.
This equation states that if the observer congruence has both vorticity
and shear/expansion then $\tilde{R}_{ijkl}$ does not obey the algebraic
Bianchi identities for a 3D curvature tensor.

Note this remarkable aspect: all the terms in the Maxwell equations
(\ref{eq:DivE1+3}), (\ref{eq:DivB1+3}) and (\ref{eq:CurlE2}) have
a gravitational counterpart in (\ref{eq:DivG1+3}), (\ref{eq:Bianchi1+3TimeTime})
and (\ref{eq:Bianchi1+3SpaceTime}), respectively, substituting $\{\vec{E},\vec{B}\}\rightarrow\{\vec{G},\vec{H}\}$
(up to some numerical factors). As for (\ref{eq:CurlB3}), there are
clear gravitational analogues in (\ref{eq:CurlH}) to the terms $\vec{G}\times\vec{B}$
and the current $4\pi\vec{j}$, but not to the remaining terms. It
should nevertheless be noted that, as shown in Sec.~\ref{sec:Linear-gravitoelectromagnetism}
below, in the post-Newtonian regime (or in the ``GEM limit'' of
linearized theory), the term $2\tilde{\nabla}\theta$ of (\ref{eq:CurlH})
embodies a contribution analogous to the displacement current term
$\dot{\vec{E}}$ of (\ref{eq:CurlB3}). The gravitational equations
in turn contain, as one might expect, terms with no parallel in electromagnetism,
most of them involving the shear/expansion tensor $K_{(\alpha\beta)}$.

Eqs.~(\ref{eq:DivG1+3})-(\ref{eq:Bianchi1+3TimeSpace}) are the
inertial fields based version of the tidal tensor equations (\ref{eq:EgravTrace})-(\ref{eq:BianchiIdTidalTensors})
of Sec. \ref{sec:Tidal tensor analogy}. Finally, it is also worth
mentioning that these equations have been expressed in tetrad formalism
also in the literature, e.g. \cite{EMMbook,ElstUggla}, albeit in
a different language; we note that, for congruence adapted frames
($\vec{\Omega}=\vec{\omega}$), zero cosmological constant, and perfect
fluids, Eqs. (6.45)-(6.47) of \cite{EMMbook} correspond, respectively,
to Eqs. \eqref{eq:Bianchi1+3TimeTime}-\eqref{eq:Bianchi1+3TimeSpace}
above; and (6.50)-(6.51) therein to \eqref{eq:DivG1+3}-\eqref{eq:CurlH}.

\subsubsection{Special cases: quasi-Maxwell regime (1+3 formalism), and hypersurface
orthogonal observers (3+1 formalism)\label{sub:Special-cases:-quasi-Maxwell}}

Eqs.~(\ref{eq:DivG1+3})-(\ref{eq:Bianchi1+3TimeSpace}) encompass
two notable regimes in the literature: (i) the ``quasi-Maxwell''
regime of Sec.~\ref{sub:quasi-Maxwell}, corresponding to stationary
fields, and a frame adapted to a \emph{rigid} congruence of stationary
observers, which is obtained by setting $K_{(\alpha\beta)}$ and all
time derivatives to zero; and (ii) the case of a frame adapted to
an hypersurface orthogonal (i.e., vorticity free) congruence, obtained
by setting $\vec{H}=0$ in the equations above, leading to what is
sometimes dubbed the ``3+1'' splitting, which is closely related
to the well known ADM formalism (e.g.~\cite{ADM,Gourgoulhon,Gravitation}).
Note that these special limits correspond to the ones equally labeled
(i) and (ii) in Sec. \ref{sub:The-derivative-operator}. Let us start
by case (i), also known as the ``1+3 formalism'' (e.g. \cite{ZonozPRD})
or threading picture \cite{The many faces,GEM User Manual} for stationary
spacetimes, which is where the similarity with the electromagnetism
gets closer, since, as we have seen in the previous section, most
of the differing terms between the gravitational field Eqs.~(\ref{eq:DivG1+3}),(\ref{eq:CurlH}),
(\ref{eq:Bianchi1+3TimeTime}), (\ref{eq:Bianchi1+3SpaceTime}), and
their electromagnetic counterparts (\ref{eq:DivE1+3})-(\ref{eq:CurlE2}),
involve $K_{(\alpha\beta)}$. The field equations in this regime are
given in Table {\small{{{{\ref{Tab:analogy1+3}}}}}}. Therein
we drop the hats in the indices; as discussed in Sec.~\ref{sub:The-derivative-operator}
(notable case (ii)), in this regime one can identify the quotient
space with a 3-D Riemannian manifold $(\Sigma,\mathbf{h})$ whose
metric $h_{ij}$ measures the \emph{fixed} distance between neighboring
observers. The indices $i,j,...$ are raised and lowered by $h_{ij}$,
$\vec{G}$ and $\vec{H}$ can be interpreted as vector fields on $(\Sigma,\mathbf{h})$,
the derivatives $\tilde{\nabla}_{i}$ become the usual covariant derivatives
with respect to $h_{ij}$, and $\tilde{R}_{ij}$ is its Ricci tensor.
The gravitational field equations in Table \ref{Tab:analogy1+3} exhibit
a striking similarity with their electromagnetic counterparts, in
spite of some natural differences that remain --- numerical factors,
the source and quadratic terms in (\ref{Tab:analogy1+3}.4b) with
no electromagnetic counterpart. We note in particular that, by simply
replacing $\{\vec{E},\vec{B}\}\rightarrow\{\vec{G},\vec{H}\}$ in
(\ref{Tab:analogy1+3}.5a)-(\ref{Tab:analogy1+3}.8a), one obtains,
up to some numerical factors, Eqs.~(\ref{Tab:analogy1+3}.5b), (\ref{Tab:analogy1+3}.7b)-(\ref{Tab:analogy1+3}.8b).
Of course, the electromagnetic terms involving products of GEM fields
with EM fields, are mimicked in gravity by second order terms in the
GEM fields. This is intrinsic to the non-linear nature of the gravitational
field, and may be thought of as manifesting the fact that the gravitational
field sources itself. It is interesting to note in this\pagebreak{}
\begin{table}[h]
\caption{{\small{{{{\label{Tab:analogy1+3}The gravito-electromagnetic analogy
based on inertial GEM fields.}}}}}}

\centering{}{\small{{{{\setlength{\arrayrulewidth}{0.8pt}}}}}}%
\begin{tabular}{>{\centering}p{39.6ex}c>{\centering}p{41.9ex}c}
\hline 
\multicolumn{2}{c}{{\small{{{{\raisebox{3.2ex}{}\raisebox{0.3ex}{Electromagnetism}}}}}}} & \multicolumn{2}{c}{{\small{{{{\raisebox{3.2ex}{}\raisebox{0.3ex}{Gravity}}}}}}}\tabularnewline
\hline 
{\small{{{{\raisebox{3.5ex}{}\raisebox{0.5ex}{Lorentz Force:
}}}}}}  &  & {\small{{{{\raisebox{3.5ex}{}\raisebox{0.5ex}{Geodesic Equation
(}$\vec{H}=\vec{\Omega}+\vec{\omega}$):}}}}}  & \tabularnewline
{\small{{{{\raisebox{6.5ex}{}\raisebox{2.8ex}{$\left(\nabla_{\mathbf{U}}\mathbf{U}\right)^{\hat{i}}=\frac{q}{m}\left[U^{\hat{0}}E^{\hat{i}}+(\vec{U}\times\vec{B})^{\hat{i}}\right]$}}}}}}  & {\small{{{{\raisebox{2.8ex}{(\ref{Tab:analogy1+3}.1a)}}}}}}  & {\small{{{{\raisebox{6ex}{}\raisebox{2.8ex}{${\displaystyle \frac{\tilde{D}\vec{U}}{d\tau}=U^{\hat{0}}\left[U^{\hat{0}}\vec{G}+\vec{U}\times\vec{H}-\sigma_{\ \hat{j}}^{\hat{i}}U^{\hat{j}}\mathbf{e}_{\hat{i}}-\frac{1}{3}\theta\vec{U}\right]}$}}}}}}  & {\small{{{{\raisebox{2.8ex}{(\ref{Tab:analogy1+3}.1b)}}}}}}\tabularnewline
\hline 
{\small{{{{\raisebox{3ex}{}Precession of magnetic dipole:}}}}}  &  & {\small{{{{\raisebox{3ex}{}Gyroscope ``precession'':}}}}}  & \tabularnewline
{\small{{{{\raisebox{5.5ex}{}\raisebox{2ex}{${\displaystyle \frac{D\vec{S}}{d\tau}=\vec{\mu}\times\vec{B}}$}}}}}}  & {\small{{{{\raisebox{2ex}{(\ref{Tab:analogy1+3}.2a)}}}}}}  & {\small{{{{\raisebox{2ex}{~~${\displaystyle \frac{d\vec{S}}{d\tau}=\vec{S}\times\vec{\Omega}}$}}}}}}  & {\small{{{{\raisebox{2ex}{(\ref{Tab:analogy1+3}.2b)}}}}}}\tabularnewline
\hline 
\multicolumn{4}{c}{{\small{{{{\raisebox{4.5ex}{}\raisebox{0.8ex}{Stationary fields,
rigid, congruence adapted frame: $\vec{\Omega}=\vec{\omega}=\vec{H}/2$
(quasi-Maxwell formalism)}}}}}}}\tabularnewline
\hline 
{\small{{{{\raisebox{3ex}{}Force on magnetic dipole:}}}}}  &  & {\small{{{{Force on gyroscope:}}}}}  & \tabularnewline
{\small{{{{\raisebox{6.5ex}{}\raisebox{2.5ex}{${\displaystyle \vec{F}_{{\rm EM}}=\tilde{\nabla}(\vec{B}\cdot\vec{\mu})-\frac{1}{2}\vec{\mu}(\tilde{\nabla}\cdot\vec{B})-\frac{1}{2}(\vec{\mu}\cdot\vec{H})\vec{E}}$}}}}}}  & {\small{{{{\raisebox{2ex}{(\ref{Tab:analogy1+3}.3a)}}}}}}  & {\small{{{{\raisebox{6.5ex}{}\raisebox{2.5ex}{${\displaystyle \vec{F}_{{\rm G}}=\frac{1}{2}\left[\tilde{\nabla}(\vec{H}\cdot\vec{S})-\vec{S}(\tilde{\nabla}\cdot\vec{H})-2(\vec{S}\cdot\vec{H})\vec{G}\right]}$}}}}}}  & {\small{{{{\raisebox{2ex}{(\ref{Tab:analogy1+3}.3b)}}}}}}\tabularnewline
\hline 
{\small{{{{\raisebox{3ex}{}Maxwell Source Equations}}}}}  &  & {\small{{{{\raisebox{3ex}{}Einstein Equations}}}}}  & \tabularnewline
{\small{{{{\raisebox{3ex}{}$F_{\,\,\,\,;\beta}^{\alpha\beta}=4\pi j^{\alpha}$}}}}}  &  & {\small{{{{$R_{\mu\nu}=8\pi\left(T_{\mu\nu}-\frac{1}{2}g_{\mu\nu}T_{\,\,\,\alpha}^{\alpha}\right)$}}}}}  & \tabularnewline
{\small{{{{\raisebox{3.8ex}{}$\bullet$ Time Component:}}}}}  &  & {\small{{{{$\bullet$ Time-Time Component:}}}}}  & \tabularnewline
{\small{{{{\raisebox{3ex}{}$\tilde{\nabla}\cdot\vec{E}=4\pi\rho_{c}+\vec{H}\cdot\vec{B}$}}}}}  & {\small{{{{(\ref{Tab:analogy1+3}.4a)}}}}}  & {\small{{{{\raisebox{3ex}{}$\tilde{\nabla}\cdot\vec{G}=-4\pi(2\rho+T_{\ \alpha}^{\alpha})+{\vec{G}}^{2}+\frac{1}{2}{\vec{H}}^{2}$}}}}}  & {\small{{{{(\ref{Tab:analogy1+3}.4b)}}}}}\tabularnewline
{\small{{{{\raisebox{3.8ex}{}$\bullet$ Space Components:}}}}}  &  & {\small{{{{$\bullet$ Time-Space Components:}}}}}  & \tabularnewline
{\small{{{{\raisebox{3ex}{}$\tilde{\nabla}\times\vec{B}=\vec{G}\times\vec{B}+4\pi\vec{j}$}}}}}  & {\small{{{{(\ref{Tab:analogy1+3}.5a)}}}}}  & {\small{{{{$\tilde{\nabla}\times\vec{H}=2\vec{G}\times\vec{H}-16\pi\vec{J}$}}}}}  & {\small{{{{(\ref{Tab:analogy1+3}.5b)}}}}}\tabularnewline
 &  & {\small{{{{\raisebox{3.8ex}{}$\bullet$ Space-Space Component:}}}}}  & \tabularnewline
{\small{{{{\raisebox{6.5ex}{}\raisebox{3.3ex}{$No\,\, electromagnetic\,\, analogue$}}}}}}  &  & {\small{{{{\raisebox{3.2ex}{$\tilde{\nabla}_{i}G_{j}-G_{i}G_{j}+\frac{1}{2}{\vec{H}}^{2}h_{ij}+\tilde{R}_{ij}=8\pi\left(\frac{1}{2}h_{ij}T_{\ \alpha}^{\alpha}+T_{ij}\right)$}~~}}}}}  & {\small{{{{\raisebox{0.3ex}{(\ref{Tab:analogy1+3}.6)}}}}}}\tabularnewline
\hline 
{\small{{{{\raisebox{3ex}{}Bianchi Identity}}}}}  &  & {\small{{{{\raisebox{3ex}{}Algebraic Bianchi Identity}}}}}  & \tabularnewline
{\small{{{{\raisebox{3ex}{}$\ \star F_{\ \ \ ;\beta}^{\alpha\beta}=0\quad(\Leftrightarrow F_{[\alpha\beta;\gamma]}=0\ )$}}}}}  &  & {\small{{{{\raisebox{3ex}{}$\star R_{\ \ \ \gamma\beta}^{\gamma\alpha}=0\quad(\Leftrightarrow R_{[\alpha\beta\gamma]\delta}=0)$}}}}}  & \tabularnewline
{\small{{{{\raisebox{3.8ex}{}$\bullet$ Time Component:}}}}}  &  & {\small{{{{$\bullet$ Time-Time (or Space-Space) Component:}}}}}  & \tabularnewline
{\small{{{{\raisebox{3ex}{}$\tilde{\nabla}\cdot\vec{B}=-\vec{H}\cdot\vec{E}$}}}}}  & {\small{{{{(\ref{Tab:analogy1+3}.7a)}}}}}  & {\small{{{{$\tilde{\nabla}\cdot\vec{H}=-\vec{H}\cdot\vec{G}$}}}}}  & {\small{{{{(\ref{Tab:analogy1+3}.7b)}}}}}\tabularnewline
{\small{{{{\raisebox{3.8ex}{}$\bullet$ Space Components:}}}}}  &  & {\small{{{{$\bullet$ Space-Time Components:}}}}}  & \tabularnewline
{\small{{{{\raisebox{4.5ex}{}\raisebox{1.5ex}{$\tilde{\nabla}\times\vec{E}=\vec{G}\times\vec{E}$
}}}}}}  & {\small{{{{\raisebox{1.5ex}{(\ref{Tab:analogy1+3}.8a) }}}}}}  & {\small{{{{\raisebox{1.5ex}{$\tilde{\nabla}\times\vec{G}=0$}}}}}}  & {\small{{{{\raisebox{1.5ex}{(\ref{Tab:analogy1+3}.8b)}}}}}}\tabularnewline
\hline 
\end{tabular}
\end{table}
context that the term $2\vec{G}\times\vec{H}$ in Eq.~(\ref{Tab:analogy1+3}.5b),
sourcing the curl of the gravitomagnetic field, resembles the electromagnetic
Poynting vector $\vec{p}_{EM}=\vec{E}\times\vec{B}/4\pi$; and the
contribution ${\vec{G}}^{2}+{\vec{H}}^{2}/2$ in Eq.~(\ref{Tab:analogy1+3}.5b),
sourcing the divergence of the gravitoelectric field, resembles the
electromagnetic energy density $\rho_{EM}=({\vec{E}}^{2}+{\vec{B}}^{2})/8\pi$.
For these reasons these quantities are dubbed in e.g.~\cite{Oliva,ZonozBell,ZonozPRD}
gravitational ``energy density'' and ``energy current density'',
respectively. It is also interesting that, in the asymptotic limit,
$\vec{p}_{G}\equiv-\vec{G}\times\vec{H}/4\pi$ corresponds to the
time-space components of the Landau-Lifshitz \cite{LandauLifshitz}
pseudo-tensor $t^{\mu\nu}$ \cite{Kaplan}. One should however bear
in mind that, by contrast with their electromagnetic counterparts,
these quantities are artifacts of the reference frame, with no physical
significance from a \emph{local} point of view (see related discussion
in Sec.~\ref{sub:Matte's-equations-vs}).

Let us turn now to case (ii); taking a vorticity-free congruence of
observers (i.e., $\vec{\omega}=\vec{H}=0$), the Einstein Eqs.~(\ref{eq:DivG1+3})-(\ref{eq:SpaceSpace1+3})
can be written as, respectively, 
\begin{eqnarray}
16\pi\rho & = & \tilde{R}+\theta^{2}-K^{(\hat{i}\hat{j})}K_{(\hat{i}\hat{j})}\ ;\label{eq:R00Hyp}\\
8\pi\vec{J} & = & \tilde{\nabla}\theta-\tilde{\nabla}_{\hat{j}}K^{(\hat{j}\hat{i})}\vec{e}_{\hat{i}}\;;\label{eq:R0iHyp}\\
\dot{K}_{(\hat{i}\hat{j})} & = & G_{\hat{i}}G_{\hat{j}}-\tilde{\nabla}_{\hat{i}}G_{\hat{j}}-\tilde{R}_{ij}-\theta K_{(\hat{i}\hat{j})}+8\pi\left(T_{\hat{i}\hat{j}}-\frac{1}{2}\delta_{\hat{i}\hat{j}}T_{\ \alpha}^{\alpha}\right)\ .\label{eq:RijHyp}
\end{eqnarray}
This regime corresponds to case (i) discussed in Sec. \ref{sub:The-derivative-operator};
thus $\tilde{\nabla}_{i}$ are covariant derivatives with respect
to the metric $h_{ij}$ induced on the hypersurfaces $\Sigma_{t}$
orthogonal to $u^{\alpha}$, and $\tilde{R}$ and $\tilde{R}_{ij}$
are, respectively, their 3-D Ricci scalar and tensor. Eq.~(\ref{eq:R00Hyp})
is known in the framework of the ADM formalism \cite{ADM,Gourgoulhon,BaumgarteShapiro}
as the ``Hamiltonian constraint''. Since this equation is the tetrad
time-time component of Eq.~(\ref{eq:EinsteinField}a), it can either
be directly obtained from the latter by noting that, when $\omega^{\alpha}=0$,
$K_{(\alpha\beta)}$ is the extrinsic curvature of the hypersurfaces
orthogonal to the congruence, and employing the scalar Gauss equation
(e.g.~Eq.~(2.95) of \cite{Gourgoulhon}); or from Eq.~(\ref{eq:DivG1+3})
above, computing $R=R_{\ \hat{i}}^{\hat{i}}-R_{\hat{0}\hat{0}}$ from
Eqs.~(\ref{eq:R00}), (\ref{eq:R0i}) (with $\vec{H}=\vec{0}$),
substituting into (\ref{eq:DivG1+3}), and then using (\ref{eq:EinsteinField}a)
to eliminate $R$. Eq.~(\ref{eq:R0iHyp}) follows directly from Eq.~(\ref{eq:CurlH})
by making $\vec{H}=\vec{0}$, and is known as the ``momentum constraint''
\cite{Gourgoulhon,BaumgarteShapiro}. The space-space Eq.~(\ref{eq:RijHyp})
is the dynamical equation for the extrinsic curvature. It can be put
in the usual forms in the literature as follows: 1) noting from Eq.
(3.17) of \cite{Gourgoulhon} that $G_{\alpha}=-(1/N)\tilde{\nabla}_{\alpha}N$,
where $N$ is the lapse function; 2) noting that, since $\Gamma_{\hat{0}\hat{j}}^{\hat{i}}=\Omega_{\ \hat{j}}^{\hat{i}}=0$,
$\dot{K}_{(\hat{i}\hat{j})}=\nabla_{\mathbf{u}}K_{(\hat{i}\hat{j})}$,
and then using%
\footnote{Our conventions relate to the ones in \cite{Gourgoulhon} by identifying
our $\{K_{(ij)},\theta\}$ with $\{-K_{ij},-K\}$ therein.%
} Eq.~(3.42) of \cite{Gourgoulhon} to express this in terms of the
Lie derivative $\mathcal{L}_{\mathbf{m}}K_{(\hat{i}\hat{j})}$ along
the ``normal evolution vector'' $\mathbf{m}=N\mathbf{u}$, obtaining
Eq. (4.17) of \cite{Gourgoulhon}, or in terms of $\mathcal{L}_{\mathbf{m}+\bm{\beta}}K_{(\hat{i}\hat{j})}$,
for some suitable \emph{spatial} ``shift'' vector $\bm{\beta}$
$(\beta^{\alpha}u_{\alpha}=0)$, leading to the best known ``ADM''
form Eq. (4.64) of \cite{Gourgoulhon}. Eqs.~(\ref{eq:R00Hyp})-(\ref{eq:R0iHyp})
have little resemblance to their \emph{physical} electromagnetic counterparts
(\ref{eq:DivE1+3})-(\ref{eq:CurlB3}) (for $\vec{H}=\vec{B}=\vec{0}$);
but in this framework a different (purely \emph{formal}) analogy is
sometimes drawn (e.g.~\cite{BaumgarteShapiro}): a parallelism between
Eqs.~(\ref{eq:R00Hyp})-(\ref{eq:R0iHyp}) and the two electromagnetic
constraints (for Lorentz frames in flat spacetime) $\partial_{i}E^{i}=4\pi\rho_{c}$
and $\partial_{i}B^{i}=0$, and between the ADM evolution equations
for $K_{(ij)}$ and for the spatial metric, written in a coordinate
system adapted to the foliation (e.g.~Eqs.~(4.63)-(4.64) of \cite{Gourgoulhon}),
and the dynamical equations for the curls of $\vec{B}$ and $\vec{E}$.

\subsection{Relation with tidal tensor formalism\label{sub:Relation1+3_TTensors}}

The analogy based on the gravito-electromagnetic fields $\vec{G}$
and $\vec{H}$ is intrinsically different from the gravito-electromagnetic
analogy based on tidal tensors of Sec. \ref{sec:Tidal tensor analogy};
the latter stems from tensor equations, whereas the former are fields
of inertial forces, i.e., artifacts of the reference frame. A relationship
between the two formalisms exists nevertheless, as in an arbitrary
frame one can express the gravitational tidal tensors in terms of
the GEM fields, using the expressions for the tetrad components of
Riemann tensor Eqs.~(\ref{eq:R0ij0})-(\ref{eq:Rjkl0}). This relationship
is in many ways illuminating, as we shall see; it is one of the main
results in this work, due the importance of using the two formalisms
together in practical applications, to be presented elsewhere~(e.g.
\cite{PaperInvariantes}). In an arbitrary frame one can express the
gravitational tidal tensors in terms of the GEM fields, using the
expressions for the tetrad components of Riemann tensor (\ref{eq:R0ij0})-(\ref{eq:Rijkl}).
The expressions obtained are to be compared with the analogous electromagnetic
situation, i.e., the electromagnetic tidal tensors computed from the
fields as measured in an arbitrarily accelerating, rotating, and shearing
frame (in flat or curved spacetime).

We start by the electromagnetic tidal tensors; using the abbreviated
notation $E_{\alpha\beta}\equiv(E^{u})_{\alpha\beta}=F_{\alpha\gamma;\beta}u^{\gamma}$,
$B_{\alpha\beta}\equiv(B^{u})_{\alpha\beta}=\star F_{\alpha\gamma;\beta}u^{\gamma}$,
cf. Table \ref{analogy}, it follows that 
\[
E_{\alpha\gamma}=E_{\alpha;\gamma}-F_{\alpha\beta}u_{\ ;\gamma}^{\beta}\,\,;\qquad B_{\alpha\gamma}=B_{\alpha;\gamma}-\star F_{\alpha\beta}u_{\ ;\gamma}^{\beta}\,\,.
\]
Using decompositions (\ref{eq:Fdecomp}), and Eq.~(\ref{eq:DTetrad}),
we obtain the tetrad components ($E_{\hat{0}\hat{i}}=B_{\hat{0}\hat{i}}=0)$:
\begin{eqnarray}
E_{\hat{i}\hat{j}} & = & \tilde{\nabla}_{\hat{j}}E_{\hat{i}}-\epsilon_{\hat{i}}^{\ \hat{l}\hat{m}}B_{\hat{m}}K_{\hat{l}\hat{j}}\;;\label{Eij1+3}\\
B_{\hat{i}\hat{j}} & = & \tilde{\nabla}_{\hat{j}}B_{\hat{i}}+\epsilon_{\hat{i}}^{\ \hat{l}\hat{m}}E_{\hat{m}}K_{\hat{l}\hat{j}}\;;\label{eq:Bij1+3}\\
E_{\hat{i}\hat{0}} & = & \dot{E}_{\hat{i}}+(\vec{\Omega}\times\vec{E})_{\hat{i}}+(\vec{G}\times\vec{B})_{\hat{i}}\;;\label{eq:Ei01+3}\\
B_{\hat{i}\hat{0}} & = & \dot{B}_{\hat{i}}+(\vec{\Omega}\times\vec{B})_{\hat{i}}-(\vec{G}\times\vec{E})_{\hat{i}}\;,\label{Bi01+3}
\end{eqnarray}
or, using $K_{ij}=\omega_{ij}+K_{(ij)}$, and choosing a congruence
adapted frame ($\vec{\omega}=\vec{\Omega}=\vec{H}/2$), 
\begin{eqnarray}
E_{\hat{i}\hat{j}} & = & \tilde{\nabla}_{\hat{j}}E_{\hat{i}}-\frac{1}{2}\left[\vec{B}\cdot\vec{H}\delta_{\hat{i}\hat{j}}-B_{\hat{j}}H_{\hat{i}}\right]-\epsilon_{\hat{i}}^{\ \hat{l}\hat{m}}B_{\hat{m}}K_{(\hat{l}\hat{j})}\;;\label{Eijadapted}\\
B_{\hat{i}\hat{j}} & = & \tilde{\nabla}_{\hat{j}}B_{\hat{i}}+\frac{1}{2}\left[\vec{E}\cdot\vec{H}\delta_{\hat{i}\hat{j}}-E_{\hat{j}}H_{\hat{i}}\right]+\epsilon_{\hat{i}}^{\ \hat{l}\hat{m}}E_{\hat{m}}K_{(\hat{l}\hat{j})}\;;\label{eq:Bijadapted}\\
E_{\hat{i}\hat{0}} & = & \dot{E}_{\hat{i}}+\frac{1}{2}(\vec{H}\times\vec{E})_{\hat{i}}+(\vec{G}\times\vec{B})_{\hat{i}}\;;\label{eq:Ei0adapted}\\
B_{\hat{i}\hat{0}} & = & \dot{B}_{\hat{i}}+\frac{1}{2}(\vec{H}\times\vec{B})_{\hat{i}}-(\vec{G}\times\vec{E})_{\hat{i}}\;.\label{Bi0adapted}
\end{eqnarray}

Let us compute the gravitational tidal tensors. From the definitions
of $\mathbb{E}_{\alpha\beta}$ and $\mathbb{H}_{\alpha\beta}$ in
Table \ref{analogy}, and using the tetrad components of the Riemann
tensor, Eqs.~(\ref{eq:R0ij0})-(\ref{eq:Rjkl0}), we obtain ($\mathbb{E}_{\hat{0}\hat{\alpha}}=\mathbb{E}_{\hat{\alpha}\hat{0}}=\mathbb{H}_{\hat{0}\hat{\alpha}}=\mathbb{H}_{\hat{\alpha}\hat{0}}=0$):
\begin{eqnarray}
\mathbb{E}_{\hat{i}\hat{j}} & = & -\tilde{\nabla}_{\hat{j}}G_{\hat{i}}+G_{\hat{i}}G_{\hat{j}}-\dot{K}_{\hat{i}\hat{j}}+K_{\hat{l}\hat{j}}\Omega_{\ \hat{i}}^{\hat{l}}+\Omega_{\ \hat{j}}^{\hat{l}}K_{\hat{i}\hat{l}}-K_{\ \hat{j}}^{\hat{l}}K_{\hat{i}\hat{l}}\;;\label{Eij}\\
\mathbb{H}_{\hat{i}\hat{j}} & = & -\tilde{\nabla}_{\hat{j}}\omega_{\hat{i}}+\delta_{\hat{i}\hat{j}}\tilde{\nabla}\cdot\vec{\omega}+2G_{\hat{j}}\omega_{\hat{i}}+\epsilon_{\hat{i}}^{\ \hat{l}\hat{m}}\tilde{\nabla}_{\hat{l}}K_{(\hat{j}\hat{m})}\;.\label{Hij}
\end{eqnarray}
For a congruence adapted frame these expressions become: 
\begin{eqnarray}
\mathbb{E}_{\hat{i}\hat{j}} & = & -\tilde{\nabla}_{\hat{j}}G_{\hat{i}}+G_{\hat{i}}G_{\hat{j}}+\frac{1}{4}\left({\vec{H}}^{2}\delta_{\hat{i}\hat{j}}-H_{\hat{j}}H_{\hat{i}}\right)+\frac{1}{2}\epsilon_{\hat{i}\hat{j}\hat{k}}\dot{H}^{\hat{k}}+\epsilon_{\ \hat{j}\hat{m}}^{\hat{l}}H^{\hat{m}}K_{(\hat{i}\hat{l})}\nonumber \\
 &  & -\dot{K}_{(\hat{i}\hat{j})}-\delta^{\hat{l}\hat{m}}K_{(\hat{i}\hat{l})}K_{(\hat{m}\hat{j})}\;;\label{eq:EijGravadapted}\\
\mathbb{H}_{\hat{i}\hat{j}} & = & -\frac{1}{2}\left[\tilde{\nabla}_{\hat{j}}H_{\hat{i}}+(\vec{G}\cdot\vec{H})\delta_{\hat{i}\hat{j}}-2G_{\hat{j}}H_{\hat{i}}\right]+\epsilon_{\hat{i}}^{\ \hat{l}\hat{m}}\tilde{\nabla}_{\hat{l}}K_{(\hat{j}\hat{m})}\;.\label{eq:Hijadapted}
\end{eqnarray}
In (\ref{eq:Hijadapted}) we substituted $\tilde{\nabla}\cdot H=-\vec{G}\cdot\vec{H}$
using Eq.~(\ref{eq:Bianchi1+3TimeTime}). Note the formal similarities
with the electromagnetic analogues (\ref{Eijadapted})-(\ref{eq:Bijadapted}).
All the terms present in $E_{ij}$ and $B_{ij}$, except for the last
term of the latter, have a correspondence in their gravitational counterparts
$\mathbb{E}_{ij}$, $\mathbb{H}_{ij}$, substituting \textcolor{black}{$\{\vec{E},\vec{B}\}\rightarrow-\{\vec{G},\vec{H}\}$}
and correcting some factors of 2. However, the gravitational tidal
tensors contain additional terms, which (together with the differing
numerical factors) encode the crucial differences in the tidal dynamics
of the two interactions. The fourth and fifth terms in (\ref{eq:EijGravadapted})
have the role of canceling out the antisymmetric part of $\tilde{\nabla}_{\hat{j}}G_{\hat{i}}$,
that is, canceling out the contribution of the curl of $\vec{G}$
to the gravitoelectric tidal tensor, as can be seen from Eq.~(\ref{eq:Bianchi1+3SpaceTime}).
Note in particular the term involving $\dot{H}^{i}$, which has no
counterpart in the electric tidal tensor (\ref{Eijadapted}); in Eq.~(\ref{eq:Bianchi1+3SpaceTime}),
that term shows up ``inducing'' the curl of $\vec{G}$, in a role
analogous to $\dot{B}^{i}$ in the equation (\ref{eq:CurlE2}) for
$\tilde{\nabla}\times\vec{E}$, which might lead one to think about
gravitational induction effects in analogy with Faraday's law of electromagnetism.
The fact that it is being subtracted in (\ref{eq:EijGravadapted}),
means, however, that the curl of $\vec{G}$ does not translate into
\emph{physical,} \emph{covariant} forces. For instance, it does not
induce rotation in a set of free neighboring particles (see Eq.~(\ref{eq:TidalGR})
above and discussion therein), nor does it torque an extended rigid
body, as shown in the companion paper \cite{Gyros}.

There are some interesting special regimes where the relation between
the tidal tensor and the inertial fields formalism becomes simpler.
One is the ``quasi-Maxwell'' regime of Sec. \ref{sub:quasi-Maxwell},
i.e., \emph{stationary spacetime}s, and a \emph{frame adapted to a
rigid (i.e., shear and expansion-free) congruence of stationary observers}.
The gravitational tidal tensors \emph{as measured in such frame} can
be expressed entirely in terms of the gravitoelectric $\vec{G}$ and
gravitomagnetic $\vec{H}$ fields; the non-vanishing components are:
\begin{eqnarray}
\mathbb{E}_{ij} & = & -\tilde{\nabla}_{j}G_{i}+G_{i}G_{j}+\frac{1}{4}\left({\vec{H}}^{2}h_{ij}-H_{j}H_{i}\right);\label{EijQM}\\
\mathbb{H}_{ij} & = & -\frac{1}{2}\left[\tilde{\nabla}_{j}H_{i}+(\vec{G}\cdot\vec{H})h_{ij}-2G_{j}H_{i}\right].\label{HijQM}
\end{eqnarray}
The hats in the indices of these expressions are dropped (as we did
in Table \ref{Tab:analogy1+3}) because, as discussed in Secs. \ref{sub:The-derivative-operator}
and \ref{sub:Special-cases:-quasi-Maxwell}, in this regime $h_{ij}$
is a canonical metric on the quotient space, whose Levi-Civita connection
is $\tilde{\nabla}_{j}$; so the $i,j$ (raised and lowered by $h_{ij}$)
may refer to an arbitrary (possibly coordinate) basis in this manifold,
not necessarily tetrad components.

The non-vanishing components of the electromagnetic tidal tensors
are, under the same conditions, 
\begin{equation}
E_{ij}=\tilde{\nabla}_{j}E_{i}-\frac{1}{2}\left[\vec{B}\cdot\vec{H}h_{ij}-B_{j}H_{i}\right]\quad{\rm (a)}\qquad E_{i0}=\frac{1}{2}(\vec{H}\times\vec{E})_{i}+(\vec{G}\times\vec{B})_{i}\quad{\rm (b)}\label{eq:EabQM}
\end{equation}

\begin{equation}
B_{ij}=\tilde{\nabla}_{j}B_{i}+\frac{1}{2}\left[\vec{E}\cdot\vec{H}h_{ij}-E_{j}H_{i}\right]\quad{\rm (a)}\qquad B_{i0}=\frac{1}{2}(\vec{H}\times\vec{B})_{i}-(\vec{G}\times\vec{E})_{i}\quad{\rm (b)}\label{eq:BabQM}
\end{equation}
Thus again, even in the stationary regime, the electromagnetic tidal
tensors have non-vanishing time components, unlike their gravitational
counterparts. The spatial parts, however, are very similar in form;
note that \textcolor{black}{replacing $\{\vec{E},\vec{B}\}\rightarrow-\{\vec{G},\vec{H}/2\}$
in (\ref{eq:BabQM}), the time components vanish, and one} \textcolor{black}{\emph{almost}}\textcolor{black}{{}
obtains the space part (\ref{HijQM}), apart from the factor of 2
in the third term; and that a similar substitution in (\ref{eq:EabQM})}
\textcolor{black}{\emph{almost}}\textcolor{black}{{} leads to (\ref{EijQM}),
apart from the term $G_{i}G_{j}$, which has no electromagnetic counterpart.
The gravitational and electromagnetic tidal tensors are nevertheless
very different, even in this regime; namely in their symmetries. $E_{ij}$
is not symmetric, whereas $\mathbb{E}_{ij}$ is (the second and third
terms in (\ref{EijQM}) are obviously symmetric; and that the first
one also is can be seen from Eq.~(\ref{Tab:analogy1+3}.8b) of Table
\ref{Tab:analogy1+3}). As for the magnetic tidal tensors, note that,
by virtue of Eq.~(\ref{Tab:analogy1+3}.5b), the last term of (\ref{HijQM})
ensures that, in vacuum, the antisymmetric part $H_{[i;j]}$ (i.e.,
the curl of $\vec{H}$) is subtracted from $H_{i;j}$ in (\ref{Hij}),
thus keeping $\mathbb{H}_{ij}$ symmetric, by contrast with $B_{ij}$.
This can be seen explicitly by noting that} \textcolor{black}{\emph{in
vacuum}} \textcolor{black}{(\ref{HijQM}) can be put in the equivalent
form 
\[
\mathbb{H}_{ij}=-\frac{1}{2}\left[H_{i;j}-H_{[i;j]}+(\vec{G}\cdot\vec{H})h_{ij}-2G_{(j}H_{i)}\right]\;,
\]
where we used $H_{[i;j]}=2G_{[j}H_{i]}$, as follows from Eq.~(\ref{Tab:analogy1+3}.5b).}

Another interesting regime to consider is the weak field limit, where
the non-linearities of the gravitational field are negligible, and
compare with electromagnetism in inertial frames. From Eqs.~(\ref{Eijadapted})-(\ref{Bi0adapted}),
the non-vanishing components of the electromagnetic tidal tensors
measured by observers at rest in an inertial frame are: 
\[
E_{ij}=E_{i,j}\;;\qquad E_{i0}=\dot{E}_{i}\;;\qquad B_{ij}=B_{i,j}\;;\qquad B_{i0}=\dot{B}_{i}\;,
\]
i.e., they reduce to ordinary derivatives of the electric and magnetic
fields. The \emph{linearized} gravitational tidal tensors are, from
Eqs. (\ref{eq:EijGravadapted})-(\ref{eq:Hijadapted}): 
\begin{equation}
\mathbb{E}_{ij}\approx-G_{i,j}+\frac{1}{2}\epsilon_{ijk}\dot{H}^{k}-\dot{K}_{(ij)}\;;\ \quad{\rm (a)}\qquad\mathbb{H}_{ij}\approx-\frac{1}{2}H_{i,j}+\epsilon_{i}^{\ lm}K_{(jm),l}\;.\quad{\rm (b)}\label{eq:Eij_Hij_linear}
\end{equation}
Thus, even in the linear regime, the gravitational tidal tensors cannot,
in general, be regarded as derivatives of the gravitoelectromagnetic
fields $\vec{G}$ and $\vec{H}$. Noting, from Eq.~(\ref{eq:KijLinear})
below, that $K_{(ij)}$ is the time derivative of the spatial metric,
we see that \emph{only if the fields are time independent} in the
chosen frame do we have $\mathbb{E}_{ij}\approx-G_{i,j}$, $\mathbb{H}_{ij}\approx-\frac{1}{2}H_{i,j}$.

\subsection{Force on a gyroscope\label{sub:Force-quasi-Maxwell}}

In the framework of the inertial GEM fields, there is also an analogy
relating the gravitational force on a gyroscope and the electromagnetic
force on a magnetic dipole. This is different from the analogy based
on tidal tensors, and not as general. We start with equations (\ref{analogy}.2)
of Table~\ref{analogy}, which tell us that the forces are determined
by the magnetic/gravitomagnetic tidal tensors as \emph{seen by the
particle}. For the spatial part of the forces, only the space components
of the tidal tensors, as measured in the particle's proper frame,
contribute. Comparing Eqs.~(\ref{eq:Bijadapted}) and (\ref{eq:Hijadapted}),
which yield the tidal tensors in terms of the electromagnetic/gravitoelectromagnetic
fields, we see that a close formal analogy is possible only when $K_{(\alpha\beta)}=0$
in the chosen frame. Thus, a close analogy between the forces in this
formalism can hold only when the particle is at rest with respect
to a congruence for which $K_{(\alpha\beta)}=0$; that is, a rigid
congruence. The rigidity requirement can be satisfied only in special
spacetimes \cite{MasonPooe}; it is ensured in the ``quasi-Maxwell''
regime --- that is, stationary spacetimes, and congruences tangent
to time-like Killing vector fields therein.

Let us start by the electromagnetic problem --- a magnetic dipole
at rest in a rigid, but arbitrarily accelerating and rotating frame.
Since the dipole is at rest in that frame, we have $\mu^{\alpha}=(0,\mu^{i})$;
hence the spatial part of the force is $F_{EM}^{i}=B^{ji}\mu_{j}$.
Substituting (\ref{eq:BabQM}a) in this expression yields the force
exerted on the dipole, in terms of the electric and magnetic fields
\emph{as measured in this frame}: 
\begin{equation}
\vec{F}_{EM}=\tilde{\nabla}(\vec{B}\cdot\vec{\mu})+\frac{1}{2}\left[\vec{\mu}(\vec{E}\cdot\vec{H})-(\vec{\mu}\cdot\vec{H})\vec{E}\right]\;.\label{eq:FEM_QM0}
\end{equation}
Using $\vec{H}\cdot\vec{E}=-\tilde{\nabla}\cdot\vec{B}$, cf.~Eq.
(\ref{Tab:analogy1+3}.7a), we can re-write this expression as 
\begin{equation}
\vec{F}_{EM}=\tilde{\nabla}(\vec{B}\cdot\vec{\mu})-\frac{1}{2}\left[\vec{\mu}(\tilde{\nabla}\cdot\vec{B})+(\vec{\mu}\cdot\vec{H})\vec{E}\right]\;.\label{eq:FEM_QM}
\end{equation}

Consider now the analogous gravitational situation: a gyroscope at
rest (i.e., with zero 3-velocity, $U^{i}=0$) with respect to stationary
observers (arbitrarily accelerated and rotating) in a stationary gravitational
field. If the Mathisson-Pirani condition is employed (see \cite{Gyros}
for details), the force exerted on it is described by Eq.~(\ref{analogy}.2b)
of Table \ref{analogy}; using (\ref{HijQM}) we write this force
in terms of the GEM fields: 
\begin{equation}
\vec{F}_{G}=\frac{1}{2}\left[\tilde{\nabla}(\vec{H}\cdot\vec{S})+\vec{S}(\vec{G}\cdot\vec{H})-2(\vec{S}\cdot\vec{H})\vec{G}\right]\;.\label{eq:FNatario0}
\end{equation}
From Eq.~(\ref{Tab:analogy1+3}.7b) we have $\vec{G}\cdot\vec{H}=-\tilde{\nabla}\cdot\vec{H}$;
substituting yields \cite{Natario}: 
\begin{equation}
\vec{F}_{G}=\frac{1}{2}\left[\tilde{\nabla}(\vec{H}\cdot\vec{S})-\vec{S}(\tilde{\nabla}\cdot\vec{H})-2(\vec{S}\cdot\vec{H})\vec{G}\right]\;.\label{eq:FNatario}
\end{equation}
\textcolor{black}{Note that replacing $\{\vec{\mu},\vec{E},\vec{B}\}\rightarrow\{\vec{S},\vec{G},\vec{H}/2\}$
in Eq.~(\ref{eq:FEM_QM0}) one} \textcolor{black}{\emph{almost}}\textcolor{black}{{}
obtains (\ref{eq:FNatario0}), except for a factor of 2 in the last
term. The last term of (\ref{eq:FNatario0})-(\ref{eq:FNatario}),
in this framework, can be interpreted as the ``weight'' of the dipole's
energy~\cite{Natario}. It plays, together with Eq.~(}\ref{Tab:analogy1+3}\textcolor{black}{.5b),
a crucial role in the dynamics, as it cancels out the contribution
of the curl of $\vec{H}$ to the force, ensuring that it is given
by a contraction of $S^{\alpha}$ with a} \textcolor{black}{\emph{symmetric}}\textcolor{black}{{}
tensor $\mathbb{H}_{\alpha\beta}$ (see the detailed discussion in
Sec.~\ref{sub:Relation1+3_TTensors}). This contrasts with the electromagnetic
case, where the curl of $\vec{B}$ is manifest in $B_{\alpha\beta}$
(which has an antisymmetric part) and in the force $F_{EM}^{\alpha}$. }

\textcolor{black}{The expression (\ref{eq:FNatario}) was first found
in \cite{Natario}, where it was compared to the force on a magnetic
dipole as measured in the} \textcolor{black}{\emph{inertial}}\textcolor{black}{{}
frame} \textcolor{black}{\emph{momentarily}}\textcolor{black}{{}
comoving with it, in which case the last two terms of (\ref{eq:FEM_QM})
vanish; herein we add expression (\ref{eq:FEM_QM}), which is its
electromagnetic counterpart for analogous conditions (the frame where
the particle is at rest can be arbitrarily accelerating and rotating),
and shows that the analogy is even stronger.}

\section{``Ultra-stationary'' spacetimes\label{sec:Ultrastationary}}

\textit{\emph{Ultra-stationary spacetimes}} are stationary spacetimes
admitting \emph{rigid} geodesic time-like congruences. In the coordinate
system adapted to such congruence, the metric is generically obtained
by taking $\Phi=0$ in Eq.~(\ref{eq:Stationary}), leading to, 
\begin{equation}
ds^{2}=-\left(dt-\mathcal{A}_{i}(x^{k})dx^{i}\right)^{2}+h_{ij}(x^{k})dx^{i}dx^{j}\ .\label{ultrastationary}
\end{equation}
Examples of these spacetimes are the Som-Raychaudhuri metrics \cite{SomRaychaudhuri},
the van Stockum interior solution \cite{stockum}, and the Gödel \cite{godel}
spacetime; see \cite{CHPreprint} for their discussion in this context.
This is an interesting class of spacetimes in the context of GEM,
due to the close \emph{similarity} with electrodynamics, which was
explored in an earlier work \cite{CHPRD} by one of the authors: 1)
they are exactly mapped \cite{Drukker:2003mg,CHPRD}, via the Klein-Gordon
equation, into curved 3-spaces with a ``magnetic'' field; 2) their
gravitomagnetic tidal tensor is linear \cite{CHPRD} (just like in
the case of electromagnetism), and, up to a factor, matches the covariant
derivative of the magnetic field of the electromagnetic analogue.
A link between these two properties was suggested%
\footnote{In the earlier work Refs. \cite{CHPRD,CHPreprint} by one of the authors
(to whom the \emph{exact} GEM fields analogy of Sec.~\ref{sec:3+1}
was not yet known), it was suggested that the above mapping could
be interpreted as arising from the similarity of magnetic tidal forces
manifest in relations (\ref{ttemg}). It seems, however, to be much
more related to the analogy based on GEM ``vector'' fields manifest
in Eqs.~(\ref{eq:Hultra}) and (\ref{eq:GeoUltra}). Even though
the exact correspondence (\ref{ttemg}) reinforces in some sense the
analogy, tidal forces do not seem to be the underlying principle behind
the mapping, since: i) there is no electromagnetic counterpart to
the non-vanishing gravitoelectric tidal tensor $\mathbb{E}_{\alpha\beta}$;
ii) the Klein-Gordon equation $\Box\Phi=m^{2}\Phi$ and the Hamiltonian
in Sec.~IV of \cite{CHPRD} are for a (free) monopole particle, which
feels no tidal forces. Thus one would expect it to reveal coordinate
artifacts such as the fields $\vec{G,\ }\vec{H}$, not physical tidal
forces.%
} in \cite{CHPRD}; however, the non-vanishing gravitoelectric tidal
tensor (while no electric field is present in the map) was a question
left unanswered. Herein, putting together the knowledge from the tidal
tensor and the inertial force formalisms (Secs.~\ref{sec:Tidal tensor analogy}
and \ref{sec:3+1}), we revisit these spacetimes and shed new light
on these issues.

Eqs.~(\ref{eq:GEMFieldsQM}) yield the GEM fields corresponding to
the frame adapted to the rest observers, of 4-velocity $u^{\alpha}=\delta_{0}^{\alpha}$
(in the coordinate system of \eqref{ultrastationary}). They tell
us that the gravitoelectric field vanishes, $\vec{G}=0$, which is
consistent with the fact that no electric field arises in the mapping
above; and that the gravitomagnetic field $\vec{H}$ is \emph{linear}
in the metric potentials: 
\begin{equation}
\vec{H}=\tilde{\nabla}\times\vec{\mathcal{A}}.\label{eq:Hultra}
\end{equation}

These properties can be interpreted as follows. The fact that $\vec{G}=0$
means that the rest observers are freely falling (as their acceleration
$a^{\alpha}=-G^{\alpha}$ is zero); the very special property of these
spacetimes is that such \emph{geodesic} congruence is rigid, i.e.
has no shear/expansion, allowing the metric to be time independent
in a coordinate system associated to those observers (unlike the situation
in general, e.g.~the Kerr or Schwarzschild spacetimes). The gravitomagnetic
field, on the other hand, does not vanish in this frame, which means
in this context (since the frame is congruence adapted, see Sec.~\ref{sub:The-reference-frame}
and Eq. (\ref{eq:GEM Fields Cov})), that the congruence has vorticity.
The equation of motion for a free particle in this frame, cf.~Eq.~(\ref{geoQM}),
reduces to 
\begin{equation}
\frac{\tilde{D}\vec{U}}{d\tau}=U^{\hat{0}}\vec{U}\times\vec{H}\,\,,\label{eq:GeoUltra}
\end{equation}
similar to the equation of motion of a charged particle under the
action of a magnetic field; and since $\vec{H}$ is linear in the
metric, the similarity with the electromagnetic analogue is indeed
close.

Let us now examine the tidal effects. This type of spacetimes have
a very special property: the gravitomagnetic tidal tensor measured
by the observers $u^{\alpha}=\delta_{0}^{\alpha}$ is \emph{linear}
in the fields (and thus in the metric potentials), cf.~Eq.~(\ref{HijQM}),
and, just like in the electromagnetic analogue, it is given by the
covariant derivative of $\vec{H}$ with respect to the spatial metric
$h_{ij}$: 
\begin{equation}
\mathbb{H}_{ij}=-\frac{1}{2}\tilde{\nabla}_{j}H_{i}=-\frac{1}{2}\tilde{\nabla}_{j}(\tilde{\nabla}\times\vec{\mathcal{A}})_{i}\label{ttemg}
\end{equation}
($\mathbb{H}_{0j}=\mathbb{H}_{00}=\mathbb{H}_{j0}$ for these observers).
This reinforces the similarity with electromagnetism. The gravitoelectric
tidal tensor is, however, non-zero, as seen from Eq.~(\ref{EijQM}):
\begin{equation}
\mathbb{E}_{ij}=\frac{1}{4}\left({\vec{H}}^{2}h_{ij}-H_{j}H_{i}\right)\ ,\label{eq:EijUltra}
\end{equation}
even though $\vec{G}=\vec{0}$. This should not be surprising, for
the following reasons: i) it is always possible to make $\vec{G}$
vanish by choosing freely falling observers (this is true in an arbitrary
spacetime), but that does not eliminate the tidal effects, as they
arise from the curvature tensor; ii) in the case of ultrastationary
spacetimes, $\mathbb{E}_{\alpha\beta}$ is actually a non-linear tensor
in $\vec{H}$, which merely reflects the fact that, except on very
special circumstances, $\mathbb{E}_{\alpha\beta}$ cannot be thought
of as simply a covariant derivative of some gravitoelectric field
$\vec{G}$.

The tidal tensor (\ref{eq:EijUltra}) exhibits other interesting properties.
It vanishes along the direction of the gravitomagnetic field $H^{\alpha}$:
let $X^{\alpha}$ be a spatial vector (with respect to $u^{\alpha}$,
$X^{\alpha}u_{\alpha}=0$); if it is parallel to $H^{\alpha}$, then
$\mathbb{E}_{\ \beta}^{\alpha}X^{\beta}=0$. That is, the tidal force,
or the relative acceleration of two neighboring test particles of
4-velocity $u^{\alpha}$, connected by $X^{\alpha}$, vanishes. If
$X^{\alpha}$ is orthogonal to the gravitomagnetic field, $H^{\alpha}X_{\alpha}=0$,
then it is an \emph{eigenvector} of $\mathbb{E}_{\ \beta}^{\alpha}$,
with eigenvalue ${\vec{H}}^{2}$. Thus, in the two dimensional subspace
(on the rest space $u^{i}=0$) spanned by the vectors orthogonal to
$H^{\alpha}$, the tidal force $-\mathbb{E}_{\ \beta}^{\alpha}X^{\beta}$
is \emph{proportional} to the separation vectors $X^{\alpha}$. Next
we will physically interpret this for the special case of the Gödel
universe.

\subsection{The Gödel Universe}

The Gödel universe is a solution corresponding to an \emph{homogeneous}
rotating dust with negative cosmological constant. The homogeneity
implies that the dust rotates around every point. The line element
can be put in the form (\ref{ultrastationary}), with 
\begin{equation}
\mathcal{A}_{i}dx^{i}=e^{\sqrt{2}\omega x}dy\ ,\ \ \ \ \gamma_{ij}dx^{i}dx^{j}=dx^{2}+\frac{1}{2}e^{2\sqrt{2}\omega x}dy^{2}+dz^{2}\ ,\label{eq:GodelMetric}
\end{equation}
where $\omega$ is a constant. The gravitomagnetic field is \emph{uniform},
$\vec{H}=\tilde{\nabla}\times\vec{\mathcal{A}}=2\omega\vec{e}_{z}$;
hence, by virtue of (\ref{ttemg}), the gravitomagnetic tidal tensor
vanishes, $\mathbb{H}_{\alpha\beta}=0$. For this reason, this universe
has been interpreted in \cite{CHPRD,CHPreprint} as being analogous
to an uniform magnetic field in the curved 3-manifold with metric
$\gamma_{ij}$, and the homogeneous rotation physically interpreted
in analogy with a gas of charged particles subject to a uniform magnetic
field --- as in that case one likewise has Larmor orbits around any
point.

Now we will interpret its gravitoelectric tidal tensor. In the coordinate
system of (\ref{eq:GodelMetric}) it reads, for the rest ($u^{i}=0$)
observers, 
\[
\mathbb{E}_{ij}=\omega^{2}\left(\gamma_{ij}-\delta_{i}^{z}\delta_{j}^{z}\right)\,\,.
\]
It vanishes along $z$, and is isotropic in the spatial directions
$x,y$ orthogonal to $\vec{H}$. It is similar to the Newtonian tidal
tensor $\partial_{i}\partial_{j}V$ 
of a potential $V=\omega^{2}(x^{2}+y^{2})/2$, corresponding to a
2-D harmonic oscillator, which is the potential of the Newtonian analogue
of the Gödel Universe~\cite{OzsvathGodel}: a uniform, infinitely
long and wide cylinder of dust rotating \emph{rigidly} with angular
velocity $\omega$. The potential $V$ is such that the gravitational
attraction exactly balances the centrifugal force on each fluid element
of the rotating cylinder. %
The rigid rotation causes a curious effect in the Newtonian system.
Consider a Cartesian coordinate system $\mathcal{S}$ with origin
at the axis of rotation of the cylinder, and let $\vec{r}$ be the
position vector of an arbitrary dust particle. Its equation of motion
is $\dot{\vec{r}}=\vec{\omega}\times\vec{r}$. Now take a particular
dust particle at position $\vec{r}_{0}$, %
and consider the Cartesian coordinate system $\mathcal{S}'$ originating
and \emph{comoving} with it. The position vector relative to $\mathcal{S}'$
is $\vec{r}'=\vec{r}-\vec{r}_{0}$. Hence the equation of motion of
an arbitrary dust particle with respect to $\mathcal{S}'$ reads 
\[
\dot{\vec{r}}'=\dot{\vec{r}}-\dot{\vec{r}}_{0}=\vec{\omega}\times(\vec{r}-\vec{r}_{0})=\vec{\omega}\times\vec{r}'\ ,
\]
which is formally identical to the equation in $\mathcal{S}$, replacing
$\vec{r}'$ by $\vec{r}$. That is, in the frame $\mathcal{S}'$,
the fluid is seen to be rigidly rotating about the new origin $\vec{r}'=0$
(or $\vec{r}=\vec{r}_{0}$, in the coordinates of $\mathcal{S}$).
Since the cylinder is infinite, the picture in the frame $\mathcal{S}'$
is indistinguishable from the one at $\mathcal{S}$. We see therefore
that any point $\vec{r}$ rotating rigidly with angular velocity $\vec{\omega}$
in the frame $\mathcal{S}$ can be an axis of rotation for the fluid
indistinguishable from the ``original one''.

Therefore, whilst the gravitomagnetic field and tidal tensor, as well
as the mapping via Klein-Gordon equation in \cite{CHPRD}, link to
the magnetic analogue of the Gödel universe, the gravitoelectric tidal
tensor links to the Newtonian analogue, both yielding consistent models
to picture the homogeneous rotation of this universe.

\section{Linear gravitoelectromagnetism\label{sec:Linear-gravitoelectromagnetism}}

The oldest and best known gravito-electromagnetic analogies are the
ones based on linearized gravity, which have been worked out by many
authors throughout the years, see e.g.~\cite{Gravitation and Inertia,Gravitation and Spacetime,Carroll,Harris1991,Tucker Clark,Ruggiero:2002hz,General Relativity,Ciufolini Nature Review,Wald et al 2010,Wald,QuentinBailey,IorioReview}.
As is usually presented, one considers a metric given by small perturbations
$|\varepsilon_{\alpha\beta}|\ll1$ around Minkowski spacetime, $g_{\alpha\beta}=\eta_{\alpha\beta}+\varepsilon_{\alpha\beta}$,
and from the components $\varepsilon_{\alpha\beta}$ one defines the
3-vectors $\vec{G}$ and $\vec{H}$, in terms of which one writes
the gravitational equations. Let us write the line element of such
metric in the general form 
\begin{equation}
ds^{2}=-\left(1+2\Phi\right)dt^{2}+2\mathcal{A}_{j}dtdx^{j}+\left(\delta_{ij}+2\xi_{ij}\right)dx^{i}dx^{j}\ .\label{eq:Linearmetric}
\end{equation}

If ones considers stationary perturbations, as is more usual (e.g.
\cite{Gravitation and Inertia,General Relativity,Ciufolini Nature Review,Wald et al 2010,Wald,QuentinBailey,IorioReview}),
the GEM fields are (up to numerical factors in the different definitions)
$\vec{G}=-\nabla\Phi$, $\vec{H}=\nabla\times\vec{\mathcal{A}}$,
where, \emph{in this section} (and only herein!), $\nabla_{i}\equiv\partial/\partial x_{i}$
(equaling the covariant derivative operator associated to the background
Euclidean metric $\delta_{ij}$). These fields are straightforwardly
related to the ones in Sec.~\ref{sec:3+1}: they are just, \emph{to
linear order}, minus the acceleration and twice the vorticity of the
zero 3-velocity observers ($u^{i}=0$) with respect to the coordinate
system used in (\ref{eq:Linearmetric}) (they can be called ``static
observers''). Thus they are simply a linear approximation to the
quasi-Maxwell fields in Eqs.~(\ref{eq:GEMFieldsQM}).

If the fields depend on time, different definitions of the fields
exist in the literature, as a complete, one to one GEM analogy based
on inertial fields, holding simultaneously for the geodesic equation
and for the field equations, is not possible, as we shall see below
(cf.~also \cite{Gravitation and Spacetime,Harris1991,Tucker Clark,CHPreprint,PaperIAU}).
So if one chooses to write one of them in an electromagnetic like
form, the other will contain extra terms. We stick to defining $\vec{G}$
and $\vec{H}$ by minus the acceleration and twice the vorticity of
the observer congruence (i.e.~the same definitions given in Sec.~\ref{sub: 1+3 Geodesics}
for congruence adapted frames, only this time linearized), which seems
to make more sense from a physical point of view, as with these definitions
the fields appear in the equation of geodesics playing roles formally
analogous to the electric and magnetic fields in the Lorentz force.
That amounts to define: 
\[
\vec{G}=-\nabla\Phi-\frac{\partial\vec{\mathcal{A}}}{\partial t}\;;\quad\vec{H}=\nabla\times\vec{\mathcal{A}}\;.
\]
The space part of the linearized equation for the geodesics, in the
coordinate basis $\mathbf{e}_{\alpha}\equiv\partial_{\alpha}$ associated
to the coordinate system in (\ref{eq:Linearmetric}), is obtained
from the corresponding exact equation (\ref{eq:Geo3+1}), for orthonormal
tetrads, as follows%
\footnote{One could also obtain it directly from the covariant version (\ref{eq:FGEMCov}),
\eqref{eq:FGEMCov2}, by setting therein $\Omega^{\alpha}=\omega^{\alpha}=H^{\alpha}/2$,
noting that, to linear order $\Gamma_{0k}^{i}=\epsilon_{\ jk}^{i}H^{j}/2+\partial\xi_{\ k}^{i}/\partial t$,
and using (\ref{eq:KijLinear}), as done below. %
}. One first notes that the coordinate triad of basis vectors $\mathbf{e}_{i}$
are connecting vectors between the $u^{i}=0$ observers; thus they
co-rotate with the congruence, and therefore the orthonormal tetrad
which follows $\mathbf{e}_{\alpha}$ as close as possible is the \emph{congruence
adapted} tetrad (obtained by setting $\vec{\Omega}=\vec{\omega}=\vec{H}/2$,
cf. Sec. \ref{sub:The-reference-frame}); i.e., a tetrad such that
$\mathbf{e}_{\hat{0}}\propto\mathbf{e}_{0}$ (for one to be dealing
with the same observers) and that $\mathbf{e}_{\hat{i}}$ co-rotates
with the $\mathbf{e}_{i}$, but without enduring the shear and expansion
effects of the former (since the $\mathbf{e}_{\hat{\alpha}}$ remain
orthonormal). Let $e_{\ \hat{\alpha}}^{\beta}$ denote the transformation
matrix between $\mathbf{e}_{\alpha}$ and $\mathbf{e}_{\hat{\alpha}}$:
$\mathbf{e}_{\hat{\alpha}}=e_{\ \hat{\alpha}}^{\beta}\mathbf{e}_{\beta}$.
To linear order, $e_{\ \hat{\alpha}}^{\beta}$, and its inverse $e_{\ \alpha\ }^{\hat{\beta}}$,
are given by: 
\begin{equation}
\begin{array}{c}
{\displaystyle \mathbf{e}_{\hat{0}}=(1-\Phi)\mathbf{e}_{0}\;;\qquad\mathbf{e}_{\hat{i}}=\mathbf{e}_{i}-\xi_{i}^{\ j}\mathbf{e}_{j}-\mathcal{A}_{i}\mathbf{e}_{0}}\;;\\
\\
{\displaystyle \mathbf{e}_{0}=(1+\Phi)\mathbf{e}_{\hat{0}}\;;\qquad\mathbf{e}_{i}=\mathbf{e}_{\hat{i}}+\xi_{i}^{\ \hat{j}}\mathbf{e}_{\hat{j}}}+\mathcal{A}_{i}\mathbf{e}_{\hat{0}}\;.
\end{array}\label{eq:Linear-ehat}
\end{equation}
Thus, $U^{\hat{i}}=e_{\ \alpha\ }^{\hat{i}}U^{\alpha}=U^{i}+\xi_{i}^{\ j}U_{j}$;
using $U^{i}=dx^{i}/d\tau$, substituting into (\ref{eq:Geo3+1}),
linearizing in the perturbations and keeping lowest order terms in
$U^{i}$, and noting that, to linear order, 
\begin{equation}
K_{(ij)}\equiv u_{(i;j)}=\sigma_{ij}+\frac{1}{3}\theta\delta_{ij}\approx\frac{\partial\xi_{ij}}{\partial t};\qquad\theta=K_{\ i}^{i}=\frac{\partial\xi_{\ i}^{i}}{\partial t}\;,\label{eq:KijLinear}
\end{equation}
the equation for the geodesics reads: 
\begin{equation}
\frac{d\vec{U}}{dt}=\vec{G}+\vec{U}\times\vec{H}-2\frac{\partial\xi_{\ j}^{i}}{\partial t}U^{j}\vec{e}_{i}\;.\label{eq:Geolinear}
\end{equation}
That is, the extra term, compared to the Lorentz force of electromagnetism,
comes \emph{from the time derivative of the spatial metric} (which
is true also in the exact case, as we have seen in Sec.~\ref{sub: 1+3 Geodesics}).
Noting that $d\vec{U}/dt\approx d^{2}\vec{x}/dt^{2}-\vec{v}\partial\Phi/\partial t$,
with $\vec{v}=d\vec{x}/dt$, we can also write this result as 
\begin{equation}
\frac{d^{2}\vec{x}}{dt^{2}}=\vec{G}+\vec{v}\times\vec{H}-2\frac{\partial\xi_{\ j}^{i}}{\partial t}v^{j}\vec{e}_{i}+\frac{\partial\Phi}{\partial t}\vec{v}\;.\label{eq:Geolinear2}
\end{equation}

The gravitational field equations in this regime are obtained by linearizing
(\ref{eq:DivG1+3})-(\ref{eq:Bianchi1+3TimeSpace}) and substituting
relations (\ref{eq:KijLinear}): 
\begin{equation}
\begin{array}{c}
{\displaystyle \nabla\cdot\vec{G}=-4\pi(2\rho+T_{\ \alpha}^{\alpha})-\frac{\partial^{2}\xi_{\ i}^{i}}{\partial t^{2}}}\;;\qquad({\rm i)}\qquad\qquad\qquad{\displaystyle \nabla\times\vec{G}=-\frac{\partial\vec{H}}{\partial t}}\;;\qquad{\rm (ii)}\\
{\displaystyle \nabla\cdot\vec{H}=0\;;\qquad({\rm iii)}\qquad\qquad\qquad\qquad{\displaystyle \nabla\times\vec{H}=}-16\pi\vec{J}+4\frac{\partial}{\partial t}\xi_{j}^{\ [j,k]}\vec{e}_{k}\;;\qquad{\rm (iv)}}\\
{\displaystyle G_{j,i}+\frac{1}{2}\epsilon_{ijk}\frac{\partial H^{k}}{\partial t}+\frac{\partial^{2}}{\partial t^{2}}\xi_{ij}+2\xi_{\ (j,i)k}^{k}-\nabla^{2}\xi_{ij}-\xi_{\ k,ij}^{k}=8\pi\left(T_{ij}+\frac{1}{2}\delta_{ij}T_{\ \alpha}^{\alpha}\right)}\;.\qquad{\rm (v)}
\end{array}\label{eq:LinearFieldEqs}
\end{equation}
Eqs.~(\ref{eq:LinearFieldEqs}i), (\ref{eq:LinearFieldEqs}iv), and
(\ref{eq:LinearFieldEqs}v), are, respectively, the time-time, time-space,
and space-space components of Einstein's equations with sources (\ref{eq:EinsteinField}a);
Eqs.~(\ref{eq:LinearFieldEqs}iii) and (\ref{eq:LinearFieldEqs}ii)
are, respectively the time-time and space-time components of the identities
(\ref{eq:EinsteinField}b). To obtain (\ref{eq:LinearFieldEqs}v)
from the exact Eq.~(\ref{eq:SpaceSpace1+3}), we note that $\tilde{R}_{ij}$
reads, to linear order 
\[
\tilde{R}_{ij}\simeq\Gamma_{ij,k}^{k}-\Gamma_{kj,i}^{k}\simeq2\xi_{\ (j,i)k}^{k}-\nabla^{2}\xi_{ij}-\xi_{\ k,ij}^{k}
\]
As for the time-space component of the identity (\ref{eq:EinsteinField}b),
i.e., Eq.~(\ref{eq:Bianchi1+3TimeSpace}), it yields the trivial,
at linear order, equation $\star\tilde{R}_{\ ji}^{j}=0$.

Eqs.~(\ref{eq:LinearFieldEqs}) encompass two particularly important
regimes: the ``GEM limit'', and gravitational radiation. Starting
by the latter, in a source free region ($T^{\alpha\beta}=0$) one
can, as is well known, through gauge transformations (employing the
harmonic gauge condition, and further specializing to the transverse
traceless, or radiation, gauge, see e.g.~\cite{General Relativity})
make $\vec{\mathcal{A}}=\Phi=\xi_{\ i}^{i}=\xi_{\ \ ,j}^{ij}=0$;
with this choice, the only non trivial equation left is (\ref{eq:LinearFieldEqs}v),
yielding the 3-D wave equation $\partial^{2}\xi_{ij}/\partial t^{2}=\nabla^{2}\xi_{ij}$.

The GEM regime is obtained making $\xi_{ij}=-\Phi\delta_{ij}$ (which
effectively neglects radiation); in this case, the traceless shear
of the congruence of zero 3-velocity observers ($u^{i}=0$ in the
coordinates system of (\ref{eq:Linearmetric})) vanishes, $\sigma_{\alpha\beta}=0$,
and we have $u_{(i;j)}=\theta\delta_{ij}/3=-\delta_{ij}\partial\Phi/\partial t$.
This is also the case for the post-Newtonian regime (e.g.~\cite{The many faces,DSX,JantzenThomas,Nordtvedt1988,Nordvedt1973}).
Moreover, the source is assumed to be non-relativistic, so that the
contribution of the pressure and stresses in Eq.~(\ref{eq:LinearFieldEqs})
is negligible: $2\rho+T_{\ \alpha}^{\alpha}\approx\rho$. The two
versions of the equation for the geodesics, (\ref{eq:Geolinear})
and (\ref{eq:Geolinear2}), then read, respectively, 
\begin{equation}
\frac{d\vec{U}}{dt}=\vec{G}+\vec{U}\times\vec{H}+2\frac{\partial\Phi}{\partial t}\vec{v}\;;\qquad\frac{d^{2}\vec{x}}{dt^{2}}=\vec{G}+\vec{v}\times\vec{H}+3\frac{\partial\Phi}{\partial t}\vec{v}\label{eq:GeoLinGEM}
\end{equation}
and Eqs.~(\ref{eq:LinearFieldEqs}) above become 
\begin{equation}
\begin{array}{c}
{\displaystyle \nabla\cdot\vec{G}=-4\pi\rho+3\frac{\partial^{2}\Phi}{\partial t^{2}}}\;;\qquad({\rm i)}\qquad\qquad\qquad{\displaystyle \nabla\times\vec{G}=-\frac{\partial\vec{H}}{\partial t}}\;;\qquad{\rm (ii)}\\
{\displaystyle \nabla\cdot\vec{H}=0\;;\qquad({\rm iii)}\qquad\qquad{\displaystyle \nabla\times\vec{H}=}-16\pi\vec{J}+4\frac{\partial\vec{G}}{\partial t}-4\frac{\partial^{2}\vec{\mathcal{A}}}{\partial t^{2}}\;;\qquad{\rm (iv)}}\\
{\displaystyle \frac{\partial}{\partial t}\mathcal{A}_{(i,j)}-\left(\frac{\partial^{2}\Phi}{\partial t^{2}}-\nabla^{2}\Phi\right)\delta_{ij}}=-4\pi\rho\delta_{ij}\;.\qquad{\rm (v)}
\end{array}\label{eq:LinearFieldEqsGEM}
\end{equation}
In some works, e.g.~\cite{Ruggiero:2002hz}, the gravitoelectric
field is given a different definition: $\vec{G}'=-\nabla\Phi-\frac{1}{4}\partial\vec{\mathcal{A}}/\partial t$.
With this definition, and choosing the harmonic gauge condition, which
implies $\nabla\cdot\vec{\mathcal{A}}=-4\partial\Phi/\partial t$,
the non-Maxwellian term in Eq.~(\ref{eq:LinearFieldEqsGEM}i) disappears;
but, on the other hand, a ``non-Lorentzian'' term appears in the
equations for the geodesics, where in the place of $\vec{G}$ in Eqs.
(\ref{eq:Geolinear})-(\ref{eq:Geolinear2}), we would have instead
$\vec{G}'-\frac{3}{4}\partial\vec{\mathcal{A}}/\partial t$. As for
the non-Maxwellian term in Eq.~(\ref{eq:LinearFieldEqsGEM}iv), it
is neglected in the post-Newtonian regime \cite{DSX,The many faces}.

The presence of the terms $\partial\vec{H}/\partial t$ and $\partial\vec{G}/\partial t$,
``inducing'' curls in $\vec{G}$ and $\vec{H}$, respectively, analogous
to the induction terms of electromagnetism, leads to the question
of whether one can talk about gravitational induction effects in analogy
with electrodynamics. Indeed, there is a debate concerning the applicability
and physical content of this analogy for time-dependent fields, see
e.g.~\cite{CHPRD} and references therein. Although a discussion
of the approaches to this issue in the literature is outside the scope
of this work, still there are some points that can be made based on
the material herein. If one considers a time dependent gravitational
field, such as the one generated by a moving point mass, e.g.~Eq.~(2.10)
of \cite{PaperIAU}, one finds that indeed the corresponding gravitoelectric
field $\vec{G}$ is different from the one of a point mass at rest,
and has a curl. That is, the acceleration $-\vec{G}$ of the congruence
of observers at rest with respect to the background inertial frame
(the ``post-Newtonian grid'', e.g.~\cite{JantzenThomas}), acquires
a curl when the source moves with respect to that frame. From Eq.~(\ref{eq:LinearFieldEqsGEM}ii),
one can think about this curl as induced by the time-varying gravitomagnetic
field $\vec{H}$, see e.g.~\cite{Nordtvedt1988}. These fields are
well suited to describe the \emph{apparent} Newtonian and Coriolis-like
accelerations of particles in geodesic motion, as shown by Eq.~(\ref{eq:GeoLinGEM})
above (one must just bear in mind that in the case of time-dependent
fields, the motion is not determined solely by $\vec{G}$ and $\vec{H}$;
there is an additional term with no analogue in the Lorentz force
law, which leads to important differences). However, the latter are
artifacts of the reference frame; the \emph{physical} (i.e., tidal)
forces tell a different story, as one does not obtain the correct
tidal forces by differentiation of $\vec{G}$ and $\vec{H}$ (as is
the case with electrodynamics). Namely, the curls of the GEM fields
do not translate into these forces. The linearized gravitoelectric
tidal tensor, Eq.~(\ref{eq:Eij_Hij_linear}a), reads in the GEM regime
($K_{(ij)}=-\delta_{ij}\partial\Phi/\partial t$), 
\begin{equation}
\mathbb{E}_{ij}\approx-G_{i,j}+\frac{1}{2}\epsilon_{ijk}\frac{\partial H^{k}}{\partial t}-\frac{\partial\Phi}{\partial t}\delta_{ij}=-G_{(i,j)}-\frac{\partial\Phi}{\partial t}\delta_{ij}\;,\label{eq:EijGEM}
\end{equation}
where we see that the curl (\ref{eq:LinearFieldEqsGEM}ii) is subtracted
from the derivative of $\vec{G}$. That is, only the symmetrized derivative
$G_{(i,j)}$ describes physical, covariant forces. This is manifest
in the fact that the curl of $\vec{G}$ does not induce a rotation
on a set of neighboring particles (the gravitational field only shears
the set, see Sec.~\ref{sub:Gravity-vs-Electromagnetism} and Eq.~(\ref{eq:TidalGR})
therein), nor does it torque a rigid test body, see \cite{Gyros}.
Note that in electromagnetism this rotation and torque are tidal manifestations
of Faraday's law of induction. Likewise, the curl of $\vec{H}$ is
not manifest in the gravitomagnetic tidal effects (e.g., the force
on a gyroscope); the linearized gravitomagnetic tidal tensor (\ref{eq:Eij_Hij_linear}b)
reads, in this regime: 
\begin{equation}
\mathbb{H}_{ij}\approx-\frac{1}{2}\left[H_{i,j}-2\epsilon_{ijl}^{\ \ }\left(\frac{\partial G^{l}}{\partial t}-\frac{\partial^{2}\mathcal{A}^{l}}{\partial t^{2}}\right)\right]\;,\label{eq:HijGEM}
\end{equation}
where again we can see that the induction contribution $4\partial\vec{G}/\partial t$
(and also the one of the term $\partial^{2}\vec{\mathcal{A}}/\partial t^{2}$)
to the curl of $\vec{H}$ is subtracted from the derivative of $\vec{H}$.
The physical consequences are explored in \cite{Gyros}: in electromagnetism,
due to vacuum equation $\nabla\times\vec{B}=\partial\vec{E}/\partial t$,
there is a non-vanishing force on a magnetic dipole, $F_{EM}^{i}=B^{\beta i}\mu_{\beta}$
($=\nabla^{i}(\vec{\mu}\cdot\vec{B})$ in the comoving inertial frame,
cf. Eq. (\ref{eq:FEM_QM0})), whenever it moves in a non-homogeneous
field; this is because the electric field measured by the particle
is time-varying, and so $\nabla\times\vec{B}\ne0\Rightarrow B_{ij}\ne0\Rightarrow\vec{F}_{EM}\ne0$.
That is not necessarily the case in gravity. In vacuum, from Eqs.~(\ref{eq:LinearFieldEqsGEM}iv)
and (\ref{eq:HijGEM}), we have $\mathbb{H}_{ij}=-H_{(i,j)}/2$, and
the gravitational force on a gyroscope, cf.~Eq.~(\ref{analogy}.2b)
of Table \ref{analogy}, is $F_{G}^{i}=\frac{1}{2}H^{(i,j)}S_{j}$.
Thus no analogous induction effect is manifest in the force, and in
fact spinning particles in non-homogeneous gravitational fields can
move along geodesics, as exemplified in \cite{Gyros}.

As for the equation of motion for the gyroscope's spin vector, from
Eq.~(\ref{eq:Fermi-Walker}) we get, in terms of components in the
coordinate system associated to (\ref{eq:Linearmetric}), 
\begin{equation}
\frac{dS^{i}}{dt}=-\Gamma_{0j}^{i}S^{j}=\frac{1}{2}(\vec{S}\times\vec{H})^{i}-\frac{\partial\Phi}{\partial t}S^{i}\;.\label{PrecessLinear}
\end{equation}
Comparing with the equation for the precession of a magnetic dipole
(with respect to an inertial frame), $d\vec{S}/d\tau=\vec{\mu}\times\vec{B}$,
there is a factor of $1/2$, and an additional term. The origin of
the former is explained in Sec.~\ref{sub:3+1 Gyroscope-precession}:
it is due to the fact the field $\vec{H}$, causing the Coriolis (or
gravitomagnetic) acceleration of test particles via Eq.~(\ref{eq:Geolinear}),
is distinct from the field causing the gyroscope precession in (\ref{PrecessLinear});
in general they are \emph{independent}. $\vec{H}$ is the sum of the
vorticity $\vec{\omega}$ of the observer congruence with the angular
velocity of rotation $\vec{\Omega}$ of the frame's spatial triads
relative to Fermi-Walker transport; and to Eq.~(\ref{PrecessLinear})
only the latter part contributes. In the case of a congruence adapted
frame ($\vec{\Omega}=\vec{\omega})$, which is the problem at hand
(the frame is adapted to the congruence of $u^{i}=0$ observers),
this originates the relative factor of $1/2$. Note also that the
same factor shows up also in the force on the gyroscope discussed
above, but in this case by the opposite reason: to $\vec{F}_{G}$
the vorticity $\vec{\omega}$ is the only part of $\vec{H}$ that
contributes, cf.~Eq.~(\ref{Hij}). The second term in (\ref{PrecessLinear})
merely reflects the fact that the basis vectors $\mathbf{e}_{i}$
expand; if, using expressions (\ref{eq:Linear-ehat}), one transforms
to the orthonormal basis $S^{i}=e_{\ \hat{i}}^{i}S^{\hat{i}}$, and
substitutes into (\ref{PrecessLinear}), that term vanishes, as expected
from the exact result (\ref{eq:Spin3+1v2}).

If the field is stationary, we have a one to one correspondence with
electromagnetism \emph{in inertial frames}. Eq.~(\ref{eq:LinearFieldEqs}v)
above becomes identical to (\ref{eq:LinearFieldEqs}i), and then we
are left with a set of four equations --- Eqs.~(\ref{eq:LinearFieldEqs}i)-(\ref{eq:LinearFieldEqs}iv)
with the time dependent terms dropped --- similar, up to some factors,
to the time-independent Maxwell equations in an inertial frame. These
equations can also be obtained by linearization of Eqs.~(\ref{Tab:analogy1+3}4b)-(\ref{Tab:analogy1+3}4b)
of Table \ref{Tab:analogy1+3}. The space part of the equation of
the geodesics: $d^{2}\vec{x}/dt^{2}=\vec{G}+\vec{v}\times\vec{H}$,
cf.~Eq.~(\ref{eq:Geolinear2}) above, is also similar to the Lorentz
force in a Lorentz frame. The equation for the evolution of the spin
vector of a gyroscope, in the coordinate basis, becomes simply $d\vec{S}/d\tau=\vec{S}\times\vec{H}/2$,
which gives the precession relative to the background Minkowski frame,
and is similar to the precession of a magnetic dipole in a magnetic
field. The force on a gyroscope whose center of mass it \emph{at}
\emph{rest} is $\vec{F}_{G}=\nabla(\vec{S}\cdot\vec{H})/2$, similar
to the force $\vec{F}_{EM}=\nabla(\vec{\mu}\cdot\vec{B})$ on a magnetic
dipole at rest in a Lorentz frame; the same for the differential precession
of gyroscopes/dipoles at rest: for a spatial separation vector $\delta x^{\alpha}$
they read, respectively, $\delta\vec{\Omega}_{G}=-\nabla(\delta\vec{x}\cdot\vec{H})/2$
and $\delta\vec{\Omega}_{EM}=-\nabla(\delta\vec{x}\cdot\vec{B})$.

\section{The \emph{formal} analogy between gravitational \textit{tidal tensors}
and electromagnetic \textit{fields\label{sec:Weyl analogy}}}

There is a set of analogies, based on exact expressions, relating
the Maxwell tensor $F^{\alpha\beta}$ and the Weyl tensor $C_{\alpha\beta\gamma\delta}$.
These analogies rest on the fact that: 1) they both irreducibly decompose
into an electric and a magnetic type spatial tensors; 2) these tensors
obey differential equations --- Maxwell's equations and the so called
``higher order'' gravitational field equations --- which are formally
analogous to a certain extent \cite{Maartens:1997fg,EMMbook,bel,general relativity and cosmology,ellis:97};
and 3) they form invariants in a similar fashion \cite{matte,bel,Bonnor:1995zf,cherubini:02}.
In this section we will briefly review these analogies and clarify
their physical content in the light of the previous approaches.

The Maxwell tensor splits, with respect to a unit time-like vector
$u^{\alpha}$, into its electric, $E^{\alpha}\equiv(E^{u})^{\alpha}=F_{\ \beta}^{\alpha}u^{\beta}$,
and magnetic, $B^{\alpha}\equiv(B^{u})^{\alpha}=\star F_{\ \ \beta}^{\alpha}u^{\beta}$,
parts; i.e., the electric and magnetic fields as measured by the observers
of 4-velocity $u^{\alpha}$. These are spatial vectors: $E^{\alpha}u_{\alpha}=B^{\alpha}u_{\alpha}=0$,
thus possessing 3+3 independent components, which completely encode
the 6 independent components of $F_{\mu\nu}$, as can be seen explicitly
in the decompositions (\ref{eq:Fdecomp}). In spite of their dependence
on $u^{\alpha}$, one can use $E^{\alpha}$ and $B^{\beta}$ to define
two tensorial quantities which are $u^{\alpha}$ independent, namely
\begin{equation}
E^{\alpha}E_{\alpha}-B^{\alpha}B_{\alpha}=-\frac{F_{\alpha\beta}F^{\alpha\beta}}{2}\ ,\qquad E^{\alpha}B_{\alpha}=-\frac{\star F_{\alpha\beta}F^{\alpha\beta}}{4}\ ;\label{EMinv}
\end{equation}
these are the only algebraically independent invariants one can define
from the Maxwell tensor.

The Weyl tensor has a formally similar decomposition: with respect
to a unit time-like vector $u^{\alpha}$, it splits irreducibly into
its electric, $\mathcal{E}_{\alpha\beta}\equiv(\mathcal{E}^{u})_{\alpha\beta}=C_{\alpha\gamma\beta\sigma}u^{\gamma}u_{\textrm{ }}^{\sigma}$,
and magnetic, $\mathcal{H}_{\alpha\beta}\equiv(\mathcal{H}^{u})_{\alpha\beta}=\star C_{\alpha\gamma\beta\sigma}u^{\gamma}u_{\textrm{ }}^{\sigma}$,
parts. These two spatial tensors, both of which are symmetric and
traceless (hence have 5 independent components each), completely encode
the 10 independent components of the Weyl tensor, as can be seen by
writing \cite{Maartens:1997fg} 
\begin{equation}
C_{\alpha\beta}^{\ \ \ \gamma\delta}=4\left\{ 2u_{[\alpha}u^{[\gamma}+g_{[\alpha}^{\ [\gamma}\right\} \mathcal{E}_{\beta]}^{\ \delta]}+2\left\{ \epsilon_{\alpha\beta\mu\nu}u^{[\gamma}\mathcal{H}^{\delta]\mu}u^{\nu}+\epsilon^{\gamma\delta\mu\nu}u_{[\alpha}\mathcal{H}_{\beta]\mu}u_{\nu}\right\} \ \label{decompWeyl}
\end{equation}
(in vacuum, this equals decomposition (\ref{Bel})). Again, in spite
of their dependence on $u^{\alpha}$, one can use $\mathcal{E}_{\alpha\beta}$
and $\mathcal{H}_{\alpha\beta}$ to define the two tensorial quantities
which are $U^{\alpha}$ independent,

\begin{equation}
\mathcal{E}^{\alpha\beta}\mathcal{E}_{\alpha\beta}-\mathcal{H}^{\alpha\beta}\mathcal{H}_{\alpha\beta}=\frac{C_{\alpha\beta\mu\nu}C^{\alpha\beta\mu\nu}}{8}\ ,\ \ \ \ \ \mathcal{E}^{\alpha\beta}\mathcal{H}_{\alpha\beta}=\frac{\star C_{\alpha\beta\mu\nu}C^{\alpha\beta\mu\nu}}{16}\ ,\label{inv3}
\end{equation}
\textcolor{black}{which are} \textcolor{black}{\emph{formally}} \textcolor{black}{analogous
to the electromagnetic scalar invariants (\ref{EMinv}). Note however
that, by contrast with the latter, these are not the only independent
scalar invariants one can construct from} $C_{\alpha\beta\mu\nu}$;
there are also two cubic invariants, see e.g. \cite{bel,WyllePRD,PaperInvariantes,McIntosh et al 1994,Lozanovski:99,Zakharov}.

As stated above, these tensors obey also differential equations which
have some formal similarities with Maxwell's; such equations, dubbed
the ``higher order field equations'', are obtained from the Bianchi
identities \textbf{$R_{\sigma\tau[\mu\nu;\alpha]}=0$}. These, together
with the field equations (\ref{eq:EinsteinField}a), lead to:

\begin{equation}
C_{\ \nu\sigma\tau;\mu}^{\mu}=8\pi\left(T_{\nu[\tau;\sigma]}-\frac{1}{3}g_{\nu[\tau}T_{;\sigma]}\right)\ ,\label{fe4}
\end{equation}

Expressing $C_{\alpha\beta\delta\gamma}$ in terms of $\mathcal{E}_{\alpha\beta}$
and $\mathcal{H}_{\alpha\beta}$ using (\ref{decompWeyl}), and taking
time and space projections of (\ref{fe4}) using the projectors (\ref{eq:Time_Space_Proj}),
we obtain, assuming a \emph{perfect fluid}, the set of equations 
\begin{equation}
\begin{array}{c}
{\displaystyle {\tilde{\nabla}^{\mu}\mathcal{E}_{\nu\mu}=\frac{8\pi}{3}\tilde{\nabla}_{\nu}\rho+3\omega^{\mu}\mathcal{H}_{\nu\mu}+\epsilon_{\nu\alpha\beta}\sigma_{\ \gamma}^{\alpha}\mathcal{H}^{\beta\gamma}\ ;}\spa{0.4cm}}\\
{\displaystyle {{\rm {curl}}\mathcal{H}_{\mu\nu}=\nabla_{\mathbf{u}}^{\perp}\mathcal{E}_{\mu\nu}+\mathcal{E}_{\mu\nu}\theta-3\sigma_{\tau\langle\mu}\mathcal{E}_{\nu\rangle}^{\ \tau}-\omega^{\tau}\epsilon_{\tau\rho(\mu}\mathcal{E}_{\nu)}^{\ \rho}-2a^{\rho}\epsilon_{\rho\tau(\mu}\mathcal{H}_{\nu)}^{\ \tau}+4\pi(\rho+p)\sigma_{\mu\nu}\ ;}}
\end{array}\label{MarteensG1}
\end{equation}

\begin{equation}
\begin{array}{c}
{\displaystyle {\tilde{\nabla}^{\mu}\mathcal{H}_{\nu\mu}=-8\pi(\rho+p)\omega_{\nu}-3\omega^{\mu}\mathcal{E}_{\nu\mu}-\epsilon_{\nu\alpha\beta}\sigma_{\ \gamma}^{\alpha}\mathcal{E}^{\beta\gamma}\ ;}\spa{0.4cm}}\\
{\displaystyle {{\rm {curl}}\mathcal{E}_{\mu\nu}=-\nabla_{\mathbf{u}}^{\perp}\mathcal{H}_{\mu\nu}-\mathcal{H}_{\mu\nu}\theta+3\sigma_{\tau\langle\mu}\mathcal{H}_{\nu\rangle}^{\ \tau}+\omega^{\tau}\epsilon_{\tau\rho(\mu}\mathcal{H}_{\nu)}^{\ \rho}-2a^{\rho}\epsilon_{\rho\tau(\mu}\mathcal{E}_{\nu)}^{\ \tau}\ ,}}
\end{array}\label{MaartensG2}
\end{equation}
where, following the definitions in \cite{Maartens:1997fg,EMMbook},
$\epsilon_{\mu\nu\rho}\equiv\epsilon_{\mu\nu\rho\tau}u^{\tau}$, ${\rm curl}A_{\alpha\beta}\equiv\epsilon_{\ \ (\alpha}^{\mu\nu}A_{\beta)\nu;\mu}$,
and the index notation $\langle\mu\nu\rangle$ stands for the spatially
projected, symmetric and trace free part of a rank two tensor: 
\[
A_{\langle\mu\nu\rangle}\equiv h_{(\mu}^{\ \ \alpha}h_{\nu)}^{\beta}A_{\alpha\beta}-\frac{1}{3}h_{\mu\nu}h_{\alpha\beta}A^{\alpha\beta}\ ,
\]
with $h_{\ \beta}^{\alpha}\equiv(h^{u})_{\ \beta}^{\alpha}$, cf.
Eq. (\ref{eq:Time_Space_Proj}). $\nabla_{\mathbf{u}}^{\perp}$ and
$\tilde{\nabla}$ (which in the equations above we could have written
as well $\nabla^{\perp}$, for they are the same along the spatial
directions) are the derivative operators whose action on a spatial
vector is defined in Eqs. (\ref{eq:DdtDFdt}) and (\ref{eq:NablaTildaCov}),
respectively%
\footnote{To make contact with the notation in {[}35,36{]}, we note that the
restriction to the spatial directions of both $\tilde{\nabla}$ and
$\nabla^{\perp}$ yields the 3-D connection ``$\bar{\nabla}$''
of {[}36{]} (``$D$'' of {[}35{]}).%
}. For a rank two \emph{spatial} tensor $A^{\alpha\beta}$, we have
$\nabla_{\mathbf{u}}^{\perp}A^{\alpha\beta}=h_{\ \mu}^{\alpha}h_{\ \nu}^{\beta}\nabla_{\mathbf{u}}A^{\mu\nu}$
and $\tilde{\nabla}_{\alpha}A^{\alpha\beta}=h_{\ \mu}^{\alpha}h_{\ \nu}^{\beta}\nabla_{\alpha}A^{\mu\nu}$.
As before, the quantities $\theta\equiv u_{\ ;\alpha}^{\alpha}$,
$\sigma_{\mu\nu}\equiv h_{\ \mu}^{\alpha}h_{\ \nu}^{\beta}u_{\alpha;\beta}-\theta h_{\mu\nu}/3$,
$\omega^{\alpha}\equiv\epsilon_{\ \beta\gamma}^{\alpha}u_{\gamma;\beta}/2$
and $a^{\alpha}$ are, respectively, the expansion, shear, vorticity
and acceleration of the congruence of observers with 4-velocity $u^{\alpha}$.

The analogous electromagnetic equations are the ones in Sec.~\ref{sub:Maxwell-equations-1+3},
which we can re-write as 
\begin{equation}
\tilde{\nabla}_{\mu}E^{\mu}=4\pi\rho_{c}+2\omega_{\mu}B^{\mu}\;;\label{eq:Gaus3+1}
\end{equation}
\begin{equation}
\epsilon^{\alpha\gamma\beta}B_{\beta;\gamma}=\nabla_{\mathbf{u}}^{\perp}E^{\alpha}-\sigma_{\ \beta}^{\alpha}E^{\beta}+\frac{2}{3}\theta E^{\alpha}-\epsilon_{\ \beta\gamma}^{\alpha}\omega^{\beta}E^{\gamma}+\epsilon_{\ \beta\gamma}^{\alpha}B^{\beta}a^{\gamma}+4\pi j^{\langle\alpha\rangle}\;;\label{eq:CurlBMaartens}
\end{equation}
\begin{equation}
\tilde{\nabla}_{\mu}B^{\mu}=-2\omega_{\mu}E^{\mu}\;;\label{eq:DivB3+1}
\end{equation}
\begin{equation}
\epsilon^{\alpha\gamma\beta}E_{\beta;\gamma}=-\nabla_{\mathbf{u}}^{\perp}B^{\alpha}+\sigma_{\ \beta}^{\alpha}B^{\beta}-\frac{2}{3}\theta B^{\alpha}+\epsilon_{\ \beta\gamma}^{\alpha}\omega^{\beta}B^{\gamma}+\epsilon_{\ \ }^{\alpha\mu\sigma}E_{\mu}a_{\sigma}\;.\label{eq:CurlEMaartens}
\end{equation}
Eqs.~(\ref{eq:Gaus3+1}) and (\ref{eq:DivB3+1}) follow from Eqs.
(\ref{eq:GaussLawCov}) and (\ref{eq:DivBCov}), respectively, by
noting that, for an arbitrary spatial vector $A^{\alpha}$, 
\[
A_{\ ;\beta}^{\beta}=\left(\top_{\beta}^{\gamma}+h_{\beta}^{\gamma}\right)\left(\top_{\lambda}^{\beta}+h_{\lambda}^{\beta}\right)A_{\ ;\gamma}^{\lambda}=\left(\top_{\lambda}^{\gamma}+h_{\lambda}^{\gamma}\right)A_{\ ;\gamma}^{\lambda}=A^{\beta}a_{\beta}+\tilde{\nabla}_{\alpha}A^{\alpha}.
\]
Eqs.~(\ref{eq:CurlBMaartens}) and (\ref{eq:CurlEMaartens}) follow
from Eqs.~(\ref{eq:CurlB1}) and (\ref{eq:CurlE1}) by decomposing
$K_{(\alpha\beta)}=\sigma_{\alpha\beta}+\theta h_{\alpha\beta}/3$.

It is worth mentioning that the \emph{exact} wave equations for $E^{\alpha}$
and $B^{\alpha}$ in this formalism were obtained in \cite{Tsagas2005Waves},
Eqs. (39)-(40) therein%
\footnote{The wave equations in \cite{Tsagas2005Waves,TsagasOther} are obtained
using also the Ricci identities $2\nabla_{[\gamma}\nabla_{\beta]}X_{\alpha}=R_{\delta\alpha\beta\gamma}X^{\delta}$,
which couple the electromagnetic fields to the curvature tensor; this
coupling is shown to lead to amplification phenomena, suggested therein
as a possible explanation for the observed (and unexplained) large-scale
cosmic magnetic fields.%
}. As for the \emph{exact} wave equations for $\mathcal{E}_{\alpha\beta}$
and $\mathcal{H}_{\alpha\beta}$, they have not, to our knowledge,
been derived in the literature; only in some approximations, such
as in e.g. \cite{ellis:97,Dunsby:1998hd}, or the linear regime of
the next section.

\subsection{Matte's equations vs Maxwell equations. Tidal tensor interpretation
of gravitational radiation. \label{sub:Matte's-equations-vs}}

\begin{table}[H]
\caption{\label{tab:MatteMaxwell}Formal analogy between Maxwell's equations
(differential equations for electromagnetic \emph{fields}) and Matte's
equations (differential equations for gravitational \emph{tidal} tensors)}

\begin{tabular}{cccccccc}
\hline 
\raisebox{3.5ex}{}\raisebox{0.5ex}{Electromagnetism}  &  &  &  &  & \raisebox{0.5ex}{Linearized Gravity}  &  & \tabularnewline
\hline 
\hline 
\raisebox{3.5ex}{}Maxwell's Equations  &  &  &  &  & Matte's Equations  &  & \tabularnewline
\raisebox{3ex}{} $E_{\,,i}^{i}=0$  &  & (\ref{tab:MatteMaxwell}.1a)  &  &  & $\mathbb{E}_{\,\,\,\,,i}^{ij}=0$  &  & (\ref{tab:MatteMaxwell}.1b)\tabularnewline
\raisebox{3ex}{} $B_{\,,i}^{i}=0$  &  & (\ref{tab:MatteMaxwell}.2a)  &  &  & $\mathbb{H}_{\,\,\,\,,i}^{ij}=0$  &  & (\ref{tab:MatteMaxwell}.2b)\tabularnewline
\raisebox{4.5ex}{} ${\displaystyle \epsilon^{ikl}E_{l,k}=-\frac{\partial B^{i}}{\partial t}}$  &  & (\ref{tab:MatteMaxwell}.3a)  &  &  & ${\displaystyle \epsilon^{ikl}\mathbb{E}_{\,\,\, l,k}^{j}=-\frac{\partial\mathbb{H}^{ij}}{\partial t}}$  &  & (\ref{tab:MatteMaxwell}.3b)\tabularnewline
\raisebox{7.3ex}{}\raisebox{2.2ex}{${\displaystyle \epsilon^{ikl}B_{l,k}=\frac{\partial E^{i}}{\partial t}}$}  &  & \raisebox{2.2ex}{(\ref{tab:MatteMaxwell}.4a)}  &  &  & ~~\raisebox{2.2ex}{${\displaystyle \epsilon^{ikl}\mathbb{H}_{\,\,\, l,k}^{j}=\frac{\partial\mathbb{E}^{ij}}{\partial t}}$}~~  &  & \raisebox{2.2ex}{(\ref{tab:MatteMaxwell}.4b)}\tabularnewline
\hline 
\raisebox{3.5ex}{}Wave equations  &  &  &  &  & Wave equations  &  & \tabularnewline
\raisebox{5ex}{} ${\displaystyle \left(\frac{\partial^{2}}{\partial t^{2}}-\partial^{k}\partial_{k}\right)E^{i}=0}$  &  & (\ref{tab:MatteMaxwell}.5a)  &  &  & ${\displaystyle \left(\frac{\partial^{2}}{\partial t^{2}}-\partial^{k}\partial_{k}\right)\mathbb{E}_{ij}=0}$  &  & (\ref{tab:MatteMaxwell}.5b)\tabularnewline
\raisebox{8ex}{}\raisebox{2.5ex}{${\displaystyle \left(\frac{\partial^{2}}{\partial t^{2}}-\partial^{k}\partial_{k}\right)B^{i}=0}$}  &  & \raisebox{2.5ex}{(\ref{tab:MatteMaxwell}.6a)}  &  &  & ~~\raisebox{2.5ex}{${\displaystyle \left(\frac{\partial^{2}}{\partial t^{2}}-\partial^{k}\partial_{k}\right)\mathbb{H}_{ij}=0}$}~~  &  & \raisebox{2.5ex}{(\ref{tab:MatteMaxwell}.6b)}\tabularnewline
\hline 
\end{tabular}
\end{table}

In vacuum, the Bianchi identities become: 
\begin{equation}
R_{\sigma\tau[\mu\nu;\alpha]}=0;\quad(a)\qquad R_{\ \alpha\beta\gamma;\mu}^{\mu}=0\quad(b)\label{eq:BianchiVacuum}
\end{equation}
(the second equation following from the first and from vacuum equation
$R_{\mu\nu}=0$). The formal analogy with Eqs.~(\ref{eq:MaxwellFieldEqs}),
for $j^{\alpha}=0$, is now more clear \cite{bel}. In a nearly Lorentz
frame where $u^{i}=0$, and to linear order in the metric potentials,
Eqs.~(\ref{MarteensG1})-(\ref{MaartensG2}), for vacuum, become
Eqs.~(\ref{tab:MatteMaxwell}.1b)-(\ref{tab:MatteMaxwell}.4b) of
Table \ref{tab:MatteMaxwell}, which are \emph{formally} similar to
Maxwell's equations in a Lorentz frame (\ref{tab:MatteMaxwell}.1a)-(\ref{tab:MatteMaxwell}.4a).
The analogy in Eqs.~(\ref{tab:MatteMaxwell}.1)-(\ref{tab:MatteMaxwell}.4)
was first found by Matte \cite{matte}, and further studied by some
other authors \cite{bel,CampbellMorganAjp,Tchrakian}. Taking curls
of Eqs.~(\ref{tab:MatteMaxwell}.3a)-(\ref{tab:MatteMaxwell}.4a)
we obtain the wave equations for the electromagnetic fields; and taking
curls of (\ref{tab:MatteMaxwell}.3b)-(\ref{tab:MatteMaxwell}.4b),
we obtain gravitational waves as wave equations for gravitational
tidal tensors.

Hence, to this degree of accuracy, vacuum gravitational waves can
be cast as a pair of oscillatory tidal tensors $\mathbb{E}_{\alpha\beta}$,
$\mathbb{H}_{\alpha\beta}$, propagating in space by mutually inducing
each other, just like the pair of fields $E^{\alpha}$, $B^{\alpha}$,
in the case of the electromagnetic waves. Also, just like $E^{\alpha}$
and $B^{\alpha}$ are equal in magnitude and mutually orthogonal for
a purely radiative field, the same applies to the waves in (\ref{tab:MatteMaxwell}.5b)-(\ref{tab:MatteMaxwell}.6b)
of Table \ref{tab:MatteMaxwell}. In the electromagnetic case this
implies that the two invariants (\ref{EMinv}) vanish; likewise, the
gravitational invariants (\ref{inv3}) also vanish for a solution
corresponding to pure gravitational radiation according to Bel's second
criterion (cf.~e.g.~\cite{Zakharov} p. 53) --- a definition based
on ``super-energy'', see below.

An interesting aspect of this formulation of gravitational radiation,
contrasting with the more usual approaches in the literature, e.g.
\cite{General Relativity,Gravitation,Gravitation and Inertia,CordaTidalTensors}
--- which consist of equations for the propagation of gauge fields
(the components of the metric tensor), having no local physical significance
(only their second derivatives may be related to physically measurable
quantities, see in this respect \cite{HawkingAPJ1966}) --- is that
Eqs.~(\ref{tab:MatteMaxwell}.5b)-(\ref{tab:MatteMaxwell}.6b) are
equations for the propagation of \emph{tensors of physical forces,}
with direct translation in physical effects: the relative acceleration
of two neighboring test particles via geodesic deviation equation
(\ref{analogy}.1b) of Table \ref{analogy}, the force on a spinning
test particle, via Mathisson-Papapetrou-Pirani Eq.~(\ref{analogy}.2b),
or the relative precession of two nearby gyroscopes, via Eqs.~(\ref{eq:RelPrecGrav0})-(\ref{eq:RelPrecGrav}).

It is instructive to note this contrast: whereas in electromagnetic
radiation it is the vector fields that propagate, gravitational radiation
is a purely tidal effect, i.e., traveling tidal tensors not subsidiary
to any associated (electromagnetic-like, or Newtonian-like) vector
field; it is well known that there are no vector waves in gravity
(see e.g.~\cite{DeserFranklin,Gravitation,Tucker Clark}; such waves
would carry negative energy if they were to exist, cf.~\cite{Gravitation}
p. 179). We have seen in Sec.~\ref{sub:Relation1+3_TTensors} that,
except for the very special case of the linear regime in weak, \emph{stationary}
fields (and static observers therein), the gravitational tidal tensors
cannot be cast as derivatives of some vector field. In the electromagnetic
case there are of course also tidal effects associated to the wave;
but their dynamics follows trivially%
\footnote{We thank J. Penedones for discussions on this point.%
} from Eqs. (\ref{tab:MatteMaxwell}.3a)-(\ref{tab:MatteMaxwell}.4a)
of Table \ref{tab:MatteMaxwell}; to this accuracy, the tidal tensors
as measured by the background static observers are just $E_{ij}=E_{i,j}$,
$B_{ij}=B_{i,j}$; hence the equations of their evolution (i.e., the
``electromagnetic higher order equations'') are: 
\begin{equation}
\epsilon_{i}^{\ kl}E_{jl,k}=0\;;\qquad\epsilon_{i}^{\ kl}B_{jl,k}=0\;;\label{eq:HigherOrderEMEqs1}
\end{equation}

\begin{equation}
\epsilon_{i}^{\ kl}E_{lj,k}=\epsilon_{i}^{\ kl}E_{lk,j}=-\frac{\partial B_{ij}}{\partial t}\;;\qquad\epsilon_{i}^{\ kl}B_{lj,k}=\epsilon_{i}^{\ kl}B_{lk,j}=\frac{\partial E_{ij}}{\partial t}\;.\label{eq:HigherOrderEMEqs2}
\end{equation}
These four equations are the \emph{physical} analogues of the pair
of gravitational Eqs.~(\ref{tab:MatteMaxwell}.3b)-(\ref{tab:MatteMaxwell}.4b);
we have two more equations in electromagnetism, since $E_{ij}$ and
$B_{ij}$ are not symmetric. Eqs.~(\ref{eq:HigherOrderEMEqs1}),
and the first equality in Eqs.~(\ref{eq:HigherOrderEMEqs2}), come
from the fact that derivatives in flat spacetime commute; therefore
$\epsilon_{i}^{\ kl}E_{jl,k}=\epsilon_{i}^{\ kl}E_{j,[lk]}=0$ and
$E_{lj,k}=E_{l,jk}=E_{lk,j}$. Thus, Eqs.~(\ref{eq:HigherOrderEMEqs2}),
which are the only ones that contain dynamics, are obtained by simply
differentiating Eqs.~(\ref{tab:MatteMaxwell}.3a)-(\ref{tab:MatteMaxwell}.4a)
with respect to $x^{j}$. The wave equations for the electromagnetic
tidal tensors follow likewise from differentiating Eqs.~(\ref{tab:MatteMaxwell}.5a)-(\ref{tab:MatteMaxwell}.6a)
with respect to $x^{j}$. Note that the fact that, in gravity, $\mathbb{H}_{j[l,k]}\ne0$,
is again related to the fact that, even in the linear regime, the
gravitational tidal tensors are not derivatives of some vector fields.

\subsubsection{Super-energy}

Using the analogies herein as a guiding principle, a gravitational
4-index tensor $T^{\alpha\beta\gamma\delta}$ --- the Bel-Robinson,
or ``super-energy'' tensor, see e.g. \cite{SenovillaBigPaper,Maartens:1997fg}
--- constructed from the curvature tensor in a way formally analogous
to the way the energy-momentum tensor of the electromagnetic field
($T_{{\rm EM}}^{\alpha\beta}$) is constructed from $F^{\alpha\beta}$,
has been proposed. The motivation is to find \emph{local, covariant},
quantities alternative to the gravitational energy and momentum given
by the Landau-Lifshitz pseudo-tensor \cite{LandauLifshitz} (which
can only have a meaning in a global sense, and in asymptotically flat
spacetimes); the former however do not have the same dimensions, which
has been posing difficulties in their physical interpretation. For
a discussion on this issue and on the possible relation between energy
and super-energy (which is still an open problem) we refer to \cite{SenovillaBigPaper}
p. 31 (and references therein), and \cite{HawkingAPJ1966,PiraniSE,Garecki,KomarSuperenergy,Teyssandier,BelIHP,AlfonsoSE}.
Herein we would just like to point out that the viewpoint that gravitational
waves are characterized by a flow of super-energy fits well with their
interpretation as a pair of propagating tidal tensors, since, as can
be seen comparing Eqs. (23)-(24) to (40)-(41) of \cite{Maartens:1997fg},
the ``super-energy'' scalar $W\equiv T^{\alpha\beta\gamma\delta}u_{\alpha}u_{\beta}u_{\gamma}u_{\delta}$
and the ``super-Poynting'' vector $\mathcal{P}^{\langle\alpha\rangle}\equiv-T^{\langle\alpha\rangle\beta\gamma\delta}u_{\beta}u_{\gamma}u_{\delta}$
(as measured by some observer $u^{\alpha}$), when written explicitly
in terms of tidal tensors, are formally analogous to electromagnetic
field energy $\rho_{EM}\equiv T_{EM}^{\alpha\beta}u_{\alpha}u_{\beta}$
and Poynting vector $p_{EM}^{\langle\alpha\rangle}\equiv-T_{EM}^{\langle\alpha\rangle\beta}u_{\beta}$,
only with $\mathbb{E}_{\alpha\beta}$, $\mathbb{H}_{\alpha\beta}$
in the place of the electromagnetic fields $E^{\alpha}$, $B^{\alpha}$.

\subsection{The relationship with the other GEM analogies}

The analogy drawn in this section is between the electromagnetic fields
and the electric and magnetic parts of the Weyl tensor: $\{E^{\alpha},\, B^{\alpha}\}\leftrightarrow\{\mathcal{E}_{\mu\nu},\,\mathcal{H}_{\mu\nu}\}$.
It is clear, from the discussion of the physical meaning of $\{\mathbb{E}_{\mu\nu},\,\mathbb{H}_{\mu\nu}\}$
in Sec.~\ref{sec:Tidal tensor analogy}, and from the discussion
in Sec.~\ref{sec:3+1} of the dynamical gravitational counterparts
of $\{E^{\alpha},\, B^{\alpha}\}$, that this analogy is a purely
formal one. It draws a parallelism between electromagnetic fields
(whose dynamical gravitational analogues are the GEM inertial fields
$\{\vec{G},\vec{H}\}$ of Sec.~\ref{sec:3+1}), with gravitational
\emph{tidal} fields, which, as shown in Sec.~\ref{sec:Tidal tensor analogy},
are the physical analogues \emph{not} of $\{E^{\alpha},\, B^{\alpha}\}$,
but instead of the electromagnetic \emph{tidal} tensors $\{E_{\mu\nu,\ }B_{\mu\nu}\}$
(these, in an inertial frame, are \emph{derivatives} of the $E^{\alpha}$
and $B^{\alpha}$, cf. Eqs. (\ref{Eij1+3})-(\ref{Bi01+3})). This
sheds light on some conceptual difficulties in the literature regarding
the physical content of the analogy and in particular the physical
interpretation of the tensor $\mathcal{H}_{\mu\nu}$, see \cite{CHPreprint}
for details. It is also of crucial importance for the correct understanding
of physical meaning of the curvature invariants, and their implications
on the motion of test particles, which will be subject of detailed
study elsewhere \cite{PaperInvariantes}.

\section{When can gravity be similar to electromagnetism?\label{sec:When-can-gravity}}

The gravitational and electromagnetic interactions have many intrinsic
differences, perhaps the most basic of them being that the equivalence
principle between inertial mass and gravitational mass/charge has
no counterpart in electrodynamics (in the multipole language of Sec.
\ref{sec:Tidal tensor analogy}, no covariant gravitational force
is exerted on a monopole particle, by contrast with the electromagnetic
Lorentz force). But other important differences exist and are manifest
in the approaches herein; at the same time striking similarities emerged.

A crucial point to realize is that the two exact physical gravito-electromagnetic
analogies --- the tidal tensor analogy of Sec.~\ref{sec:Tidal tensor analogy},
and the inertial GEM fields analogy of Sec.~\ref{sec:3+1} --- do
not rely on a close physical similarity between the interactions;
the gravitational objects $\{\vec{G},\vec{H},\mathbb{E}_{\alpha\beta},\mathbb{H}_{\alpha\beta}\}$,
despite playing \emph{analogous dynamical roles} to the ones played
by the objects $\{\vec{E},\vec{B},E_{\alpha\beta},B_{\alpha\beta}\}$
in electromagnetism, are themselves in general very different from
the latter, even for seemingly analogous setups (e.g.~the EM field
of spinning charge, and the gravitational field of a spinning mass).
In this sense, these analogies have a different status compared to
the popular GEM analogy based on linearized theory, which, \emph{in
order to hold}, require a degree of similarity between the interactions
to which the former two are not bound.

What the tidal tensor formalism of Sec.~\ref{sec:Tidal tensor analogy},
together with the inertial fields formalism of Sec.~\ref{sec:3+1},
provide, is a ``set of tools'' to determine under which precise
conditions a similarity between the gravitational and electromagnetic
interactions may be expected.

The key differences between electromagnetic and gravitational tidal
tensors are: a) they do not exhibit, generically, the same symmetries;
b) gravitational tidal tensors are spatial whereas the electromagnetic
ones are not; c) electromagnetic tidal tensors are linear in the corresponding
fields, whereas the gravitational ones are not.

The electromagnetic tidal tensors, for a given observer, only have
the same symmetries and time-projections as the gravitational ones
when the Maxwell tensor is covariantly constant along the observer's
worldline; that is implied by Eqs.~(\ref{analogy}.8) and (\ref{analogy}.5)
of Table \ref{analogy}. This restricts the eligible setups to intrinsically
stationary fields (i.e., whose time-dependence, if it exists, can
be gauged away by a change of frame), and to a special class of observers
therein; for electromagnetic fields in flat spacetime, those observers
must be \emph{static} in the inertial frame where the fields are explicitly
time-independent. This is an important point that is worth discussing
in some detail. Consider the two basic analogous fields, the Coulomb
field of a point charge, and the Schwarzschild gravitational field.
Consider also in the latter observers $\mathcal{O}$ in circular motion:
4-velocity $U^{\alpha}=(U^{0},0,0,U^{\phi})$, angular velocity $\Omega=U^{\phi}/U^{0}$.
The worldlines of these observers are tangent to Killing vector fields:
$U^{\alpha}\parallel\xi^{\alpha};\ \mathcal{L}_{\xi}g_{\alpha\beta}=0$.
One can say (e.g.~\cite{Gravitation,SemerakStationary}) that they
see a constant spacetime geometry; for this reason they are called
``stationary observers''. Now consider observers in circular motion
around a Coulomb charge. Despite moving along worldlines tangent to
vector fields which are symmetries of the electromagnetic field: $U^{\alpha}\parallel\xi^{\alpha};\ \mathcal{L}_{\xi}F_{\alpha\beta}=0$,
the observers $U^{\alpha}$ do \emph{not} see a covariantly constant
field: $F_{\alpha\beta;\gamma}U^{\gamma}\ne0$, which by virtue of
Eqs.~(\ref{analogy}.5a), (\ref{analogy}.8a), implies that the electromagnetic
tidal tensors have an antisymmetric part (in particular the spatial
part $B_{[ij]}\ne0$), and thus means that they cannot be similar
to their gravitational counterparts. This is a natural consequence
of Maxwell's equations, and can be easily understood as follows. The
magnetic tidal tensor measured by $\mathcal{O}$ is a covariant derivative
of the magnetic field as measured in the inertial frame momentarily
comoving with it: $B_{\alpha\beta}\equiv\star F_{\alpha\gamma;\beta}U^{\gamma}={B_{\alpha;\beta}}|_{U=const}=(B_{MCRF})_{\alpha;\beta}$.
Now, $B_{[ij]}\ne0$ means that $\vec{B}_{MCRF}$ has a curl; which
is to be expected, since in the MCRF the electric field is time-dependent
(constant in magnitude but \emph{varying} in direction), which, by
virtue of Maxwell's equation $\nabla\times\vec{B}=\partial\vec{E}/\partial t=\gamma\vec{E}\times\vec{\Omega}$
(holding in the MCRF, and for which (\ref{analogy}.5a) is a covariant
form) induces a curl in $\vec{B}$.%

Even if one considers static observers in stationary fields, so that
the gravitational and electromagnetic tidal tensors have the same
symmetries, still one may not see a close similarity between the interactions.
The electromagnetic tidal tensors are linear in the electromagnetic
fields, and the latter themselves \emph{linear} in the electromagnetic
4-potential $A^{\alpha}=(\phi,\vec{A}$), whereas the gravitational
tidal tensors are \emph{non-linear} in the GEM fields, as shown by
Eqs.~(\ref{Eij})-(\ref{Hij}), the gravitomagnetic field $\vec{H}$
being itself \emph{non-linear} in the metric potentials $\Phi,\ \vec{\mathcal{A}}$.
This means that one can expect a similarity between tidal tensors
in two limiting cases --- linearized theory, and the ultrastationary
spacetimes considered in Sec.~\ref{sec:Ultrastationary}, where $\Phi=\vec{G}=0$,
and, therefore, cf.~Eqs.~(\ref{Hij}) and (\ref{eq:GEMFieldsQM}),
the exact gravitomagnetic tidal tensor is linear (both in the metric
and in the GEM fields): $\mathbb{H}_{\hat{i}\hat{j}}=-\tilde{\nabla}_{\hat{j}}H_{\hat{i}}/2=-\tilde{\nabla}_{\hat{j}}(\tilde{\nabla}\times\vec{\mathcal{A}})_{\hat{i}}/2$.
We have seen in Sec.~\ref{sec:Ultrastationary} that there is indeed
an exact mapping (via the Klein-Gordon equation) between the dynamics
in these spacetimes and an electromagnetic setup.

In what concerns concrete effects, the precise conditions (namely
regarding the time dependence of the fields) for occurrence of a gravito-electromagnetic
similarity are specific to the type of effect. For the tidal effects
(which imply physical, \emph{covariant} gravitational forces) such
as the force on a spinning particle or the worldline deviation of
two neighboring particles, it is the tidal tensors \emph{as measured
by the test particles} (4-velocity $U^{\alpha}$) that determine the
effects, cf.~Eqs.~(\ref{analogy}.1)-(\ref{analogy}.2); which means
that \emph{it is along the particle's worldline that the constancy
of the fields is required}. This basically implies that the similarity
only occurs at the instant when the particles are at rest in stationary
fields, so it does not hold in a dynamical situation. In the case
of the correspondence between the Lorentz force, Eq.~(\ref{eq:Lorentz}),
and the geodesic equation formulated as an inertial force (which is
a reference frame effect), we see from Eq.~(\ref{eq:Geo3+1}) that
the requirement is that the frame is rigid, i.e.~$\sigma_{\alpha\beta}=\theta=0$;
as explained in Sec.~\ref{sub: 1+3 Geodesics}, this amounts to saying
that the spatial part of the metric (in the coordinates associated
to such frame) must be \emph{time-independent}. This can also be stated
in the following manner, generalizing to the exact case the conclusion
obtained in \cite{PaperIAU} in the context of the post-Newtonian
approximation: in the case of the GEM analogy for the geodesic equation,
the \emph{stationarity of the fields is required in the observer's
frame} (not in the test particle's frame! The test particles can move
along arbitrary worldlines). As for the gyroscope ``precession''
(\ref{eq:Spin3+1v2}) and the correspondence with the precession of
a magnetic dipole (\ref{eq:DipolePrec}), there is no restriction
on the time dependence of the fields.

\section{Conclusion}

In this work we collected and further developed different gravito-electromagnetic
analogies existing in the literature, and clarified the connection
between them. A detailed summary of the material in this paper is
given in the introduction; herein we conclude by briefly summarizing
the main outcome of each approach, and their applicability. The analogies
split into two classes: physical and purely formal. In the second
category falls the analogy between the electric and magnetic parts
of the Weyl and Maxwell tensors, discussed in Sec \ref{sec:Weyl analogy}.
The physical analogies are divided into two classes: exact analogies,
and the best known post-Newtonian and linearized theory approaches.
Exact physical analogies are the analogy between the electromagnetic
fields and the inertial fields of Sec.~\ref{sec:3+1}, and the tidal
tensor analogy of Sec.~\ref{sec:Tidal tensor analogy}.

These analogies are useful from a practical point of view, as they
provide a familiar formalism and insight from electromagnetic phenomena
to describe otherwise more complicated gravitational problems. Indeed,
there is a number of fundamental equations, summarized in Table \ref{tab:SummaryAnalogies},
which can be obtained from the electromagnetic counterparts by simple
application of the analogy. But the existence of these analogies,
especially the exact, physical ones, is also interesting from the
theoretical point of view, unveiling intriguing similarities --- both
in the tidal tensor, and in the inertial field formalism, manifest
in Tables \ref{analogy} and \ref{Tab:analogy1+3}, respectively ---
and enlightening differences.

The tidal tensor formalism is primarily suited for a transparent comparison
between the two interactions, since it is based on mathematical objects
describing covariant \emph{physical} forces common to both theories.
Comparing the tidal tensors of both sides is straightforward from
Eqs. (\ref{analogy}.3)-(\ref{analogy}.7) of Table \ref{analogy}.
Fundamental differences are encoded in their symmetries and time projections;
herein we explored them in terms of the worldline deviation of (monopole)
test particles; and in the companion paper \cite{Gyros}, in terms
of the dynamics of spinning multipole test particles. The latter is
perhaps the most natural application of the formalism; but it can
be useful in many other applications, namely gravitational radiation
(as discussed in Sec.~\ref{sub:Matte's-equations-vs}), and whenever
one wishes to study the physical aspects of spacetime curvature.
\begin{table}[H]
\centering{}

\caption{\label{tab:SummaryAnalogies}What can be computed by direct application
of the GEM analogies}

\begin{tabular}{|l|c|}
\hline 
\raisebox{4.6ex}{}\raisebox{1ex}{~~~~~~~~~~~~~~~~~~~~~\textbf{Result}}  & \raisebox{1ex}{\textbf{Approach}}\tabularnewline
\hline 
\hline 
$\bullet$\raisebox{4ex}{}~Geodesic deviation equation (\ref{analogy}.1b)
of Table \ref{analogy}:  & \tabularnewline
~~-Replacing $\{q,E_{\alpha\beta}\}\rightarrow\{m,-\mathbb{E}_{\alpha\beta}\}$
in (\ref{analogy}.1a).  & \tabularnewline
$\bullet$\raisebox{4ex}{}~Force on a gyroscope (\ref{analogy}.1b):  & \textbf{Tidal tensor analogy}\tabularnewline
~~-Replacing $\{\mu^{\alpha},B_{\alpha\beta}\}\rightarrow\{S^{\alpha},-\mathbb{H}_{\alpha\beta}\}$
in (\ref{analogy}.1a).  & (Exact, general results)\tabularnewline
$\bullet$\raisebox{4ex}{}~Gravitational field equations (\ref{analogy}.3b)-(\ref{analogy}.4b),
(\ref{analogy}.6b)-(\ref{analogy}.7b):  & \tabularnewline
~~-Replacing $\{E_{\alpha\beta},B_{\alpha\beta}\}\rightarrow\{\mathbb{E}_{\alpha\beta},\mathbb{H}_{\alpha\beta}\}$
in Eqs.~(\ref{Et})-(\ref{Banti}),  & \tabularnewline
\raisebox{4ex}{}\raisebox{1.7ex}{~~\emph{~and} $\rho_{c}\rightarrow2\rho+T_{\ \alpha}^{\alpha}$
in (\ref{Et}), $j^{\alpha}\rightarrow2J^{\alpha}$ in (\ref{Banti}).}  & \tabularnewline
\hline 
$\bullet$\raisebox{4ex}{}~Geodesic Equation (\ref{geoQM}) (\emph{stationary}
fields)  & \tabularnewline
~~-Replacing $\{q,\vec{E},\vec{B}\}\rightarrow\{m,\vec{G},\vec{H}\}$
in (\ref{eq:Lorentz}), multiplying by $\gamma$.  & \tabularnewline
$\bullet$\raisebox{4ex}{}~Gyroscope ``precession'' Eq.~(\ref{eq:Spin3+1v2})
(\emph{arbitrary} fields):  & \tabularnewline
~~-Replacing $\{\vec{\mu},\vec{B}\}\rightarrow\{\vec{S},\vec{H}/2\}$
in (\ref{eq:DipolePrec}).  & \textbf{Inertial ``GEM fields'' analogy}\tabularnewline
$\bullet$\raisebox{4ex}{}~Force on gyroscope Eq.~(\ref{eq:FNatario0})
(\emph{stationary} fields,  & (Exact results, require special frames)\tabularnewline
~~\emph{~}particle's worldline tangent to time-like Killing vector):  & \tabularnewline
~~-Replacing $\{\vec{\mu},\vec{E},\vec{B}\}\rightarrow\{\vec{S},\vec{G},\vec{H}/2\}$
in (\ref{eq:FEM_QM0}), factor  & \tabularnewline
\raisebox{4ex}{}\raisebox{1.7ex}{~~~of 2 in the last term.}  & \tabularnewline
\hline 
$\bullet$\raisebox{4ex}{}~Higher order field equations (\ref{tab:MatteMaxwell}.1b)-(\ref{tab:MatteMaxwell}.4b):  & \tabularnewline
~~-Replacing $\{\vec{E},\vec{B}\}\rightarrow\{\mathbb{E}_{ij},\mathbb{H}_{ij}\}$
in Eqs.~(\ref{tab:MatteMaxwell}.1a)-(\ref{tab:MatteMaxwell}.4a).  & \textbf{Weyl-Maxwell tensors analogy}\tabularnewline
$\bullet$~\raisebox{4ex}{}Equations of gravitational waves (\ref{tab:MatteMaxwell}.5b)-(\ref{tab:MatteMaxwell}.6b):  & \raisebox{1.7ex}{(Results for linearized theory)}\tabularnewline
\raisebox{4ex}{}\raisebox{1.7ex}{~~-Replacing $\{\vec{E},\vec{B}\}\rightarrow\{\mathbb{E}_{ij},\mathbb{H}_{ij}\}$
in Eqs.~(\ref{tab:MatteMaxwell}.5a)-(\ref{tab:MatteMaxwell}.6a).}  & \tabularnewline
\hline 
\end{tabular}
\end{table}

The analogy based on inertial GEM fields from the 1+3 formalism, Sec.
\ref{sec:3+1}, is a very powerful formalism, with vast applications;
especially in the case of stationary spacetimes, where for arbitrarily
strong fields the equation for geodesics is cast in a form similar
to Lorentz force; many other effects related to frame-dragging can
be treated exactly with the GEM fields: gyroscope ``precession''
\cite{Gyros,The many faces,GEM User Manual,JantzenThomas,Natario},
the Sagnac effect \cite{RizziRuggieroSagnac}, the Faraday rotation
\cite{ZonozPRD}, the force on a gyroscope (Sec.~\ref{sub:Force-quasi-Maxwell}
and \cite{Natario}; note however that it is not as general as the
tidal tensor formulation of the same force); and other applications,
such as the matching of stationary solutions \cite{MenaNatario},
or describing the ``hidden momentum'' of spinning particles \cite{Gyros}.
The general formulation of GEM fields in Sec.~\ref{sec:3+1}, applying
to arbitrary fields and frames, extends the realm of applicability
of this formalism.

The well known analogies between electromagnetism and post-Newtonian
and linearized gravity, follow as a limiting case of the exact approach
in Sec.~\ref{sec:3+1}. In the case of the tidal effects, they can
be seen also as a limiting case of the tidal tensor analogy of Sec.
\ref{sec:Tidal tensor analogy} (in the sense that for weak, \emph{time-independent
fields,} the gravitational tidal tensors reduce to derivatives of
the GEM fields). Realizing this, and understanding the conditions
under which linear GEM is obtained from the rigorous, exact approaches,
is important for a correct interpretation of the physical meaning
of the quantities involved, which is not clear in the usual derivations
in the literature (this is especially the case for many works on linear
GEM), and thus prone to misconceptions (see in this respect \cite{CHPreprint,CHPRD}.
On the other hand, linear GEM is the most important in the context
of experimental physics, as it pertains all gravitomagnetic effects
detected to date \cite{Ciufolini Lageos,Ciufolini Nature Review,GPB,Nordtvedt1988,MurphyNordtvedtPRL1,SoffelKlioner},
and the ones we hope to detect in the near future \cite{LARES}.

As for the analogy between the electric and magnetic parts of the
Weyl and Maxwell tensors, its most important application is gravitational
radiation, where it provides equations for the propagation of \emph{tensors
of physical forces} (not components of the metric tensor, as in the
more usual approaches, which are pure gauge fields), with direct translation
into physical effects (via the tidal tensor formalism of Sec.~\ref{sec:Tidal tensor analogy}).
This analogy has been used to address the fundamental questions of
the content of gravitational waves, and the ``energy'' of the gravitational
field. Namely, to propose \emph{covariant}, \emph{local} local quantities
for the gravitational field analogous to the electromagnetic field
energy and momentum densities --- the ``super-energy'' and ``super-momentum''
densities encoded in the Bel tensor. The existing criteria for radiative
states \cite{FerrandoSaezSE}, states of intrinsic radiation \cite{bel,AlfonsoSE}
or pure radiation (\cite{BelSecondCrit}, see also \cite{Zakharov}
p. 53), are also solely driven by this analogy. It is also useful
for the understanding of the quadratic invariants of the curvature
tensor; indeed, it will be shown elsewhere \cite{PaperInvariantes}
that using the two approaches together --- the formal analogies of
Sec. \ref{sec:Weyl analogy} to gain insight into the invariant structure,
and the tidal tensor analogy as a physical guiding principle --- one
can explain, in the astrophysical applications of current experimental
interest, the significance of the curvature invariants and the implications
on the motion of test particles.

\section*{Acknowledgments}

We thank J. Penedones, the anonymous referees and A. Editor for useful
comments and remarks; we also thank A. García-Parrado and J. M. M.
Senovilla for correspondence and useful discussions.

\appendix

\section{Inertial Forces --- simple examples in flat spacetime\label{sub:Simple-examples-in_flat}}

\begin{figure}
\includegraphics[width=0.95\textwidth]{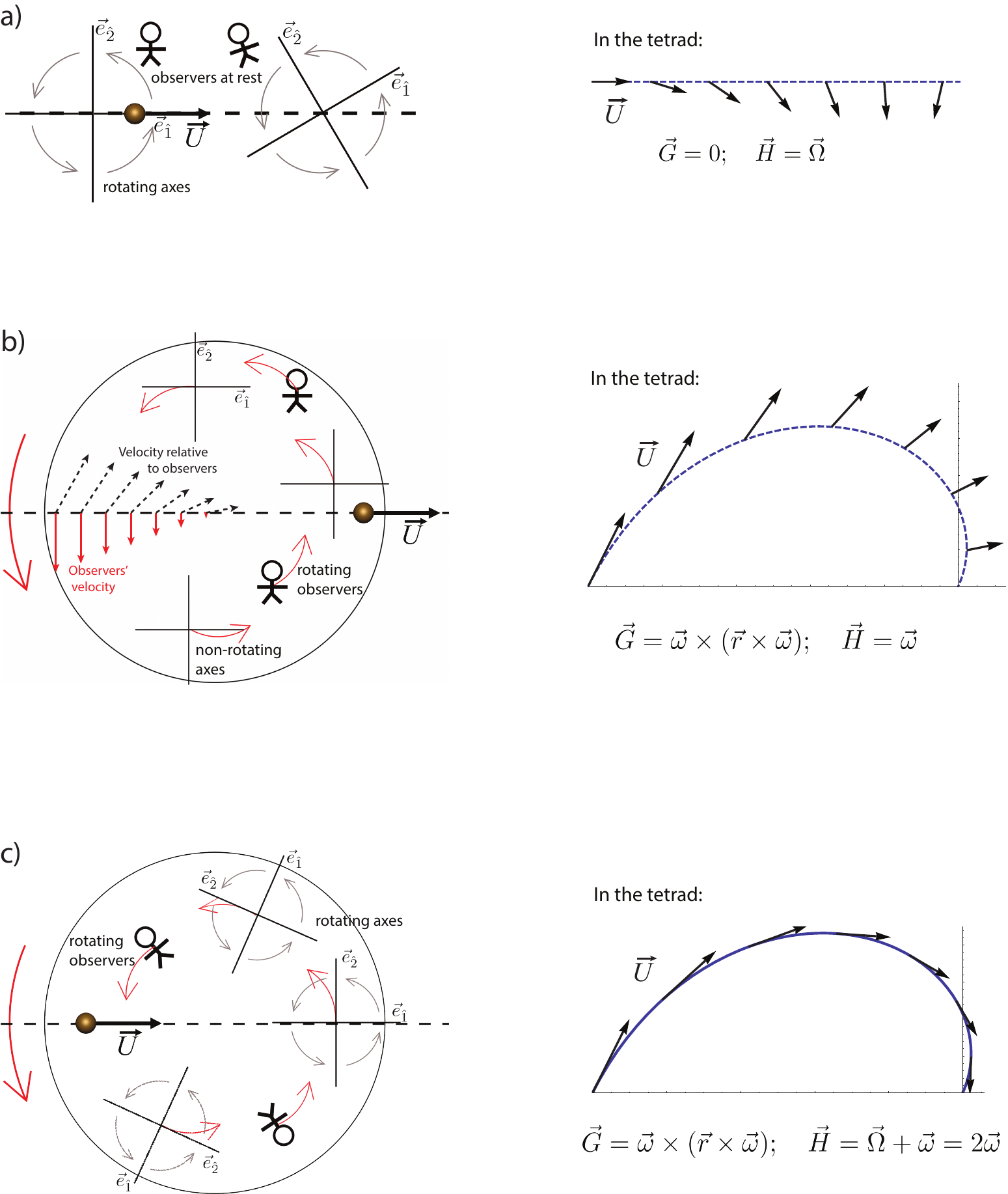}

\caption{\label{fig:Inertial}A test particle in uniform motion in flat spacetime
from the point of view of three different frames: a) a frame composed
of observers at rest, but carrying spatial triads that rotate with
uniform angular velocity $\vec{\Omega}$; b) a frame consisting of
a congruence of rigidly rotating observers (vorticity $\vec{\omega}$),
but each of them carrying a non-rotating spatial triad (i.e., that
undergoes Fermi-Walker transport); c) a rigidly rotating frame (a
frame adapted to a congruence of rigidly rotating observers); the
spatial triads co-rotate with the congruence, $\vec{\Omega}=\vec{\omega}$.
\emph{Note:} by observer's rotation we mean their circular motion
around the center; and by axes rotation we mean their rotation (relative
to FW transport) about the local tetrad's origin.}
\end{figure}

In Sec.~\ref{sub: 1+3 Geodesics} we have seen that the inertial
forces felt in a given frame arise from two \emph{independent} contributions
of different origin: the kinematics of the observer congruence (that
is, from the derivatives of the temporal basis vector of the frame,
$\mathbf{e}_{\hat{0}}=\mathbf{u}$, where $\mathbf{u}$ is the observers'
4-velocity), and the transport law for the spatial triads $\mathbf{e}_{\hat{i}}$
along the congruence. In order to illustrate these concepts with simple
examples, we shall consider, in flat spacetime, the straightline geodesic
motion of a free test particle (4-velocity $\mathbf{U}$), from the
point of view of three distinct frames: a) a frame whose time axis
is the 4-velocity of a congruence of observers at rest, but whose
spatial triads rotate uniformly with angular velocity $\vec{\Omega}$;
b) a frame composed of a congruence of rigidly rotating observers
(vorticity $\vec{\omega}$), but carrying Fermi-Walker transported
spatial triads ($\vec{\Omega}=0$); c) a rigidly rotating frame, that
is, a frame composed of a congruence of rigidly rotating observers,
carrying spatial triads co-rotating with the congruence $\vec{\Omega}=\vec{\omega}$
(i.e., ``adapted'' to the congruence, see Sec.~\ref{sub:The-reference-frame}).
This is depicted in Fig.~\ref{fig:Inertial}.

In the first case there we have a vanishing gravitoelectric field
$\vec{G}=0$, and a gravitomagnetic field $\vec{H}=\vec{\Omega}$
arising solely from the rotation (with respect to Fermi-Walker transport)
of the spatial triads; thus the only inertial force present is the
gravitomagnetic force $\vec{F}_{{\rm GEM}}=\gamma\vec{U}\times\vec{\Omega}$,
cf. Eq. (\ref{eq:Geo3+1}), with $\gamma\equiv-U^{\alpha}u_{\alpha}$.
In the frame b), there is a gravitoelectric field $\vec{G}=\vec{\omega}\times(\vec{r}\times\vec{\omega})$
due the observers acceleration, and also a gravitomagnetic field $\vec{H}=\vec{\omega}$,
which originates solely from the vorticity of the observer congruence.
That is, there is a gravitomagnetic force $\gamma\vec{U}\times\vec{\omega}$
which reflects the fact that the relative velocity $v^{\alpha}=U^{\alpha}/\gamma-u^{\alpha}$
(or $\vec{v}=\vec{U}/\gamma$, in the observer's frame, where $\vec{u}=0$)
between the test particle and the observer it is passing by changes
in time. The total inertial forces are in this frame 
\[
\vec{F}_{{\rm GEM}}=\gamma\left[\gamma\vec{\omega}\times(\vec{r}\times\vec{\omega})+\vec{U}\times\vec{\omega}\right].
\]

In the frame c), which is the relativistic version of the classical
\emph{rigid} rotating frame, one has the effects of a) and b) combined:
a gravitoelectric field $\vec{G}=\vec{\omega}\times(\vec{r}\times\vec{\omega})$,
plus a gravitomagnetic field $\vec{H}=\vec{\omega}+\vec{\Omega}=2\vec{\omega}$,
the latter leading to the gravitomagnetic force $2\gamma\vec{U}\times\vec{\omega}$,
which is the relativistic version of the well known Coriolis acceleration,
e.g.~\cite{Goldstein}. The total inertial force is in this frame
\[
\vec{F}_{{\rm GEM}}=\gamma\left[\gamma\vec{\omega}\times(\vec{r}\times\vec{\omega})+2\vec{U}\times\vec{\omega}\right]
\]
which is the relativistic generalization of the inertial force in
e.g.~Eq.~(4.91) of \cite{Goldstein}. Moreover, in this case, as
discussed in Secs. \ref{sub:The-derivative-operator} and \ref{sub:quasi-Maxwell},
the $\Gamma_{\hat{j}\hat{k}}^{\hat{i}}$ in Eq. (\ref{eq:FGEM_Def})
are the connection coefficients of the (Levi-Civita) 3-D covariant
derivative with respect to the metric $h_{ij}$ (defined by Eq. (\ref{eq:Stationary}))
defined on the space manifold associated to the quotient of the spacetime
by the congruence; $\vec{U}$ is the vector \emph{tangent to the 3-D
curve} (see Fig. \ref{fig:Inertial}c) obtained by projecting the
particle's worldline on the space manifold, and $\ \vec{F}_{{\rm GEM}}=\tilde{D}\vec{U}/d\tau$
is simply the covariant 3-D acceleration of that curve.

\subsection*{}

\end{document}